\documentclass[12pt]{article}
\usepackage{amsmath,amsthm}
\usepackage{eufrak}
\usepackage{pb-diagram,lamsarrow,pb-lams}

\newcommand{\ar}{\renewcommand{\arraystretch}{1}} % 1.0 % 0.6
\DeclareMathAlphabet{\bb}{U}{msb}{m}{n} \gdef\C{\bb C} \gdef\dZ{\bb
Z}   \gdef\dS{\bb S} \gdef\R{\bb R}
\gdef\K{\bb K} \gdef\BH{\bb H} \gdef\F{\bb F} 

 \DeclareMathOperator{\spin}{{\bf
Spin}} 

\DeclareMathOperator{\Id}{Id} 
  \DeclareMathOperator{\sExt}{{\sf CPT}}
 
\DeclareMathOperator{\bA}{{\bf A}}  
\DeclareMathOperator{\Sym}{Sym} 
 \DeclareMathOperator{\rot}{rot}
\DeclareMathOperator{\Ext}{CPT} 
 
 \DeclareMathOperator{\SL}{SL}
\DeclareMathOperator{\SO}{SO}\DeclareMathOperator{\SU}{SU}
 \DeclareMathOperator{\GO}{O}

\newcommand{\s}{\!}

\newcommand{\cA}{\mathcal{A}}

\newcommand{\cE}{\mathcal{E}}
\newcommand{\cG}{\mathcal{G}}
\newcommand{\bcE}{\boldsymbol{\mathcal{E}}}
\newcommand{\bcK}{\boldsymbol{\mathcal{K}}}

\newcommand{\cP}{{\cal P}}

\newcommand{\cH}{{\cal H}}

\newcommand{\cS}{\mathcal{S}}

\newcommand{\sA}{{\sf A}}
\newcommand{\sB}{{\sf B}}
\newcommand{\sI}{{\sf I}}

\newcommand{\sW}{{\sf W}}
\newcommand{\sE}{{\sf E}}
\newcommand{\sC}{{\sf C}}
\newcommand{\sF}{{\sf F}}
\newcommand{\sH}{{\sf H}}

\newcommand{\sS}{{\sf S}}

\newcommand{\sX}{{\sf X}}
\newcommand{\sY}{{\sf Y}}
\newcommand{\sK}{{\sf K}}

\newcommand{\bsH}{{\boldsymbol{\sf H}}}
\newcommand{\sL}{\Lambda}

\newcommand{\bP}{{\bf P}}
\newcommand{\bN}{{\bf N}}

\newcommand{\fG}{\mathfrak{G}}

\newcommand{\fg}{\mathfrak{g}}

\newcommand{\cl}{C\kern -0.2em \ell}

\newcommand{\bO}{\mbox{\bf O}}

\newcommand{\ld}{\left[}
\newcommand{\rd}{\right]}

\newtheorem{thm}{Theorem}
\begin{document}
\title{Spinor Structure and Internal Symmetries}
\author{V.~V. Varlamov\thanks{Siberian State Industrial University,
Kirova 42, Novokuznetsk 654007, Russia, e-mail:
vadim.varlamov@mail.ru}}
\date{}
\maketitle
\begin{abstract}
 Spinor structure and internal symmetries are considered within one theoretical framework based on the generalized spin and abstract Hilbert space. Complex momentum is understood as a generating kernel of the underlying spinor structure. It is shown that tensor products of biquaternion algebras are associated with the each irreducible representation of the Lorentz group. Space-time discrete symmetries $P$, $T$ and their combination $PT$ are generated by the fundamental automorphisms of this algebraic background (Clifford algebras). Charge conjugation $C$ is presented by a pseudoautomorphism of the complex Clifford algebra. This description of the operation $C$ allows one to distinguish charged and neutral particles including particle-antiparticle interchange and truly neutral particles. Spin and charge multiplets, based on the interlocking representations of the Lorentz group, are introduced. A central point of the work is a correspondence between Wigner definition of elementary particle as an irreducible representation of the Poincar\'{e} group and $\SU(3)$-description (quark scheme) of the particle as a vector of the supermultiplet (irreducible representation of $\SU(3)$). This correspondence is realized on the ground of a spin-charge Hilbert space. Basic hadron supermultiplets of $\SU(3)$-theory (baryon octet and two meson octets) are studied in this framework. It is shown that quark phenomenologies are naturally incorporated into presented scheme. The relationship between mass and spin is established. The introduced spin-mass formula and its combination with Gell-Mann--Okubo mass formula allows one to take a new look at the problem of mass spectrum of elementary particles.
\end{abstract}
{\bf Keywords}: spinor structure, internal symmetries, Clifford algebras, quarks, mass spectrum
\section{Introduction}
One of the most longstanding problem in theoretical physics is the unification of space-time and internal symmetries. Space-time symmetries (including space inversion $P$, time reversal $T$ and their combination $PT$), generated by the Poincar\'{e} group, are treated as absolutely exact transformations of space-time continuum. On the other hand, charge conjugation $C$ presents the first transformation which is not space-time symmetry (but not approximate). This operation closely relates with a complex conjugation of the Lorentz group representations. In some sense, $C$ can be treated as an internal symmetry. A wide variety of internal symmetries (which all approximate, except the color) we have from the quark phenomenology based on the $\SU(N)$-theories. The first quark model, including light $u$, $d$ and $s$ quarks, is constructed within $\SU(3)$-theory. As a rule, particles ($qqq$-baryons and $q\overline{q}$-mesons), which unified into supermultiplets of $\SU(3)$ group, have different masses. For that reason flavor $\SU(3)$-theory is an approximate symmetry. The addition of the $c$ quark (charm) to the light $u$, $d$, $s$ quarks extends the flavor $\SU(3)$-symmetry to $\SU(4)$. Due to the large mass of the $c$ quark, $\SU(4)$-symmetry is much more strongly broken than the $\SU(3)$ of the three light quarks. The addition of the $b$ quark (bottom) extends the quark model to $\SU(5)$ with a very approximate symmetry. It is obvious that the next step in extending the flavor symmetry to $\SU(6)$ is senseless, since the existence of baryons with a $t$ quark (top) is very unlikely due to the short lifetime of the top-quark. However, there is an idea (which takes its origin from sixties) to consider exact space-time transformations and approximate internal symmetries within one theoretical framework.

As is known, elementary particles can be grouped into multiplets corresponding to irreducible representations of so-called \emph{algebras of internal symmetries} (for example, multiplets of the isospin algebra $\mathfrak{su}(2)$ or multiplets of the algebra $\mathfrak{su}(3)$). Particles from the fixed multiplet have the same parity and the same spin, but they can be distinguished by the masses. Thus, an algebra of the most general symmetry cannot be defined as a direct sum $P\oplus S$ of two ideals, where $P$ is the Poincar\'{e} algebra and $S$ is the algebra of internal symmetry, since in contrary case all the particle masses of the multiplet should be coincide with each other. One possible way to avoid this obstacle is the searching of a more large algebra $L$ that contains $P$ and $S$ as subalgebras in such way that if only one generator from $S$ commutes with $P$. In this case a mass operator $p_\mu p^\mu$ is not invariant of the large algebra $L$. Hence it follows that a Cartan subalgebra $H$ should be commute with $P$, since eigenvalues of the basis elements $H_i$ of $H\in S$ are used for definition of the states in multiplets (hypercharge, $I_3$-projection of the isospin and so on). All these quantum numbers are invariant under action of the Poincar\'{e} group $\cP$. Therefore, if $L$ is the Lie algebra spanned on the basis elements of the Poincar\'{e} algebra $P$ and on the basis elements of the semisimple algebra $S$, then $L$ can be defined as a direct sum of two ideals, $L=P\oplus S$, only in the case when $\left[P,H\right]=0$, where $H$ is the Cartan subalgebra of $S$. When the algebra of internal symmetry is an arbitrary compact algebra $K$ (for example, $\mathfrak{su}(2)$, $\mathfrak{su}(3)$, $\ldots$, $\mathfrak{su}(n)$, $\ldots$) we have the following direct sum: $K=N\oplus S=N\oplus S_1\oplus S_2\oplus\ldots\oplus S_n$, where $N$ is a center of the algebra $K$, $S$ is a semisimple algebra, and $S_i$ are simple algebras (see \cite{BR77}). In this case the large algebra $L$ can be defined also as a direct sum of two ideals, $L=P\oplus K$, when $\left[P,C\right]=0$, where $C$ is a maximal commutative subalgebra in $K$. Restrictions $\left[P,H\right]=0$ and $\left[P,C\right]=0$ on the algebraic level induce restrictions on the group level. So, if $\cG$ is an arbitrary Lie group, and $\cS$ (group of internal symmetry) and $\cP=T_4\odot\SL(2,\C)$ (Poincar\'{e} group) are the subalgebras of $\cG$ such that any $g\in\cG$ has an unique decomposition in the product $g=sp$, $s\in\cS$, $p\in\cP$, and if there exists one element $g\in\cP$, $g\not\in T_4$, such that $s^{-1}gs\in\cP$ for all $s\in\cS$, then $\cG=\cP\otimes\cS$ \cite{Mic64}. The restrictions $L=P\oplus S$ (algebraic level) and $\cG=\cP\otimes\cS$ (group level) on unification of space-time and internal symmetries were formulated in sixties in the form of so-called \textit{no-go theorems}. One of the most known no-go theorem is a Coleman-Mandula theorem \cite{CM67}. The group $\cG$ is understood in \cite{CM67} as a symmetry group of $S$ matrix. The Coleman-Mandula theorem asserts that $\cG$ is necessarily locally isomorphic to the direct product of an internal symmetry group and the Poincar\'{e} group, and this theorem is not true if the local isomorphism ($\cG\simeq\cP\otimes\cS$) is replaced by a global isomorphism. In 1966, Pais \cite{Pais} wrote: "Are there any alternatives left to the internal symmetry$\otimes$Poincar\'{e} group picture in the face of the no-go theorems?".

All the no-go theorems suppose that space-time continuum is a fundamental level of reality. However, in accordance with Penrose twistor programme \cite{Pen77,PM72}, space-time continuum is a derivative construction with respect to \textit{underlying spinor structure}. Spinor structure contains in itself pre-images of all basic properties of classical space-time, such as dimension, signature, metrics and many other. In parallel with twistor approach, decoherence theory \cite{JZKGKS} claims that in the background of reality we have a \textit{nonlocal quantum substrate} (quantum domain), and all visible world (classical domain) arises from quantum domain in the result of decoherence process \cite{Zur03,Zur03b}. In this context space-time should be replaced by the spinor structure (with the aim to avoid restrictions of the no-go theorems), and all the problem of unification of space-time and internal symmetries should be transferred to a much more deep level of the quantum domain (underlying spinor structure).

This article presents one possible way towards a unification of spinor structure and internal symmetries. At first, according to Wigner \cite{Wig39} an elementary particle is defined by an irreducible unitary \emph{representation} of the Poincar\'{e} group $\cP$. On the other hand, in accordance with $\SU(3)$-theory (and also flavor-spin $\SU(6)$-theory) an elementary particle is described by a \emph{vector} of irreducible representation of the group $\SU(3)$ ($\SU(6)$, $\ldots$, $\SU(N)$, $\ldots$). For example, in a so-called `eightfold way' \cite{GN64} hadrons (baryons and mesons) are grouped within eight-dimensional regular representation $\Sym^0_{(1,1)}$ of $\SU(3)$. With the aim to make a bridge between these interpretations (between \emph{representations} of $\cP$ and \emph{vectors} of $\Sym^0_{(1,1)}$, $\Sym^0_{(1,4)}$, $\ldots$) we introduce a \emph{spin-charge} Hilbert space $\bsH^S\otimes\bsH^Q\otimes\bsH_\infty$, where the each vector of this space presents an irreducible representation of the group $\SL(2,\C)$. At this point, charge characteristics of the particles are described by a pseudoautomorphism of the spinor structure. An action of the pseudoautomorphism allows one to distinguish charged and neutral particles within separated charge multiplets. On the other hand, spin characteristics of the particles are described via a generalized notion of the spin based on the spinor structure and irreducible representations of the group $\SL(2,\C)$. The usual definition of the spin is arrived at the restriction of $\SL(2,\C)$ to the subgroup $\SU(2)$. This construction allows us to define an action and representations of internal symmetry groups $\SU(2)$, $\SU(3)$, $\ldots$ in the space $\bsH^S\otimes\bsH^Q\otimes\bsH_\infty$ by means of a central extension. In this context the $\SU(3)$-theory is considered in detail. The fermionic and bosonic octets, which compound the eightfold way of $\SU(3)$, and also their $\SU(3)/\SU(2)$-reductions into isotopic multiplets are reformulated without usage of the quark scheme. It is well known that the quark model does not explain a mass spectrum of elementary particles. The Gell-Mann--Okubo mass formula explains only mass splitting within supermultiplets of the $\SU(3)$-theory, namely, hypercharge mass splitting within supermultiplets and charge splitting within isotopic multiplets belonging to a given supermultiplet. The analogous situation takes place in the case of B\'{e}g-Singh mass formula of the flavor-spin $\SU(6)$-theory. On the other hand, in nature we see a wide variety of baryon octets (see, for example, Particle Data Group: pdg.lbl.gov), where mass distances between these octets are not explained by the quark model. Hence it follows that mass spectrum of elementary particles should be described by a such parameter that defines relation between mass and spin. With this aim in view we introduce a relation between mass and spin in the section 2. It is shown that for representations $(l,\dot{l})$ of the Lorentz group the mass is proportional to $(l+1/2)(\dot{l}+1/2)$, and particles with the same spin but distinct masses are described by different representations of Lorentz group. The introduced mass formula defines \textit{basic mass terms}, and detailed mass spectrum is accomplished by charge and hypercharge mass splitting via the Gell-Mann--Okubo mass formula (like Zeeman effect in atomic spectra).
\section{Complex momentum}
A universal covering of the proper orthochronous Lorentz group $\SO_0(1,3)$
(rotation group of the Minkowski space-time $\R^{1,3}$)
is the spinor group
\[\ar
\spin_+(1,3)\simeq\left\{\begin{pmatrix} \alpha & \beta \\ \gamma &
\delta
\end{pmatrix}\in\C_2:\;\;\det\begin{pmatrix}\alpha & \beta \\ \gamma & \delta
\end{pmatrix}=1\right\}=\SL(2,\C).
\]
%В свою очередь, из (\ref{Spin+}) и (\ref{Spin}) (см. приложение А)
%следует, что $\spin_+(1,3)$ определяется в рамках специальной группы
%Липшица $\Lip^+_{1,3}=\Lip_{1,3}\cap\cl^+_{1,3}$, следовательно,
%$\spin_+(1,3)\in\cl^+_{1,3}$. С другой стороны, имеет место
%изоморфизм $\cl^+_{1,3}\simeq\cl_{3,0}\simeq\C(2)$, где $\C(2)$ --
%алгебра комплексных бикватернионов.

Let $\fg\rightarrow T_{\fg}$ be an arbitrary linear
representation of the proper orthochronous Lorentz group
$\SO_0(1,3)$ and let $\sA_i(t)=T_{a_i(t)}$ be an infinitesimal
operator corresponding to the rotation $a_i(t)\in\SO_0(1,3)$.
Analogously, let $\sB_i(t)=T_{b_i(t)}$, where $b_i(t)\in\fG_+$ is
the hyperbolic rotation. The elements $\sA_i$ and $\sB_i$ form a basis of Lie algebra
$\mathfrak{sl}(2,\C)$ and satisfy the relations
\begin{equation}\label{Com1}
\left.\begin{array}{lll} \ld\sA_1,\sA_2\rd=\sA_3, &
\ld\sA_2,\sA_3\rd=\sA_1, &
\ld\sA_3,\sA_1\rd=\sA_2,\\[0.1cm]
\ld\sB_1,\sB_2\rd=-\sA_3, & \ld\sB_2,\sB_3\rd=-\sA_1, &
\ld\sB_3,\sB_1\rd=-\sA_2,\\[0.1cm]
\ld\sA_1,\sB_1\rd=0, & \ld\sA_2,\sB_2\rd=0, &
\ld\sA_3,\sB_3\rd=0,\\[0.1cm]
\ld\sA_1,\sB_2\rd=\sB_3, & \ld\sA_1,\sB_3\rd=-\sB_2, & \\[0.1cm]
\ld\sA_2,\sB_3\rd=\sB_1, & \ld\sA_2,\sB_1\rd=-\sB_3, & \\[0.1cm]
\ld\sA_3,\sB_1\rd=\sB_2, & \ld\sA_3,\sB_2\rd=-\sB_1. &
\end{array}\right\}
\end{equation}
Let us consider the operators
\begin{gather}
\sX_l=\frac{1}{2}i(\sA_l+i\sB_l),\quad\sY_l=\frac{1}{2}i(\sA_l-i\sB_l),
\label{SL25}\\
(l=1,2,3).\nonumber
\end{gather}
Using the relations (\ref{Com1}), we find that
\begin{equation}\label{Com2}
\ld\sX_k,\sX_l\rd=i\varepsilon_{klm}\sX_m,\quad
\ld\sY_l,\sY_m\rd=i\varepsilon_{lmn}\sY_n,\quad \ld\sX_l,\sY_m\rd=0.
\end{equation}
%Let $H$ be an \emph{energy} operator defined in the Hilbert space $\sH_\infty$. All eigenvector subspaces $\sH_E$ of $\sH_\infty$ are finite-dimensional. Further, we suppose that there is a \emph{local representation} $P$ of the group $\spin_+(1,3)$ in $\sH_\infty$, that is, all the representing operators of $P$ commute with $H$. In particular, the each eigenvector subspace $\sH_E$ of the energy operator is invariant with respect to the operators $\sX_l$, $\sY_l$.

Further, introducing generators of the form
\begin{equation}\label{SL26}
\left.\begin{array}{cc}
\sX_+=\sX_1+i\sX_2, & \sX_-=\sX_1-i\sX_2,\\[0.1cm]
\sY_+=\sY_1+i\sY_2, & \sY_-=\sY_1-i\sY_2,
\end{array}\right\}
\end{equation}
we see that
\[
\ld\sX_3,\sX_+\rd=\sX_+,\quad\ld\sX_3,\sX_-\rd=-\sX_-,\quad\ld\sX_+,\sX_-\rd=2\sX_3,
\]
\[
\ld\sY_3,\sY_+\rd=\sY_+,\quad\ld\sY_3,\sY_-\rd=-\sY_-,\quad\ld\sY_+,\sY_-\rd=2\sY_3.
\]
In virtue of commutativity of the relations (\ref{Com2})
a space of an irreducible finite-dimensional representation of the
group $\SL(2,\C)$ can be spanned on the totality of
$(2l+1)(2\dot{l}+1)$ basis vectors $\mid\!
l,m;\dot{l},\dot{m}\rangle$, where $l,m,\dot{l},\dot{m}$ are integer
or half-integer numbers, $-l\leq m\leq l$, $-\dot{l}\leq
\dot{m}\leq \dot{l}$. Therefore,
\begin{eqnarray}
&&\sX_-\mid l,m;\dot{l},\dot{m}\rangle= \sqrt{(l+m)(l-m+1)}\mid
l,m-1,\dot{l},\dot{m}\rangle
\;\;(m>-l),\nonumber\\
&&\sX_+\mid l,m;\dot{l},\dot{m}\rangle= \sqrt{(l-m)(l+m+1)}\mid
l,m+1;\dot{l},\dot{m}\rangle
\;\;(m<l),\nonumber\\
&&\sX_3\mid l,m;\dot{l},\dot{m}\rangle=
m\mid l,m;\dot{l},\dot{m}\rangle,\nonumber\\
&&\sY_-\mid l,m;\dot{l},\dot{m}\rangle=
\sqrt{(\dot{l}+\dot{m})(\dot{l}-\dot{m}+1)}\mid
l,m;\dot{l},\dot{m}-1
\rangle\;\;(\dot{m}>-\dot{l}),\nonumber\\
&&\sY_+\mid l,m;\dot{l},\dot{m}\rangle=
\sqrt{(\dot{l}-\dot{m})(\dot{l}+\dot{m}+1)}\mid
l,m;\dot{l},\dot{m}+1
\rangle\;\;(\dot{m}<\dot{l}),\nonumber\\
&&\sY_3\mid l,m;\dot{l},\dot{m}\rangle= \dot{m}\mid
l,m;\dot{l},\dot{m}\rangle.\label{Waerden}
\end{eqnarray}
From the relations (\ref{Com2}) it follows that each of the sets of
infinitesimal operators $\sX$ and $\sY$ generates the group $\SU(2)$
and these two groups commute with each other. Thus, from the
relations (\ref{Com2}) and (\ref{Waerden}) it follows that the group
$\SL(2,\C)$, in essence, is equivalent locally to the group
$\SU(2)\otimes\SU(2)$.
\subsection{Representations of $\SL(2,\C)$}
As is known \cite{GMS}, finite-dimensional (spinor) representations
of the group $\SO_0(1,3)$ in the space of symmetrical polynomials
$\Sym_{(k,r)}$ have the following form:
\begin{equation}\label{TenRep}
T_{\fg}q(\xi,\overline{\xi})=(\gamma\xi+\delta)^{l_0+l_1-1}
\overline{(\gamma\xi+\delta)}^{l_0-l_1+1}q\left(\frac{\alpha\xi+\beta}{\gamma\xi+\delta};
\frac{\overline{\alpha\xi+\beta}}{\overline{\gamma\xi+\delta}}\right),
\end{equation}
where $k=l_0+l_1-1$, $r=l_0-l_1+1$, and the pair $(l_0,l_1)$ defines
some representation of the group $\SO_0(1,3)$ in the
Gel'fand-Naimark basis.
The relation between the numbers $l_0$, $l_1$ and the number $l$
(the weight of representation in the basis (\ref{Waerden})) is given
by the following formula:
\[
(l_0,l_1)=\left(l,l+1\right).
\]
Whence it immediately follows that
\begin{equation}\label{RelL}
l=\frac{l_0+l_1-1}{2}.
\end{equation}
As is known \cite{GMS}, if an irreducible representation of the
proper Lorentz group $\SO_0(1,3)$ is defined by the pair
$(l_0,l_1)$, then a conjugated representation is also irreducible
and is defined by a pair $\pm(l_0,-l_1)$. Therefore,
\[
(l_0,l_1)=\left(-\dot{l},\,\dot{l}+1\right).
\]
Thus,
\begin{equation}\label{RelDL}
\dot{l}=\frac{l_0-l_1+1}{2}.
\end{equation}
Let
\[
\boldsymbol{S}=\boldsymbol{s}^{\alpha_1\alpha_2\ldots\alpha_k\dot{\alpha}_1\dot{\alpha}_2\ldots
\dot{\alpha}_r}=\sum \boldsymbol{s}^{\alpha_1}\otimes
\boldsymbol{s}^{\alpha_2}\otimes\cdots\otimes
\boldsymbol{s}^{\alpha_k}\otimes
\boldsymbol{s}^{\dot{\alpha}_1}\otimes
\boldsymbol{s}^{\dot{\alpha}_2}\otimes\cdots\otimes
\boldsymbol{s}^{\dot{\alpha}_r}
\]
be a spintensor polynomial, then any pair of substitutions
\[
\alpha=\begin{pmatrix} 1 & 2 & \ldots & k\\
\alpha_1 & \alpha_2 & \ldots & \alpha_k\end{pmatrix},\quad \beta=\begin{pmatrix} 1 & 2 & \ldots & r\\
\dot{\alpha}_1 & \dot{\alpha}_2 & \ldots & \dot{\alpha}_r\end{pmatrix}
\]
defines a transformation $(\alpha,\beta)$ mapping $\boldsymbol{S}$ to the following polynomial:
\[
P_{\alpha\beta}\boldsymbol{S}=\boldsymbol{s}^{\alpha\left(\alpha_1\right)\alpha\left(\alpha_2\right)\ldots
\alpha\left(\alpha_k\right)\beta\left(\dot{\alpha}_1\right)\beta\left(\dot{\alpha}_2\right)\ldots
\beta\left(\dot{\alpha}_r\right)}.
\]
The spintensor $\boldsymbol{S}$ is called a \emph{symmetric spintensor} if at any $\alpha$, $\beta$ the equality
\[
P_{\alpha\beta}\boldsymbol{S}=\boldsymbol{S}
\]
holds. The space $\Sym_{(k,r)}$ of symmetric spintensors has the dimensionality
\begin{equation}\label{Degree}
\dim\Sym_{(k,r)}=(k+1)(r+1).
\end{equation}
The dimensionality of $\Sym_{(k,r)}$ is called a \emph{degree of the representation} $\boldsymbol{\tau}_{l\dot{l}}$ of the group $\SL(2,\C)$. It is easy to see that $\SL(2,\C)$ has representations of \textbf{\emph{any degree}}.

For the each $A\in\SL(2,\C)$ we define a linear transformation of the spintensor $\boldsymbol{s}$ via the formula
\[
\boldsymbol{s}^{\alpha_1\alpha_2\ldots\alpha_k\dot{\alpha}_1\dot{\alpha}_2\ldots
\dot{\alpha}_r}\longrightarrow\sum_{\left(\beta\right)\left(\dot{\beta}\right)}A^{\alpha_1\beta_1}A^{\alpha_2\beta_2}
\cdots A^{\alpha_k\beta_k}\overline{A}^{\dot{\alpha}_1\dot{\beta}_1}\overline{A}^{\dot{\alpha}_2\dot{\beta}_2}
\cdots\overline{A}^{\dot{\alpha}_r\dot{\beta}_r}\boldsymbol{s}^{\beta_1\beta_2\ldots\beta_k\dot{\beta}_1\dot{\beta}_2\ldots
\dot{\beta}_r},
\]
where the symbols $\left(\beta\right)$ and $\left(\dot{\beta}\right)$ mean $\beta_1$, $\beta_2$, $\ldots$, $\beta_k$ and $\dot{\beta}_1$, $\dot{\beta}_2$, $\ldots$, $\dot{\beta}_r$. This representation of $\SL(2,\C)$ we denote as $\boldsymbol{\tau}_{\frac{k}{2},\frac{r}{2}}=\boldsymbol{\tau}_{l\dot{l}}$. The each \emph{irreducible} finite dimensional representation of $\SL(2,\C)$ is equivalent to one from $\boldsymbol{\tau}_{k/2,r/2}$.

When the matrices $A$ are unitary and unimodular we come to the subgroup $\SU(2)$ of $\SL(2,\C)$. Irreducible representations of $\SU(2)$ are equivalent to one from the mappings $A\rightarrow\boldsymbol{\tau}_{k/2,0}(A)$ with $A\in\SU(2)$, they are denoted as $\boldsymbol{\tau}_{k/2}$. The representation of $\SU(2)$, obtained at the contraction $A\rightarrow\boldsymbol{\tau}_{k/2,r/2}(A)$ onto $A\in\SU(2)$, is not irreducible. In fact, it is a direct product of $\boldsymbol{\tau}_{k/2}$ by $\boldsymbol{\tau}_{r/2}$, therefore, in virtue of the Clebsh-Gordan decomposition we have here a sum of representations
\[
\boldsymbol{\tau}_{\frac{k+r}{2}},\quad\boldsymbol{\tau}_{\frac{k+r}{2}-1},\quad\ldots,\quad
\boldsymbol{\tau}_{\left|\frac{k-r}{2}\right|}.
\]

\subsubsection{Definition of the spin}
We claim that \emph{any} irreducible finite dimensional representation $\boldsymbol{\tau}_{l\dot{l}}$ of the group $\SL(2,\C)$ corresponds to a \textbf{\emph{particle of the spin}} $s$, where $s=|l-\dot{l}|$ (see also \cite{Var11}). All the values of $s$ are
\[
-s,\;\;-s+1,\;\;-s+2,\;\;\ldots,\;\;s
\]
or
\begin{equation}\label{SValues}
-|l-\dot{l}|,\;\;-|l-\dot{l}|+1,\;\;-|l-\dot{l}|+2,\;\;\ldots,\;\;|l-\dot{l}|.
\end{equation}
Here the numbers $l$ and $\dot{l}$ are
\[
l=\frac{k}{2},\quad\dot{l}=\frac{r}{2},
\]
where $k$ and $r$ are factor quantities in the tensor product
\begin{equation}\label{TenAlg}
\underbrace{\C_2\otimes\C_2\otimes\cdots\otimes\C_2}_{k\;\text{times}}\bigotimes
\underbrace{\overset{\ast}{\C}_2\otimes\overset{\ast}{\C}_2\otimes\cdots\otimes
\overset{\ast}{\C}_2}_{r\;\text{times}}
\end{equation}
associated with the representation $\boldsymbol{\tau}_{k/2,r/2}$ of $\SL(2,\C)$, where $\C_2$ and complex conjugate $\overset{\ast}{\C}_2$ are biquaternion algebras. In turn, a \emph{spinspace} $\dS_{2^{k+r}}$, associated with the tensor product (\ref{TenAlg}), is
\begin{equation}\label{SpinSpace}
\underbrace{\dS_2\otimes\dS_2\otimes\cdots\otimes\dS_2}_{k\;\text{times}}\bigotimes
\underbrace{\dot{\dS}_2\otimes\dot{\dS}_2\otimes\cdots\otimes\dot{\dS}_2}_{r\;\text{times}}.
\end{equation}
Usual definition of the spin we obtain at the restriction $\boldsymbol{\tau}_{l\dot{l}}\rightarrow\boldsymbol{\tau}_{l,0}$ (or $\boldsymbol{\tau}_{l\dot{l}}\rightarrow\boldsymbol{\tau}_{0,\dot{l}}$), that is, at the restriction of $\SL(2,\C)$ to its subgroup $\SU(2)$. In this case the sequence of spin values (\ref{SValues}) is reduced to $-l$, $-l+1$, $-l+2$, $\ldots$, $l$ (or $-\dot{l}$, $-\dot{l}+1$, $-\dot{l}+2$, $\ldots$, $\dot{l}$).

The products (\ref{TenAlg}) and (\ref{SpinSpace}) define an \emph{algebraic (spinor) structure} associated with the representation $\boldsymbol{\tau}_{k/2,r/2}$ of the group $\SL(2,\C)$. Usually, spinor structures are understood as double (universal) coverings of the orthogonal groups $\SO(p,q)$. For that reason it seems that the spinor structure presents itself a derivative construction. However, in accordance with Penrose twistor programme \cite{Pen68,Pen} the spinor (twistor) structure presents a more fundamental level of reality rather then a space-time continuum. Moreover, the space-time continuum is generated by the twistor structure. This is a natural consequence of the well known fact of the van der Waerden 2-spinor formalism \cite{Wae32}, in which any vector of the Minkowski space-time can be constructed via the pair of mutually conjugated 2-spinors. For that reason it is more adequate to consider spinors as the \emph{underlying structure}\footnote{We choose $\spin_+(1,3)$ as a \emph{generating kernel} of the underlying spinor structure. However, the group $\spin_+(2,4)\simeq\SU(2,2)$ (a universal covering of the conformal group $\SO_0(2,4)$) can be chosen as such a kernel. The choice $\spin_+(2,4)\simeq\SU(2,2)$ takes place in the Penrose twistor programme \cite{Pen} and also in the Paneitz-Segal approach \cite{PS1,PS2,PS3}.}.

Further, representations $\boldsymbol{\tau}_{s_1,s_2}$ and
$\boldsymbol{\tau}_{s^\prime_1,s^\prime_2}$ are called
\emph{interlocking irreducible representations of the Lorentz group
}, that is, such representations that
$s^\prime_1=s_1\pm\frac{1}{2}$, $s^\prime_2=s_2\pm\frac{1}{2}$
\cite{Bha45,GY48}. The two most full schemes of the interlocking
irreducible representations of the Lorentz group (Bhabha-Gel'fand-Yaglom
chains) for integer and half-integer spins are shown on the Fig.\,1
and Fig.\,2. Wave equations for the fields of type $(l,0)\oplus(0,\dot{l})$ and their solutions in the form of series in hyperspherical functions were given in \cite{Var03a}-\cite{Var07}. It should be noted that $(l,0)\oplus(0,\dot{l})$ type wave equations correspond to the usual definition of the spin. In turn, wave equations for the fields of type $(l,\dot{l})\oplus(\dot{l},l)$ (arbitrary spin chains) and their solutions in the form of series in generalized hyperspherical functions were studied in \cite{Var07b}. Wave equations for arbitrary spin chains correspond to the generalized spin $s=|l-\dot{l}|$.
\begin{figure}[ht]
\[
\dgARROWPARTS=6\dgARROWLENGTH=0.5em \dgHORIZPAD=1.7em
\dgVERTPAD=2.2ex
\begin{diagram}
\node{s}\arrow[5]{e,..,-}
\node[5]{\overset{(s,0)}{\bullet}}\arrow{e,..,-}\arrow{s,..,-}
\node{\ldots}\\
\node[6]{\vdots}\arrow{s,..,-}\\
\node{2}\arrow[3]{e,..,-}
\node[3]{\overset{(2,0)}{\bullet}}\arrow{e,..,-}\arrow{s,..,-}
\node{\ldots}\arrow{s,..,-}
\node{\overset{(\frac{s+2}{2},\frac{s-2}{2})}{\bullet}}\arrow{e,..,-}\arrow{s,..,-}
\node{\ldots}\\
\node{1}\arrow[2]{e,..,-}
\node[2]{\overset{(1,0)}{\bullet}}\arrow{e,..,-}\arrow{s,..,-}
\node{\overset{(\frac{3}{2},\frac{1}{2})}{\bullet}}\arrow{e,..,-}\arrow{s,..,-}
\node{\ldots}\arrow{e,..,-}\arrow{s,..,-}
\node{\overset{(\frac{s+1}{2},\frac{s-1}{2})}{\bullet}}\arrow{e,..,-}\arrow{s,..,-}
\node{\ldots}\\
\node{0}\arrow{e,..,-}
\node{\overset{(0,0)}{\bullet}}\arrow[4]{n,-}\arrow{e,-}\arrow[4]{s,-}
\node{\overset{(\frac{1}{2},\frac{1}{2})}{\bullet}}\arrow{e,-}\arrow{s,..,-}
\node{\overset{(1,1)}{\bullet}}\arrow{e,-}\arrow{s,..,-}
\node{\ldots}\arrow{e,-}\arrow{s,..,-}
\node{\overset{(\frac{s}{2},\frac{s}{2})}{\bullet}}\arrow{e,-}\arrow{s,..,-}
\node{\ldots}\arrow{e,t,5}{s}\\
\node{-1}\arrow[2]{e,..,-}
\node[2]{\overset{(0,1)}{\bullet}}\arrow{e,..,-}
\node{\overset{(\frac{1}{2},\frac{3}{2})}{\bullet}}\arrow{e,..,-}\arrow{s,..,-}
\node{\ldots}\arrow{e,..,-}\arrow{s,..,-}
\node{\overset{(\frac{s-1}{2},\frac{s+1}{2})}{\bullet}}\arrow{e,..,-}\arrow{s,..,-}
\node{\ldots}\\
\node{-2}\arrow[3]{e,..,-}
\node[3]{\overset{(0,2)}{\bullet}}\arrow{e,..,-}\node{\ldots}\arrow{e,..,-}
\node{\overset{(\frac{s-2}{2},\frac{s+2}{2})}{\bullet}}\arrow{e,..,-}\arrow{s,..,-}
\node{\ldots}\\
\node[6]{\vdots}\arrow{s,..,-}\\
\node{-s}\arrow[5]{e,..,-}
\node[5]{\overset{(0,s)}{\bullet}}\arrow{e,..,-}\node{\ldots}
\end{diagram}
\]
%\vspace{0.5cm}
\begin{center}{\small {\bf Fig.\,1:} Interlocking representation scheme
for the fields of integer spin (Bose-scheme), $s=0,1,2,3,\ldots$.}\end{center}
\end{figure}
\begin{figure}[ht]
\[
\dgARROWPARTS=17\dgARROWLENGTH=0.5em
\dgHORIZPAD=1.7em %1.5em
\dgVERTPAD=2.2ex %2ex
\begin{diagram}
\node[2]{}\arrow[10]{s,-}\\
\node{s}\arrow[5]{e,..,-}
\node[5]{\overset{(s,0)}{\bullet}}\arrow{e,..,-}\arrow{s,..,-}
\node{\ldots}\\
\node[5]{\vdots}\arrow{s,..,-}\\
\node{\tfrac{3}{2}}\arrow[3]{e,..,-}
\node[3]{\overset{(\frac{3}{2},0)}{\bullet}}\arrow{e,..,-}\arrow{s,..,-}
\node{\ldots}\arrow{e,..,-}\arrow{s,..,-}
\node{\overset{(\frac{2s+3}{4},\frac{2s-3}{4})}{\bullet}}\arrow{e,..,-}\arrow{s,..,-}
\node{\ldots}\\
\node{\tfrac{1}{2}}\arrow[2]{e,..,-}
\node[2]{\overset{(\frac{1}{2},0)}{\bullet}}\arrow{e,..,-}\arrow[2]{s,..,-}
\node{\overset{(1,\frac{1}{2})}{\bullet}}\arrow{e,..,-}\arrow[2]{s,..,-}
\node{\ldots}\arrow{e,..,-}\arrow[2]{s,..,-}
\node{\overset{(\frac{2s+1}{4},\frac{2s-1}{4})}{\bullet}}\arrow{e,..,-}\arrow[2]{s,..,-}
\node{\ldots}\\
\node{}\arrow[7]{e,t,15}{s}\node{\bullet}\\
\node{-\tfrac{1}{2}}\arrow[2]{e,..,-}
\node[2]{\overset{(0,\frac{1}{2})}{\bullet}}\arrow{e,..,-}
\node{\overset{(\frac{1}{2},1)}{\bullet}}\arrow{e,..,-}\arrow{s,..,-}
\node{\ldots}\arrow{e,..,-}\arrow{s,..,-}
\node{\overset{(\frac{2s-1}{4},\frac{2s+1}{4})}{\bullet}}\arrow{e,..,-}\arrow{s,..,-}
\node{\ldots}\\
\node{-\tfrac{3}{2}}\arrow[3]{e,..,-}
\node[3]{\overset{(0,\frac{3}{2})}{\bullet}}\arrow{e,..,-}
\node{\ldots}\arrow{e,..,-}\arrow{s,..,-}
\node{\overset{(\frac{2s-3}{4},\frac{2s+3}{4})}{\bullet}}\arrow{e,..,-}\arrow{s,..,-}
\node{\ldots}\\
\node[5]{\vdots}\\
\node{-s}\arrow[5]{e,..,-}
\node[5]{\overset{(0,s)}{\bullet}} \arrow{e,..,-}
\node{\ldots}
\end{diagram}
\]
\vspace{0.5cm}
\begin{center}{\small {\bf Fig.\,2:} Interlocking representation scheme for the fields of half-integer spin
(Fermi-scheme),
$s=\frac{1}{2},\,\frac{3}{2},\,\frac{5}{2},\,\ldots$.}\end{center}
\end{figure}

Let us consider in detail several interlocking representations (spin lines) shown on the Fig.\,1 and Fig.\,2.
First of all, a central row (line of spin-$0$) in the scheme shown on the Fig.\,1,
\begin{equation}\label{BRow}
\dgARROWLENGTH=0.5em \dgHORIZPAD=1.7em \dgVERTPAD=2.2ex
\begin{diagram}
\node{(0,0)}\arrow{e,-}
\node{\left(\frac{1}{2},\frac{1}{2}\right)}\arrow{e,-}
\node{(1,1)}\arrow{e,-}
\node{\left(\frac{3}{2},\frac{3}{2}\right)}\arrow{e,-}\node{(2,2)}\arrow{e,-}\node{\cdots}\arrow{e,-}
\node{\left(\frac{s}{2},\frac{s}{2}\right)}\arrow{e,-} \node{\cdots}
\end{diagram}
\end{equation}
induces a sequence of algebras
\begin{multline}
\C_0\;\longrightarrow\;\C_2\otimes\overset{\ast}{\C}_2\;\longrightarrow\;
\C_2\otimes\C_2\bigotimes\overset{\ast}{\C}_2\otimes\overset{\ast}{\C}_2\;\longrightarrow\C_2\otimes\C_2
\otimes\C_2\bigotimes\overset{\ast}{\C}_2\otimes\overset{\ast}{\C}_2\otimes\overset{\ast}{\C}_2\;\longrightarrow\;\\
\longrightarrow\;\C_2\otimes\C_2\otimes\C_2\otimes\C_2\bigotimes\overset{\ast}{\C}_2\otimes\overset{\ast}{\C}_2\otimes
\overset{\ast}{\C}_2\otimes\overset{\ast}{\C}_2\;\longrightarrow\;\ldots\;\longrightarrow\\
\longrightarrow\;\underbrace{\C_2\otimes\C_2\otimes\cdots\otimes\C_2}_{s\;\text{times}}\bigotimes
\underbrace{\overset{\ast}{\C}_2\otimes\overset{\ast}{\C}_2
\otimes\cdots\otimes\overset{\ast}{\C}_2}_{s\;\text{times}}\;\longrightarrow\;\ldots
\nonumber
\end{multline}
Or,
\[
\C_0\;\longrightarrow\;\C_4\;\longrightarrow\;\C_8\;\longrightarrow\;\C_{12}\;\longrightarrow\C_{16}\;
\longrightarrow\;\ldots\;\longrightarrow\;\C_{4s}\;\longrightarrow\;\ldots
\]
With the spin-0 line we have a sequence of associated spinspaces
\[
\dS_1\;\longrightarrow\;\dS_4\;\longrightarrow\;\dS_{16}\;\longrightarrow\;\dS_{64}\;\longrightarrow\;
\dS_{256}\;\longrightarrow\;\ldots\;\longrightarrow\;\dS_{2^{2s}}\;\longrightarrow\;\ldots
\]
and also a sequence of symmetric spaces (spaces of symmetric spintensors)
\[
\Sym_{(0,0)}\;\longrightarrow\;\Sym_{(1,1)}\;\longrightarrow\;\Sym_{(2,2)}\;\longrightarrow\;\Sym_{(3,3)}\;
\longrightarrow\;\Sym_{(4,4)}\;\longrightarrow\;\ldots\;\longrightarrow\;\Sym_{(s,s)}\;\longrightarrow\;\ldots
\]
Dimensionalities of $\Sym_{(s,s)}$ (degrees of representations of $\spin_+(1,3)$ on the spin-0 line) form a sequence
\[
1\;\longrightarrow\;4\;\longrightarrow\;9\;\longrightarrow\;16\;\longrightarrow\;25\;\longrightarrow\;\ldots
\]
On the spin-0 line (the first bosonic line) we have scalar (with positive parity $P^2=1$) and pseudoscalar ($P^2=-1$) particles. Among these scalars and pseudoscalars there are particles with positive and negative charges, and also there are neutral (or truly neutral) particles. For example, the Fig.\,3 shows eight pseudoscalar mesons of the spin 0, which form the octet $B_0$ (eight-dimensional regular representation of the group $\SU(3)$). All the particles of $B_0$ belong to spin-0 line.
\begin{figure}[ht]
\unitlength=0.70mm
\begin{center}
\begin{picture}(100,90)(0,15)
\put(35,76){$\bullet$} \put(35,80){$K^0$}\put(36,70){$\overline{s}d$}
\put(40,77){\line(1,0){22}}\put(66,76){$\bullet$}
\put(59,70){$\overline{s}u$}\put(65,80){$K^+$}
\put(69.5,73){\line(1,-2){9.5}}\put(80,50){$\bullet$}
\put(81,54){$\pi^+$}\put(71,50){$\overline{d}u$}
\put(68.5,27.5){\line(1,2){10}}\put(65,25){$\bullet$}\put(59,30){$\overline{d}s$}
\put(65,20){$\overline{K}^0$}\put(40,26){\line(1,0){22}}\put(35,25){$\bullet$}
\put(36,30){$\overline{u}s$}\put(35,20){$K^-$}
\put(23.5,48){\line(1,-2){10}}\put(20,50){$\bullet$}
\put(25,50){$\overline{u}d$}\put(16,53){$\pi^-$}
\put(23,54){\line(1,2){10}}\put(50,55){$\bullet$}
\put(50,59){$\pi^0$}\put(53,55){$\overline{d}d\overline{u}u$}
\put(50,45){$\bullet$}\put(50,40){$\eta$}\put(53,45){$\overline{s}s$}
\put(0,20){\vector(0,1){70}}\put(4,90){\textbf{Strangeness}}
\put(100,90){\textbf{Mass} (MeV)}\put(105,76){496}
%\put(107,70){\vector(0,-1){10}}
\put(105,54){138}
\put(118,54){$\pi$}\put(105,46){549} \put(118,46){$\eta$}
%\put(107,42){\vector(0,-1){10}}
\put(105,25){496}
\put(-1.5,76){-}\put(-5,76){1}
\put(-1.5,50){-}\put(-6,50){0}
\put(-1.5,25){-}\put(-6,25){-1}
\end{picture}
\end{center}
\begin{center}\begin{minipage}{22pc}{\small {\bf Fig.\,3:} Octet $B_0$ of pseudoscalar mesons with associated quark structure according to $\SU(3)$-theory. $(K^-,K^+)$, $(K^0,\overline{K}^0)$ and $(\pi^-,\pi^+)$ are pairs of particles and antiparticles with respect to each other.}\end{minipage}\end{center}
\end{figure}

Further, the spin-$1/2$ line, shown on the Fig.\,2,
\begin{multline}\label{BRow2}
\left(\frac{1}{2},0\right)\;\longrightarrow\;
\left(1,\frac{1}{2}\right)\;\longrightarrow\;
\left(\frac{3}{2},1\right)\;\longrightarrow\;
\left(2,\frac{3}{2}\right)\;\longrightarrow\;\\
\longrightarrow\;\left(\frac{5}{2},2\right)\;\longrightarrow\;\cdots\longrightarrow\;
\left(\frac{2s+1}{4},\frac{2s-1}{4}\right)\;\longrightarrow\;\cdots
\end{multline}
induces a sequence of algebras
\begin{multline}
\C_2\;\longrightarrow\;\C_2\otimes\C_2\bigotimes\overset{\ast}{\C}_2\;\longrightarrow\;
\C_2\otimes\C_2\otimes\C_2\bigotimes\overset{\ast}{\C}_2\otimes\overset{\ast}{\C}_2\;\longrightarrow\C_2\otimes\C_2
\otimes\C_2\otimes\C_2\bigotimes\overset{\ast}{\C}_2\otimes\overset{\ast}{\C}_2\otimes\overset{\ast}{\C}_2\;\longrightarrow\;\\
\longrightarrow\;\C_2\otimes\C_2\otimes\C_2\otimes\C_2\otimes\C_2\bigotimes\overset{\ast}{\C}_2\otimes\overset{\ast}{\C}_2\otimes
\overset{\ast}{\C}_2\otimes\overset{\ast}{\C}_2\;\longrightarrow\;\ldots\;\longrightarrow\\
\longrightarrow\;\underbrace{\C_2\otimes\C_2\otimes\cdots\otimes\C_2}_{(2s+1)/2\;\text{times}}\bigotimes
\underbrace{\overset{\ast}{\C}_2\otimes\overset{\ast}{\C}_2
\otimes\cdots\otimes\overset{\ast}{\C}_2}_{(2s-1)/2\;\text{times}}\;\longrightarrow\;\ldots
\nonumber
\end{multline}
Or,
\[
\C_2\;\longrightarrow\;\C_6\;\longrightarrow\;\C_{10}\;\longrightarrow\;\C_{14}\;\longrightarrow\C_{18}\;
\longrightarrow\;\ldots\;\longrightarrow\;\C_{4s}\;\longrightarrow\;\ldots
\]
With the spin-$1/2$ line we have a sequence of associated spinspaces
\[
\dS_2\;\longrightarrow\;\dS_8\;\longrightarrow\;\dS_{32}\;\longrightarrow\;\dS_{128}\;\longrightarrow\;
\dS_{512}\;\longrightarrow\;\ldots\;\longrightarrow\;\dS_{2^{2s}}\;\longrightarrow\;\ldots
\]
and also a sequence of symmetric representation spaces
\begin{multline}
\Sym_{(1,0)}\;\longrightarrow\;\Sym_{(2,1)}\;\longrightarrow\;\Sym_{(3,2)}\;\longrightarrow\;\\
\longrightarrow\;\Sym_{(4,3)}\;
\longrightarrow\;\Sym_{(5,4)}\;\longrightarrow\;\ldots\;\longrightarrow\;
\Sym_{\left(\frac{2s+1}{2},\frac{2s-1}{2}\right)}\;\longrightarrow\;\ldots\nonumber
\end{multline}
with dimensions
\[
2\;\longrightarrow\;6\;\longrightarrow\;12\;\longrightarrow\;20\;\longrightarrow\;30\;\longrightarrow\;\ldots
\]
On the spin-1/2 line (the first fermionic line) we have all known particles of the spin 1/2 including leptons (neutrino, electron, muon, $\tau$-lepton, $\ldots$) and baryons. Among leptons and baryons there are particles with positive and negative charges, and also there are neutral particles. On the Fig.\,4 we have the well-known supermultiplet of $\SU(3)$-theory containing baryons of the spin 1/2 with positive parity ($P^2=1$), where a nucleon doublet $(n,p)$ is the basic building block of the all stable matter.
\begin{figure}[ht]
\unitlength=0.70mm
\begin{center}
\begin{picture}(100,90)(0,15)
\put(35,76){$\bullet$} \put(35,80){n}\put(36,70){$ddu$}
\put(40,77){\line(1,0){22}}\put(66,76){$\bullet$}
\put(59,70){$uud$}\put(65,80){p}
\put(69.5,73){\line(1,-2){9.5}}\put(80,50){$\bullet$}
\put(81,54){$\Sigma^+$}\put(71,50){$uus$}
\put(68.5,27.5){\line(1,2){10}}\put(65,25){$\bullet$}\put(59,30){$ssu$}
\put(65,20){$\Xi^0$}\put(40,26){\line(1,0){22}}\put(35,25){$\bullet$}
\put(36,30){$ssd$}\put(35,20){$\Xi^-$}
\put(23.5,48){\line(1,-2){10}}\put(20,50){$\bullet$}
\put(25,50){$dds$}\put(16,53){$\Sigma^-$}
\put(23,54){\line(1,2){10}}\put(50,55){$\bullet$}
\put(50,58){$\Sigma^0$}\put(53,55){$uds$}
\put(50,45){$\bullet$}\put(50,40){$\Lambda$}\put(53,45){$uds$}
\put(0,20){\vector(0,1){70}}\put(4,90){\textbf{Strangeness}}
\put(100,90){\textbf{Mass} (MeV)}\put(105,76){939}
\put(107,70){\vector(0,-1){10}}\put(105,54){1192}
\put(118,54){$\Sigma$}\put(105,46){1115} \put(118,46){$\Lambda$}
\put(107,42){\vector(0,-1){10}}\put(105,25){1318}
\put(-1.5,76){-}\put(-4,76){0}
\put(-1.5,50){-}\put(-6,50){-1}
\put(-1.5,25){-}\put(-6,25){-2}
\end{picture}
\end{center}
\begin{center}\begin{minipage}{22pc}{\small {\bf Fig.\,4:} Octet $F_{1/2}$ of baryons with associated quark structure according to $\SU(3)$-theory. }\end{minipage}\end{center}
\end{figure}

The dual spin-1/2 line
\begin{multline}
\left(0,\frac{1}{2}\right)\;\longrightarrow\;
\left(\frac{1}{2},1\right)\;\longrightarrow\;
\left(1,\frac{3}{2}\right)\;\longrightarrow\;
\left(\frac{3}{2},2\right)\;\longrightarrow\;\\
\longrightarrow\;\left(2,\frac{5}{2}\right)\;\longrightarrow\;\cdots\longrightarrow\;
\left(\frac{2s-1}{4},\frac{2s+1}{4}\right)\;\longrightarrow\;\cdots\nonumber
\end{multline}
induces a sequence of algebras
\begin{multline}
\overset{\ast}{\C}_2\;\longrightarrow\;\C_2\bigotimes\overset{\ast}{\C}_2\otimes\overset{\ast}{\C}_2\;\longrightarrow\;
\C_2\otimes\C_2\bigotimes\overset{\ast}{\C}_2\otimes\overset{\ast}{\C}_2\otimes\overset{\ast}{\C}_2\;\longrightarrow\C_2\otimes\C_2
\otimes\C_2\bigotimes\overset{\ast}{\C}_2\otimes\overset{\ast}{\C}_2\otimes\overset{\ast}{\C}_2\otimes\overset{\ast}{\C}_2\;\longrightarrow\;\\
\longrightarrow\;\C_2\otimes\C_2\otimes\C_2\otimes\C_2\bigotimes\overset{\ast}{\C}_2\otimes\overset{\ast}{\C}_2\otimes
\overset{\ast}{\C}_2\otimes\overset{\ast}{\C}_2\otimes\overset{\ast}{\C}_2\;\longrightarrow\;\ldots\;\longrightarrow\\
\longrightarrow\;\underbrace{\C_2\otimes\C_2\otimes\cdots\otimes\C_2}_{(2s-1)/2\;\text{times}}\bigotimes
\underbrace{\overset{\ast}{\C}_2\otimes\overset{\ast}{\C}_2
\otimes\cdots\otimes\overset{\ast}{\C}_2}_{(2s+1)/2\;\text{times}}\;\longrightarrow\;\ldots
\nonumber
\end{multline}
For the dual spin-1/2 line we have symmetric spaces
\begin{multline}
\Sym_{(0,1)}\;\longrightarrow\;\Sym_{(1,2)}\;\longrightarrow\;\Sym_{(2,3)}\;\longrightarrow\;\\
\longrightarrow\;\Sym_{(3,4)}\;
\longrightarrow\;\Sym_{(4,5)}\;\longrightarrow\;\ldots\;\longrightarrow\;
\Sym_{\left(\frac{2s-1}{2},\frac{2s+1}{2}\right)}\;\longrightarrow\;\ldots\nonumber
\end{multline}
with the same dimensions and spinspaces.

Further, with the spin-1 line (Fig.\,1)
\begin{multline}\label{BRow3}
\left(1,0\right)\;\longrightarrow\;
\left(\frac{3}{2},\frac{1}{2}\right)\;\longrightarrow\;
\left(2,1\right)\;\longrightarrow\;
\left(\frac{5}{2},\frac{3}{2}\right)\;\longrightarrow\;\\
\longrightarrow\;\left(3,2\right)\;\longrightarrow\;\cdots\longrightarrow\;
\left(\frac{s+1}{2},\frac{s-1}{2}\right)\;\longrightarrow\;\cdots
\end{multline}
we have the underlying spinor structure generated by the following sequence of algebras:
\begin{multline}
\C_2\otimes\C_2\;\longrightarrow\;\C_2\otimes\C_2\otimes\C_2\bigotimes\overset{\ast}{\C}_2\;\longrightarrow\;\\
\longrightarrow\;\C_2\otimes\C_2\otimes\C_2\otimes\C_2\bigotimes\overset{\ast}{\C}_2\otimes\overset{\ast}{\C}_2\;\longrightarrow\C_2\otimes\C_2
\otimes\C_2\otimes\C_2\otimes\C_2\bigotimes\overset{\ast}{\C}_2\otimes\overset{\ast}{\C}_2\otimes\overset{\ast}{\C}_2\;\longrightarrow\;\\
\longrightarrow\;\C_2\otimes\C_2\otimes\C_2\otimes\C_2\otimes\C_2\otimes\C_2\bigotimes\overset{\ast}{\C}_2\otimes\overset{\ast}{\C}_2\otimes
\overset{\ast}{\C}_2\otimes\overset{\ast}{\C}_2\;\longrightarrow\;\ldots\;\longrightarrow\\
\longrightarrow\;\underbrace{\C_2\otimes\C_2\otimes\cdots\otimes\C_2}_{s+1\;\text{times}}\bigotimes
\underbrace{\overset{\ast}{\C}_2\otimes\overset{\ast}{\C}_2
\otimes\cdots\otimes\overset{\ast}{\C}_2}_{s-1\;\text{times}}\;\longrightarrow\;\ldots
\nonumber
\end{multline}
Or,
\[
\C_4\;\longrightarrow\;\C_8\;\longrightarrow\;\C_{12}\;\longrightarrow\;\C_{16}\;\longrightarrow\C_{20}\;
\longrightarrow\;\ldots\;\longrightarrow\;\C_{4s}\;\longrightarrow\;\ldots
\]
With the spin-1 line we have also the following sequence of associated spinspaces
\[
\dS_4\;\longrightarrow\;\dS_{16}\;\longrightarrow\;\dS_{64}\;\longrightarrow\;\dS_{256}\;\longrightarrow\;
\dS_{1024}\;\longrightarrow\;\ldots\;\longrightarrow\;\dS_{2^{2s}}\;\longrightarrow\;\ldots
\]
In this case symmetric spaces
\begin{multline}
\Sym_{(2,0)}\;\longrightarrow\;\Sym_{(3,1)}\;\longrightarrow\;\Sym_{(4,2)}\;\longrightarrow\;\\
\longrightarrow\;\Sym_{(5,3)}\;
\longrightarrow\;\Sym_{(6,4)}\;\longrightarrow\;\ldots\;\longrightarrow\;
\Sym_{\left(s+1,s-1\right)}\;\longrightarrow\;\ldots\nonumber
\end{multline}
have dimensions
\[
3\;\longrightarrow\;8\;\longrightarrow\;15\;\longrightarrow\;24\;\longrightarrow\;35\;\longrightarrow\;\ldots
\]
On the spin-1 line we have vector bosons with positive ($P^2=1$) or negative ($P^2=-1$) parity. Among these bosons there are particles with positive and negative charges, and also there are neutral (or truly neutral) particles. For example, the Fig.\,5 shows the octet $B_1$ of vector mesons with negative parity. It is interesting to note that a quark structure of $B_1$ coincides with the quark structure of the octet $B_0$ for pseudoscalar mesons.
\begin{figure}[ht]
\unitlength=0.70mm
\begin{center}
\begin{picture}(100,90)(0,15)
\put(35,76){$\bullet$} \put(35,80){${}^\ast\!K^0$}\put(36,70){$\overline{s}d$}
\put(40,77){\line(1,0){22}}\put(66,76){$\bullet$}
\put(59,70){$\overline{s}u$}\put(65,80){${}^\ast\!K^+$}
\put(69.5,73){\line(1,-2){9.5}}\put(80,50){$\bullet$}
\put(81,54){$\rho^+$}\put(71,50){$\overline{d}u$}
\put(68.5,27.5){\line(1,2){10}}\put(65,25){$\bullet$}\put(59,30){$\overline{d}s$}
\put(65,20){${}^\ast\!\overline{K}^0$}\put(40,26){\line(1,0){22}}\put(35,25){$\bullet$}
\put(36,30){$\overline{u}s$}\put(35,20){${}^\ast\!K^-$}
\put(23.5,48){\line(1,-2){10}}\put(20,50){$\bullet$}
\put(25,50){$\overline{u}d$}\put(16,53){$\rho^-$}
\put(23,54){\line(1,2){10}}\put(50,55){$\bullet$}
\put(50,59){$\rho^0$}\put(53,55){$\overline{d}d\overline{u}u$}
\put(50,45){$\bullet$}\put(50,40){$\varphi$}\put(53,45){$\overline{s}s$}
\put(0,20){\vector(0,1){70}}\put(4,90){\textbf{Strangeness}}
\put(100,90){\textbf{Mass} (MeV)}\put(105,76){891}
%\put(107,70){\vector(0,-1){10}}
\put(105,54){763}
\put(118,54){$\rho$}\put(105,46){782} \put(118,46){$\varphi$}
%\put(107,42){\vector(0,-1){10}}
\put(105,25){891}
\put(-1.5,76){-}\put(-5,76){1}
\put(-1.5,50){-}\put(-6,50){0}
\put(-1.5,25){-}\put(-6,25){-1}
\end{picture}
\end{center}
\begin{center}\begin{minipage}{22pc}{\small {\bf Fig.\,5:} Octet $B_1$ of vector mesons with associated quark structure according to $\SU(3)$-theory. $({}^\ast K^-,{}^\ast K^+)$, $({}^\ast K^0,{}^\ast\overline{K}^0)$ and $(\rho^-,\rho^+)$ are pairs of particles and antiparticles with respect to each other.}\end{minipage}\end{center}
\end{figure}

In turn, the dual spin-1 line
\begin{multline}
\left(0,1\right)\;\longrightarrow\;
\left(\frac{1}{2},\frac{3}{2}\right)\;\longrightarrow\;
\left(1,2\right)\;\longrightarrow\;
\left(\frac{3}{2},\frac{5}{2}\right)\;\longrightarrow\;\\
\longrightarrow\;\left(2,3\right)\;\longrightarrow\;\cdots\longrightarrow\;
\left(\frac{s-1}{2},\frac{s+1}{2}\right)\;\longrightarrow\;\cdots\nonumber
\end{multline}
induces the following sequence of algebras
\begin{multline}
\overset{\ast}{\C}_2\otimes\overset{\ast}{\C}_2\;\longrightarrow\;\C_2\bigotimes\overset{\ast}{\C}_2\otimes
\overset{\ast}{\C}_2\otimes\overset{\ast}{\C}_2\;\longrightarrow\;\\
\longrightarrow\;\C_2\otimes\C_2\bigotimes\overset{\ast}{\C}_2\otimes\overset{\ast}{\C}_2\otimes\overset{\ast}{\C}_2
\otimes\overset{\ast}{\C}_2\;\longrightarrow\C_2\otimes\C_2
\otimes\C_2\bigotimes\overset{\ast}{\C}_2\otimes\overset{\ast}{\C}_2\otimes\overset{\ast}{\C}_2
\otimes\overset{\ast}{\C}_2\otimes\overset{\ast}{\C}_2\;\longrightarrow\;\\
\longrightarrow\;\C_2\otimes\C_2\otimes\C_2\otimes\C_2\bigotimes\overset{\ast}{\C}_2\otimes\overset{\ast}{\C}_2\otimes
\overset{\ast}{\C}_2\otimes\overset{\ast}{\C}_2\otimes\overset{\ast}{\C}_2
\otimes\overset{\ast}{\C}_2\;\longrightarrow\;\ldots\;\longrightarrow\\
\longrightarrow\;\underbrace{\C_2\otimes\C_2\otimes\cdots\otimes\C_2}_{s-1\;\text{times}}\bigotimes
\underbrace{\overset{\ast}{\C}_2\otimes\overset{\ast}{\C}_2
\otimes\cdots\otimes\overset{\ast}{\C}_2}_{s+1\;\text{times}}\;\longrightarrow\;\ldots
\nonumber
\end{multline}
For the dual spin-1 line we have symmetric spaces
\begin{multline}
\Sym_{(0,2)}\;\longrightarrow\;\Sym_{(1,3)}\;\longrightarrow\;\Sym_{(2,4)}\;\longrightarrow\;\\
\longrightarrow\;\Sym_{(3,5)}\;
\longrightarrow\;\Sym_{(4,6)}\;\longrightarrow\;\ldots\;\longrightarrow\;
\Sym_{\left(s-1,s+1\right)}\;\longrightarrow\;\ldots\nonumber
\end{multline}
with the same dimensions and spinspaces.

The Fig.\,1 (Bose-scheme) and Fig.\,2 (Fermi-scheme) can be unified into one interlocking scheme shown on the Fig.\,6.

\begin{figure}[ht]
\[
\dgARROWPARTS=6\dgARROWLENGTH=0.5em
\dgHORIZPAD=2.7em %1.5em
\dgVERTPAD=3.2ex %2ex
\begin{diagram}
\node{\tfrac{5}{2}}\arrow[6]{e,..,-}
\node[6]{\overset{(\frac{5}{2},0)}{\bullet}}\arrow{n,..,-}\arrow[2]{e,..,-}\arrow[2]{s,..,-}\\
\node{2}\arrow[5]{e,..,-}
\node[5]{\overset{(2,0)}{\bullet}}\arrow[2]{e,..,-}\arrow[2]{s,..,-}
\node[2]{\ldots}\arrow{e,..,-}\\
\node{\tfrac{3}{2}}\arrow[4]{e,..,-}
\node[4]{\overset{(\frac{3}{2},0)}{\bullet}}\arrow[2]{e,..,-}\arrow[2]{s,..,-}
\node[2]{\overset{(2,\frac{1}{2})}{\bullet}}\arrow[2]{e,..,-}\arrow[2]{s,..,-}\\
\node{1}\arrow[3]{e,..,-}
\node[3]{\overset{(1,0)}{\bullet}}\arrow[2]{e,..,-}\arrow[2]{s,..,-}
\node[2]{\overset{(\frac{3}{2},\frac{1}{2})}{\bullet}}\arrow[2]{e,..,-}\arrow[2]{s,..,-}
\node[2]{\ldots}\arrow{e,..,-}\\
\node{\tfrac{1}{2}}\arrow[2]{e,..,-}
\node[2]{\overset{(\frac{1}{2},0)}{\bullet}}\arrow[2]{e,..,-}\arrow[2]{s,..,-}
\node[2]{\overset{(1,\frac{1}{2})}{\bullet}}\arrow[2]{e,..,-}\arrow[2]{s,..,-}
\node[2]{\overset{(\frac{3}{2},1)}{\bullet}}\arrow[2]{e,..,-}\arrow[2]{s,..,-}\\
\node{0}\arrow{e,-}
\node{\overset{(0,0)}{\bullet}}\arrow[6]{n,-}\arrow[6]{s,-}\arrow[2]{e,-}
\node[2]{\overset{(\frac{1}{2},\frac{1}{2})}{\bullet}}\arrow[2]{e,-}\arrow[2]{s,..,-}
\node[2]{\overset{(1,1)}{\bullet}}\arrow[2]{e,-}\arrow[2]{s,..,-}
\node[2]{\ldots}\arrow{e,t,6}{s}\\
\node{-\tfrac{1}{2}}\arrow[2]{e,..,-}
\node[2]{\overset{(0,\frac{1}{2})}{\bullet}}\arrow[2]{e,..,-}
\node[2]{\overset{(\frac{1}{2},1)}{\bullet}}\arrow[2]{e,..,-}\arrow[2]{s,..,-}
\node[2]{\overset{(1,\frac{3}{2})}{\bullet}}\arrow[2]{e,..,-}\arrow[2]{s,..,-}\\
\node{-1}\arrow[3]{e,..,-}
\node[3]{\overset{(0,1)}{\bullet}}\arrow[2]{e,..,-}
\node[2]{\overset{(\frac{1}{2},\frac{3}{2})}{\bullet}}\arrow[2]{e,-}\arrow[2]{s,..,-}
\node[2]{\ldots}\arrow{e,..,-}\\
\node{-\tfrac{3}{2}}\arrow[4]{e,..,-}
\node[4]{\overset{(0,\frac{3}{2})}{\bullet}}\arrow[2]{e,..,-}
\node[2]{\overset{(\frac{1}{2},2)}{\bullet}}\arrow[2]{e,..,-}\arrow[2]{s,..,-}\\
\node{-2}\arrow[5]{e,..,-}
\node[5]{\overset{(0,2)}{\bullet}}\arrow[2]{e,..,-}
\node[2]{\ldots}\arrow{e,..,-}\\
\node{-\tfrac{5}{2}}\arrow[6]{e,..,-}
\node[6]{\overset{(0,\frac{5}{2})}{\bullet}}\arrow[2]{e,..,-}\arrow{s,..,-}
\end{diagram}
\]
\begin{center}{\small {\bf Fig.\,6:} Interlocking representations of the fields of any spin, $s=0,\,\frac{1}{2},\,1,\,\frac{3}{2},\,\ldots$.}\end{center}
\end{figure}
\subsection{Relativistic wave equations}
In 1945, Bhabha \cite{Bha45} introduced relativistic wave equations
\begin{equation}\label{1}
i\Gamma_\mu\frac{\partial\psi}{\partial x_\mu}+m\psi=0,\quad\mu=0,1,2,3
\end{equation}
that describe systems with many masses and spins\footnote{Gel'fand and Yaglom \cite{GY48} developed a general theory of such equations including infinite-component wave equations of Majorana type \cite{Maj32}.}.

With the aim to obtain a relation between mass and spin we will find a solution of the equation (\ref{1}) in the form of plane wave
\begin{equation}\label{2}
\psi(x_0,x_1,x_2,x_3)=\psi(p_0,p_1,p_2,p_3)e^{i(-p_0x_0+p_1x_1+p_2x_2+p_3x_3)}.
\end{equation}
Substituting the plane wave (\ref{2}) into (\ref{1}), we obtain
\begin{equation}\label{4}
\left(\Gamma_0p_0-\Gamma_1p_1-\Gamma_2p_2-\Gamma_3p_3\right)\psi(p)+m\psi(p)=0.
\end{equation}
Denoting $\Gamma_0p_0-\Gamma_1p_1-\Gamma_2p_2-\Gamma_3p_3$ via $\Gamma(p)$, we see from (\ref{4}) that $\psi(p)$ is an eigenvector of the matrix $\Gamma(p)$ with the eigenvalue $-m$:
\begin{equation}\label{5}
\Gamma(p)\psi(p)=-m\psi(p).
\end{equation}
Let us show that a non-null solution $\psi(p)$ of this equation exists only for the vectors $\boldsymbol{p}(p_0,p_1,p_2,p_3)$ for which the relation
\[
p^2_0-p^2_1-p^2_2-p^2_3=m^2_i
\]
holds, where $m_i=\mu^0\lambda_i$, and $\lambda_i$ are real eigenvalues of the matrix $\Gamma_0$, $\mu^0$ is a constant.

We assume that the equation (\ref{1}) is finite dimensional. In this case the equation (\ref{5}) admits a non-null solution for such and only such vectors $\boldsymbol{p}$ for which a determinant of the matrix $\Gamma(p)+mE$ is equal to zero. Obviously, $\det\left(\Gamma(p)+mE\right)$ is a polynomial on variables $p_0$, $p_1$, $p_2$, $p_3$. Denoting it via $D(p_0,p_1,p_2,p_3)=D(p)$, we see that the polynomial $D(p)$ is constant along the surfaces of transitivity of the Lorentz group, that is, $D(p)$ is constant on the hyperboloids
\[
s^2(p)=p^2_0-p^2_1-p^2_2-p^2_3=\text{const}.
\]
Hence it follows that $D(p)$ depends only on $s^2(p):$ $D(p)=\widetilde{D}\left[s^2(p)\right]$, where $\widetilde{D}(s^2)$ is a polynomial on one variable $s^2$.

Decomposing $\widetilde{D}\left[s^2(p)\right]$ on the factors,
\begin{equation}\label{6}
\widetilde{D}\left[s^2(p)\right]=c\left[s^2(p)-m^2_1\right]\left[s^2(p)-m^2_2\right]\left[s^2(p)-m^2_3\right]\cdots
\left[s^2(p)-m^2_k\right],
\end{equation}
we see that $\det\left(\Gamma(p)+mE\right)$ is equal to zero only in the case when the vector $\boldsymbol{p}$ satisfies the condition
\begin{equation}\label{7}
s^2(p)=p^2_0-p^2_1-p^2_2-p^2_3=m^2_i,
\end{equation}
where $m^2_i$ are the roots of $\widetilde{D}$. Since the numbers $p_0$, $p_1$, $p_2$, $p_3$ are real, then the roots $m^2_i$ should be real also.

Let us find now a relation between the roots $m^2_i$ and eigenvalues of the matrix $\Gamma_0$. Supposing $p_1=p_2=p_3=0$, we obtain $s^2(p)=p^2_0$ and $\widetilde{D}\left[s^2(p)\right]=\widetilde{D}(p^2_0)$. At this point, the decomposition (\ref{6}) is written as
\begin{multline}
D(p^2_0)=c\left(p^2_0-m^2_1\right)\left(p^2_0-m^2_2\right)\cdots\left(p^2_0-m^2_k\right)=\\
=c\left(p_0-m_1\right)\left(p_0+m_1\right)\left(p_0-m_2\right)\left(p_0+m_2\right)\cdots\left(p_0-m_k\right)
\left(p_0+m_k\right).\label{8}
\end{multline}
On the other hand, at $p_1=p_2=p_3=0$ the matrix $\Gamma(p)$ is equal to $\Gamma(p)=p_0\Gamma_0$ and $\det\left(\Gamma(p)+mE\right)=\det\left(p_0\Gamma_0+mE\right)$. This determinant can be represented in the form
\begin{equation}\label{9}
\det\left(p_0\Gamma_0+mE\right)=\widetilde{c}\left(p_0-\mu^0\lambda_1\right)\left(p_0-\mu^0\lambda_2\right)\cdots
\left(p_0-\mu^0\lambda_s\right),
\end{equation}
where $\lambda_1$, $\lambda_2$, $\ldots$, $\lambda_s$ are eigenvalues of the matrix $\Gamma_0$. Comparing the decompositions (\ref{8}) and (\ref{9}), we see that at the corresponding numeration the following equalities
\begin{equation}\label{10}
m_1=\mu^0\lambda_1,\quad -m_1=\mu^0\lambda_2,\quad m_2=\mu^0\lambda_3=-\mu^0\lambda_4,\quad\ldots
\end{equation}
hold. The formula (\ref{10}) gives a relation between the roots of polynomial $\widetilde{D}$ and eigenvalues of the matrix $\Gamma_0$. It is easy to see that along with the each \emph{non-null} eigenvalue $\lambda$ the matrix $\Gamma_0$ has an eigenvalue $-\lambda$ of the same multiplicity.

It is easy to verify that for representations of the type $\boldsymbol{\tau}_{0,\dot{l}}$ we obtain
\[
m_1=\mu^0\dot{\lambda}_1,\quad -m_1=\mu^0\dot{\lambda}_2,\quad m_2=\mu^0\dot{\lambda}_3=-\mu^0\dot{\lambda}_4,\quad\ldots
\]
In general case of $\boldsymbol{\tau}_{l\dot{l}}$ we have
\[
m_1=\mu^0\lambda_1\dot{\lambda}_1,\quad -m_1=\mu^0\lambda_2\dot{\lambda}_2,\quad m_2=\mu^0\lambda_3\dot{\lambda}_3=-\mu^0\lambda_4\dot{\lambda}_4,\quad\ldots
\]

Coming to infinite dimensional representations of $\SL(2,\C)$, we have at $l\rightarrow\infty$ and $\dot{l}\rightarrow\infty$ the following relation:
\begin{equation}\label{MGY}
m^{(s)}=\mu^0\left(l+\frac{1}{2}\right)\left(\dot{l}+\frac{1}{2}\right),
\end{equation}
where $s=|l-\dot{l}|$.

\subsubsection{Wave equations in the bivector space $\R^6$}
The equations (\ref{1}) are defined in the Minkowski space-time $\R^{1,3}$. With the aim to obtain an analogue of (\ref{1}) in the underlying spinor structure we use a mapping into bivector space $\R^6$. There exists a close relationship between the metric of Minkowski space-time $\R^{1,3}$ and the metric of the bivector space $\R^6$ \cite{Pet61}\footnote{Using a mapping of curvature tensor into $\R^6$, Petrov \cite{Pet61} introduced his famous classification of Einstein spaces.}:
\begin{equation}\label{Metric}
g_{ab}\longrightarrow g_{\alpha\beta\gamma\delta}\equiv
g_{\alpha\gamma}g_{\beta\delta}-g_{\alpha\delta}g_{\beta\gamma}.
\end{equation}
In the case of $\R^{1,3}$ with the metric tensor
\[
g_{\alpha\beta}=\ar\begin{pmatrix}
-1 & 0 & 0 & 0\\
0  & -1& 0 & 0\\
0  & 0 & -1& 0\\
0  & 0 & 0 & 1
\end{pmatrix}
\]
in virtue of (\ref{Metric}) we obtain the following metric tensor for the bivector space $\R^6$:
\begin{equation}\label{MetB}
g_{ab}=\ar\begin{bmatrix}
-1& 0 & 0 & 0 & 0 & 0\\
0 & -1& 0 & 0 & 0 & 0\\
0 & 0 & -1& 0 & 0 & 0\\
0 & 0 & 0 & 1 & 0 & 0\\
0 & 0 & 0 & 0 & 1 & 0\\
0 & 0 & 0 & 0 & 0 & 1
\end{bmatrix},
\end{equation}
where the order of collective indices in $\R^6$ is $23\rightarrow
0$, $10\rightarrow 1$, $20\rightarrow 2$, $30\rightarrow 3$,
$31\rightarrow 4$, $12\rightarrow 5$. After the mapping of (\ref{1}) onto $\R^6$ we obtain the following system:
\[
\sum^3_{j=1}\left(\sL^l_j\otimes\boldsymbol{1}_{2\dot{l}+1}-
\boldsymbol{1}_{2l+1}\otimes\sL^{\dot{l}}_j\right)
\frac{\partial\boldsymbol{\psi}}{\partial a_j}+
i\sum^3_{j=1}\left(\sL^l_j\otimes\boldsymbol{1}_{2\dot{l}+1}-
\boldsymbol{1}_{2l+1}\otimes\sL^{\dot{l}}_j\right)
\frac{\partial\boldsymbol{\psi}}{\partial a^\ast_j}+
m\boldsymbol{\psi}=0,
\]
\begin{equation}\label{Complex}
\sum^3_{j=1}\left(\overset{\ast}{\sL}{}^{\dot{l}}_j\otimes
\boldsymbol{1}_{2l+1}-\boldsymbol{1}_{2\dot{l}+1}\otimes
\overset{\ast}{\sL}{}^l_j\right)\frac{\dot{\boldsymbol{\psi}}}
{\partial\widetilde{a}_j}-
i\sum^3_{j=1}\left(\overset{\ast}{\sL}{}^{\dot{l}}_j\otimes
\boldsymbol{1}_{2l+1}-\boldsymbol{1}_{2\dot{l}+1}\otimes
\overset{\ast}{\sL}{}^l_j\right)\frac{\dot{\boldsymbol{\psi}}}
{\partial\widetilde{a}^\ast_j}+m
\dot{\boldsymbol{\psi}}=0,
\end{equation}
where $\fg_1=a_1$, $\fg_2=a_2$, $\fg_3=a_3$, $\fg_4=ia_1$,
$\fg_5=ia_2$, $\fg_6=ia_3$, $a^\ast_1=-i\fg_4$, $a^\ast_2=-i\fg_5$,
$a^\ast_3=-i\fg_6$, and $\widetilde{a}_j$, $\widetilde{a}^\ast_j$
are the parameters corresponding to the dual basis, $\fg_i\in\spin_+(1,3)$.
These equations describe a particle of the spin $s=|l-\dot{l}|$ and mass $m$.
In essence, the equations (\ref{Complex}) are defined in three-dimensional complex space $\C^3$. In its turn, the space $\C^3$ is isometric to a \emph{six-dimensional bivector space} $\R^6$ (a parameter space of the Lorentz group)
\subsubsection{RWE of the proton}
In nature there are a wide variety of elementary particles which different from each other by the spin and mass. The following question arises naturally when we see on the Fig.\,1--Fig.\,3. What particles correspond to irreducible representations $\boldsymbol{\tau}_{l\dot{l}}$? For example, the spin chain $(1/2,0)\longleftrightarrow(0,1/2)$ on the Fig.\,2, defined within the representation $\boldsymbol{\tau}_{1/2,0}\oplus\boldsymbol{\tau}_{0,1/2}$, leads to a linear superposition of the two spin states $1/2$ and $-1/2$, that describes electron and corresponds to the Dirac equation
\begin{equation}\label{Dirac}
\gamma_\mu\frac{\partial\psi}{\partial x_\mu}+m_e\psi=0.
\end{equation}
After the mapping of (\ref{Dirac}) into bivector space $\R^6$, we obtain
\begin{eqnarray}
\sum^3_{j=1}\overset{\ast}{\Lambda}^{0,\frac{1}{2}}_j\frac{\partial\dot{\psi}}
{\partial\widetilde{a}_j}+i\sum^3_{j=1}\overset{\ast}{\Lambda}^{0,\frac{1}{2}}_j
\frac{\partial\dot{\psi}}{\partial\widetilde{a}^\ast_j}+
m_e\psi&=&0,\nonumber\\
\sum^3_{j=1}\Lambda^{\frac{1}{2},0}_j\frac{\partial\psi}{\partial a_j}-
i\sum^3_{j=1}\Lambda^{\frac{1}{2},0}_j\frac{\partial\psi}{\partial a^\ast_j}+
m_e\dot{\psi}&=&0,\nonumber
\end{eqnarray}
where
\begin{eqnarray}
&&\sL^{\frac{1}{2},0}_1=\frac{1}{2}c_{\frac{1}{2}\frac{1}{2}}\begin{bmatrix}
0 & 1\\
1 & 0
\end{bmatrix},\quad
\sL^{\frac{1}{2},0}_2=\frac{1}{2}c_{\frac{1}{2}\frac{1}{2}}\begin{bmatrix}
0 & -i\\
i & 0
\end{bmatrix},\quad
\sL^{\frac{1}{2},0}_3=\frac{1}{2}c_{\frac{1}{2}\frac{1}{2}}\begin{bmatrix}
1 & 0\\
0 & -1
\end{bmatrix},\nonumber\\
&&\overset{\ast}{\sL}^{0,\frac{1}{2}}_1=\frac{1}{2}\dot{c}_{\frac{1}{2}\frac{1}{2}}\begin{bmatrix}
0 & 1\\
1 & 0
\end{bmatrix},\quad
\overset{\ast}{\sL}^{0,\frac{1}{2}}_2=\frac{1}{2}\dot{c}_{\frac{1}{2}\frac{1}{2}}\begin{bmatrix}
0 & -i\\
i & 0
\end{bmatrix},\quad
\overset{\ast}{\sL}^{0,\frac{1}{2}}_3=\frac{1}{2}\dot{c}_{\frac{1}{2}\frac{1}{2}}\begin{bmatrix}
1 & 0\\
0 & -1
\end{bmatrix}.\label{LDir}
\end{eqnarray}
It is easy to see that these matrices coincide with the Pauli
matrices $\sigma_i$ when $c_{\frac{1}{2}\frac{1}{2}}=2$. Moreover, there is a deep relationship between Dirac and Maxwell equations in spinor form (see, for example, \cite{Var022,Rum36}).

As is known, electron and proton have the same spin but different masses. If we replace the electron mass $m_e$ by the proton mass $m_p$ in (\ref{Dirac}) we come to the wave equation which, at first glance, can be applied for description of the proton. However, the spin chain $(1/2,0)\longleftrightarrow(0,1/2)$ has a very simple algebraic structure and, obviously, this chain is not sufficient for description of a very complicated intrinsic structure of the proton. For that reason we must find other spin chain for the proton. It is clear that a main rule for the searching of this chain is the difference of masses $m_e$ and $m_p$. It is known that $m_p/m_e\approx 1800$. With the aim to find a proton chain we use the mass formula (\ref{MGY}). Let $s=l=1/2$ and let $m_e=\mu^0\left(l+\frac{1}{2}\right)=\mu^0$ is the electron mass\footnote{It is interesting to note that from (\ref{MGY}) it follows directly that the electron mass is the minimal rest mass $\mu^0$.}, then from (\ref{MGY}) it follows that
\[
m_p=\mu_0\left(l+\frac{1}{2}\right)\left(\dot{l}+\frac{1}{2}\right).
\]
For the mass ratio $m_p/m_e$ we have
\[
\frac{m_p}{m_e}=\left(l+\frac{1}{2}\right)\left(\dot{l}+\frac{1}{2}\right).
\]
Therefore, $\left(l+\frac{1}{2}\right)\left(\dot{l}+\frac{1}{2}\right)\approx 1800$. It is easy to verify that a proton representation can be defined within the spin chain $(59/2,29)\longleftrightarrow(29,59/2)$ that corresponds to a representation $\boldsymbol{\tau}_{59/2,29}\oplus\boldsymbol{\tau}_{29,59/2}$ of the degree 3540. In the bivector space $\R^6$ RWE for the proton chain takes the form
\begin{eqnarray}
\sum^3_{j=1}\overset{\ast}{\Lambda}^{29\frac{59}{2}}_j\frac{\partial\dot{\psi}}
{\partial\widetilde{a}_j}+i\sum^3_{j=1}\overset{\ast}{\Lambda}^{29\frac{59}{2}}_j
\frac{\partial\dot{\psi}}{\partial\widetilde{a}^\ast_j}+
m_p\psi&=&0,\nonumber\\
\sum^3_{j=1}\Lambda^{\frac{59}{2}29}_j\frac{\partial\psi}{\partial a_j}-
i\sum^3_{j=1}\Lambda^{\frac{59}{2}29}_j\frac{\partial\psi}{\partial a^\ast_j}+
m_p\dot{\psi}&=&0,\nonumber
\end{eqnarray}
where $\Lambda^{\frac{59}{2}29}_j=\Lambda^{\frac{59}{2}}_j\otimes\boldsymbol{1}_{59}-\boldsymbol{1}_{60}\otimes\Lambda^{29}_j$ and $\overset{\ast}{\Lambda}^{29\frac{59}{2}}_j=\overset{\ast}{\Lambda}^{29}_j\otimes\boldsymbol{1}_{60}-\boldsymbol{1}_{59}
\otimes\overset{\ast}{\Lambda}^{\frac{59}{2}}_j$. All the non-null elements of the matrices $\Lambda^{l\dot{l}}_j$ and $\overset{\ast}{\Lambda}^{\dot{l}l}_j$ were calculated in \cite{Var07b}. Here we do not give an explicit form of the all $\Lambda^{\frac{59}{2}29}_j$ and $\overset{\ast}{\Lambda}^{29\frac{59}{2}}_j$ (in view of their big sizes). For example, an explicit form of the matrix $\Lambda^{\frac{59}{2}29}_3$ is
\begin{multline}
\Lambda^{\frac{59}{2}29}_3=\text{diag}\left({}_1\Lambda^{\frac{59}{2}29}_3,\,{}_2\Lambda^{\frac{59}{2}29}_3,\,
{}_3\Lambda^{\frac{59}{2}29}_3,\,\ldots,\,{}_{29}\Lambda^{\frac{59}{2}29}_3,\right.\\
\left.\bO_{59},\,-{}_{29}\Lambda^{\frac{59}{2}29}_3,\,\ldots\,,-{}_3\Lambda^{\frac{59}{2}29}_3,\,
-{}_2\Lambda^{\frac{59}{2}29}_3,\,-{}_1\Lambda^{\frac{59}{2}29}_3\right),\nonumber
\end{multline}
where
\begin{eqnarray}
{}_1\Lambda^{\frac{59}{2}29}_3&=&\text{diag}\left(\frac{1711}{2},\,\frac{1653}{2},\,\frac{1595}{2},\ldots,\,
\frac{29}{2},\,-\frac{29}{2},\,\ldots,\,-\frac{1595}{2},\,-\frac{1653}{2},\,-\frac{1711}{2}\right),\nonumber\\
{}_2\Lambda^{\frac{59}{2}29}_3&=&\text{diag}\left(826,\,798,\,770,\,\ldots,\,14,\,-14,\,\ldots,\,-770,\,-798,\,-826\right),
\nonumber\\
%\end{eqnarray}
%\begin{eqnarray}
{}_3\Lambda^{\frac{59}{2}29}_3&=&\text{diag}\left(\frac{1593}{2},\,\frac{1539}{2},\,\frac{1485}{2},\ldots,\,
\frac{27}{2},\,-\frac{27}{2},\,\ldots,\,-\frac{1485}{2},\,-\frac{1539}{2},\,-\frac{1593}{2}\right),\nonumber
%{}_4\Lambda^{\frac{59}{2}29}_3&=&\text{diag}\left(767,\,741,\,715,\,\ldots,\,13,\,-13,\,\ldots,\,-715,\,-741,\,-767\right),
%\nonumber
\end{eqnarray}
\[
\ldots\ldots\ldots\ldots\ldots\ldots\ldots\ldots\ldots\ldots\ldots\ldots\ldots\ldots\ldots
\]
\[
{}_{29}\Lambda^{\frac{59}{2}29}_3=\text{diag}\left(\frac{59}{2},\,\frac{57}{2},\,\frac{55}{2},\ldots,\,
\frac{1}{2},\,-\frac{1}{2},\,\ldots,\,-\frac{55}{2},\,-\frac{57}{2},\,-\frac{59}{2}\right),
\]
and $\bO_{59}$ is the 59-dimensional zero matrix.  Spinor structure, associated with the chain $(59/2,29)\longleftrightarrow(29,59/2)$, is very complicate and will be studied in a separate work.
\subsection{Wigner interpretation}
As is known, one from keystone facts of relativistic quantum field theory claims that state vectors of the quantum system form a unitary representation of the Poincar\'{e} group $\cP=T_4\odot\SL(2,\C)$, that is, the quantum system is defined by the unitary representation of $\cP$ in the Hilbert space $\sH_\infty$. In 1939, Wigner \cite{Wig39} introduced the following (widely accepted at present time) definition of elementary particle:\\
\textbf{\emph{The quantum system, described by an irreducible unitary representation of the Poincar\'{e} group, is called an elementary particle.}}

An action of $\SL(2,\C)$ on the Minkowski space-time $\R^{1,3}$ leads to a separation of $\R^{1,3}$ onto orbits $\boldsymbol{O}$. There are six types of the orbits:\\
1. $\boldsymbol{O}^+_m\,:$ $p^2_0-p^2_1-p^2_2-p^2_3=m^2$, $m>0$, $p_0>0$;\\
2. $\boldsymbol{O}^-_m\,:$ $p^2_0-p^2_1-p^2_2-p^2_3=m^2$, $m>0$, $p_0<0$;\\
3. $\boldsymbol{O}_{im}\,:$ $p^2_0-p^2_1-p^2_2-p^2_3=-m^2$, $m>0$;\\
4. $\boldsymbol{O}^+_0\,:$ $p^2_0-p^2_1-p^2_2-p^2_3=0$, $m=0$, $p_0>0$;\\
5. $\boldsymbol{O}^-_0\,:$ $p^2_0-p^2_1-p^2_2-p^2_3=0$, $m=0$, $p_0<0$;\\
6. $\boldsymbol{O}^0_0\,:$ $\boldsymbol{0}=(0,0,0,0)$, $m=0$.\\
Hence it follows that we have six types of irreducible unitary representations $U$ of the group $\cP$ related with the orbits $\boldsymbol{O}$. The each representation $U$ acts in the Hilbert space $\sH_\infty$. For example, in case of the orbit $\boldsymbol{O}^+_m$ we have a representation $U^{m,+,s}$ which describes a massive particle of the spin $s$ and mass $m$, where $s=|l-\dot{l}|$. At this point, we have infinitely many mass orbits (hyperboloids) of type $\boldsymbol{O}^+_m$ and $\boldsymbol{O}^-_m$, where the mass distribution is defined by the formula (\ref{MGY}). The representation $U^{m,+,s}$ acts in the space $\sH^{m,+,s}_\infty$. For more details about Wigner interpretation and little groups\footnote{This topics leads to deeply developed mathematical tools related with induced representations \cite{Mac68}.} see \cite{BR77}.

%\section{Complex and real representations of $\spin_+(1,3)$}
\subsection{$CPT$ group}
Within the Clifford algebras there are infinitely many (continuous) automorphisms.
Discrete symmetries $P$ and $T$ transform (reflect) space and time
(two the most fundamental notions in relativistic physics), but space and time are not
separate and independent in the Minkowski
4-dimensional space-time continuum. For that reason a transformation of one (space or
time) induces a transformation of another. Therefore, discrete symmetries
should be expressed by such transformations of the continuum, which transform
all its structure totally with a full preservation of discrete nature\footnote{As it mentioned above (see sections 1 and 2.1.1), in the well-known Penrose twistor programme \cite{Pen68,Pen} a spinor structure is understood as the underlying (more fundamental) structure with respect to Minkowski space-time. In other words, space-time continuum is not fundamental substance in the twistor approach, this is a fully derivative (in spirit of Leibnitz philosophy) entity generated by the underlying spinor structure. In this context space-time discrete symmetries $P$ and $T$ should be considered as projections (shadows) of the fundamental automorphisms belonging to the background spinor structure.}. In 1949, Schouten \cite{Sch49} introduced such
(discrete) automorphisms. In 1955, a first systematic description of these automorphisms was given by Rashevskii
\cite{Ras55}. He showed that within the Clifford algebra $\cl_{p,q}$ over the real field $\F=\R$ there exist \emph{four fundamental automorphisms}: $\cA\rightarrow\cA$ (identity), $\cA\rightarrow\cA^\star$ (involution), $\cA\rightarrow\widetilde{\cA}$ (reversion), $\cA\rightarrow\widetilde{\cA^\star}$ (conjugation), $\cA$ is an arbitrary element of $\cl_{p,q}$.
A finite group structure of the
automorphism set
$\{\Id,\,\star,\,\widetilde{\phantom{cc}},\,\widetilde{\star}\}$ was
studied in \cite{Var01} with respect to discrete symmetries which
compound $PT$ group (so-called \emph{reflection group})\footnote{Some applications of the fundamental automorphisms to discrete symmetries of quantum field theory were considered by Rashevskii in \cite{Ras55} (see also his paper \cite{Ras58}).}.

Other important discrete symmetry is the charge conjugation $C$. In contrast
with the transformations $P$, $T$, $PT$, the operation $C$ is not
space-time discrete symmetry. As is known, the Clifford algebra $\C_n$ over the complex field $\F=\C$ is associated with a complex vector
space $\C^n$.
The extraction of the subspace
$\R^{p,q}$ in the space $\C^n$ induces in the algebra $\C_n$
a pseudoautomorphism $\cA\rightarrow\overline{\cA}$ \cite{Ras55,Ras58}. Compositions of $\cA\rightarrow\overline{\cA}$
with the fundamental automorphisms allow one to extend the set
$\{\Id,\,\star,\,\widetilde{\phantom{cc}},\,\widetilde{\star}\}$ by
the pseudoautomorphisms $\cA\rightarrow\overline{\cA}$,
$\cA\rightarrow\overline{\cA^\star}$,
$\cA\rightarrow\overline{\widetilde{\cA}}$,
$\cA\rightarrow\overline{\widetilde{\cA^\star}}$ \cite{Var03}. A
finite group structure of \emph{an automorphism set}
$\{\Id,\,\star,\,\widetilde{\phantom{cc}},\,\widetilde{\star},\,
\overline{\phantom{cc}},\,\overline{\star},\,
\overline{\widetilde{\phantom{cc}}},\,\overline{\widetilde{\star}}\}$
was studied in \cite{Var03} with respect to $CPT$ symmetries.

\begin{thm}[{\rm\cite{Var03}}]\label{tpseudo}
Let $\C_n$ be a complex Clifford algebra for $n\equiv 0\s\pmod{2}$
and let $\cl_{p,q}\subset\C_n$ be its subalgebra with a real division ring
$\K\simeq\R$ when $p-q\equiv 0,2\s\pmod{8}$ and quaternionic division ring
$\K\simeq\BH$ when $p-q\equiv 4,6\s\pmod{8}$, $n=p+q$. Then in dependence
on the division ring structure of the real subalgebra $\cl_{p,q}$ the matrix
$\Pi$ of the pseudoautomorphism $\cA\rightarrow\overline{\cA}$
has the following form:\\[0.2cm]
1) $\K\simeq\R$, $p-q\equiv 0,2\s\pmod{8}$.\\[0.1cm]
The matrix $\Pi$ for any spinor representation over the ring $\K\simeq\R$
is proportional to the unit matrix.\\[0.2cm]
2) $\K\simeq\BH$, $p-q\equiv 4,6\s\pmod{8}$.\\[0.1cm]
$\Pi=\cE_{\alpha_1\alpha_2\cdots\alpha_a}$ when
$a\equiv 0\s\pmod{2}$ and
$\Pi=\cE_{\beta_1\beta_2\cdots\beta_b}$ when $b\equiv 1\s\pmod{2}$,
where $a$ complex matrices $\cE_{\alpha_t}$
and $b$ real matrices $\cE_{\beta_s}$ form a basis of the spinor
representation of the algebra $\cl_{p,q}$ over the ring $\K\simeq\BH$,
$a+b=p+q,\,0<t\leq a,\,0<s\leq b$. At this point,
\begin{eqnarray}
\Pi\dot{\Pi}&=&\phantom{-}\sI\quad\text{if $a,b\equiv 0,1\s\pmod{4}$},
\nonumber\\
\Pi\dot{\Pi}&=&-\sI\quad\text{if $a,b\equiv 2,3\s\pmod{4}$},\nonumber
\end{eqnarray}
where $\sI$ is the unit matrix.
\end{thm}
Spinor representations of the all other automorphisms from the set $\{\Id,\,\star,\,\widetilde{\phantom{cc}},\,\widetilde{\star},\,
\overline{\phantom{cc}},\,\overline{\star},\,
\overline{\widetilde{\phantom{cc}}},\,\overline{\widetilde{\star}}\}$ are defined in a similar manner.
We list these transformations and their spinor representations:
\begin{eqnarray}
\cA\longrightarrow\cA^\star,&&\quad\sA^\star=\sW\sA\sW^{-1},\nonumber\\
\cA\longrightarrow\widetilde{\cA},&&\quad\widetilde{\cA}=\sE\sA^{t}\sE^{-1},
\nonumber\\
\cA\longrightarrow\widetilde{\cA^\star},&&\quad\widetilde{\sA^\star}=
\sC\sA^{t}\sC^{-1},\quad\sC=\sE\sW,\nonumber\\
\cA\longrightarrow\overline{\cA},&&\quad\overline{\sA}=\Pi\sA^\ast\Pi^{-1},
\nonumber\\
\cA\longrightarrow\overline{\cA^\star},&&\quad\overline{\sA^\star}=
\sK\sA^\ast\sK^{-1},\quad\sK=\Pi\sW,\nonumber\\
\cA\longrightarrow\overline{\widetilde{\cA}},&&\quad
\overline{\widetilde{\sA}}=\sS\left(\sA^{t}\right)^\ast\sS^{-1},\quad
\sS=\Pi\sE,\nonumber\\
\cA\longrightarrow\overline{\widetilde{\cA^\star}},&&\quad
\overline{\widetilde{\sA^\star}}=\sF\left(\sA^\ast\right)^{t}\sF^{-1},\quad
\sF=\Pi\sC.\nonumber
\end{eqnarray}

It is easy to verify that an automorphism
set $\{\Id,\,\star,\,\widetilde{\phantom{cc}},\,\widetilde{\star},\,
\overline{\phantom{cc}},\,\overline{\star},\,
\overline{\widetilde{\phantom{cc}}},\,\overline{\widetilde{\star}}\}$
of $\C_n$ forms a finite group of order 8.
\begin{sloppypar}
Further, let $\C_n$ be a Clifford algebra over the field $\F=\C$ and
let $\Ext(\C_n)=
\{\Id,\,\star,\,\widetilde{\phantom{cc}},\,\widetilde{\star},\,
\overline{\phantom{cc}},\,\overline{\star},\,
\overline{\widetilde{\phantom{cc}}},\,\overline{\widetilde{\star}}\}$
be \emph{an automorphism group} of the algebra $\C_n$. Then
there is an isomorphism between $\Ext(\C_n)$ and a $CPT$ group of
the discrete transformations,
$\Ext(\C_n)\simeq\{1,\,P,\,T,\,PT,\,C,\,CP,\,CT,\,CPT\}\simeq
\dZ_2\otimes\dZ_2\otimes\dZ_2$. In this case, space inversion $P$,
time reversal $T$, full reflection $PT$, charge conjugation $C$,
transformations $CP$, $CT$ and the full $CPT$--transformation
correspond to the automorphism $\cA\rightarrow\cA^\star$,
antiautomorphisms $\cA\rightarrow\widetilde{\cA}$,
$\cA\rightarrow\widetilde{\cA^\star}$, pseudoautomorphisms
$\cA\rightarrow\overline{\cA}$,
$\cA\rightarrow\overline{\cA^\star}$, pseudoantiautomorphisms
$\cA\rightarrow\overline{\widetilde{\cA}}$ and
$\cA\rightarrow\overline{\widetilde{\cA^\star}}$, respectively
\cite{Var03}.\end{sloppypar}
\begin{sloppypar}
The group $\{1,\,P,\,T,\,PT,\,C,\,CP,\,CT,\,CPT\}$ at the conditions
$P^2=T^2=(PT)^2=C^2=(CP)^2=(CT)^2=(CPT)^2=1$ and commutativity of
all the elements forms an Abelian group of order 8, which is
isomorphic to a cyclic group $\dZ_2\otimes\dZ_2\otimes\dZ_2$. In
turn, the automorphism group
$\{\Id,\,\star,\,\widetilde{\phantom{cc}},\,\widetilde{\star},\,
\overline{\phantom{cc}},\,\overline{\star},\,
\overline{\widetilde{\phantom{cc}}},\,\overline{\widetilde{\star}}\}$
in virtue of commutativity $\widetilde{\left(\cA^\star\right)}=
\left(\widetilde{\cA}\right)^\star$,
$\overline{\left(\cA^\star\right)}=\left(\overline{\cA}\right)^\star$,
$\overline{\left(\widetilde{\cA}\right)}=
\widetilde{\left(\overline{\cA}\right)}$,
$\overline{\left(\widetilde{\cA^\star}\right)}=
\widetilde{\left(\overline{\cA}\right)^\star}$ and an involution
property
$\star\star=\widetilde{\phantom{cc}}\widetilde{\phantom{cc}}=
\overline{\phantom{cc}}\;\overline{\phantom{cc}}=\Id$ is also
isomorphic to $\dZ_2\otimes\dZ_2\otimes\dZ_2$:\end{sloppypar}
\[
\{1,\,P,\,T,\,PT,\,C,\,CP,\,CT,\,CPT\}\simeq
\{\Id,\,\star,\,\widetilde{\phantom{cc}},\,\widetilde{\star},\,
\overline{\phantom{cc}},\,\overline{\star},\,
\overline{\widetilde{\phantom{cc}}},\,\overline{\widetilde{\star}}\}\simeq
\dZ_2\otimes\dZ_2\otimes\dZ_2.
\]
In 2003, the $CPT$ group was
introduced \cite{Var03} in the context of an extension of
automorphism groups of Clifford algebras. The relationship between
$CPT$ groups and extraspecial groups and universal coverings of
orthogonal groups was established in \cite{Var03,Var04}. $CPT$ groups of spinor fields
in the de Sitter spaces of different signatures were studied in the
works \cite{Var05c,Var05,Var14}. $CPT$ groups for higher spin fields have
been defined in \cite{Var11} on the spinspaces associated with
representations of the spinor group $\spin_+(1,3)$.
\subsubsection{Charged particles}
In the present form of quantum field theory complex fields correspond
to charged particles. Let us consider the action of the pseudoautomorphism
$\cA\rightarrow\overline{\cA}$ on the spinors of the fundamental representation of the group
$\spin_+(1,3)\simeq\SL(2,\C)$. The matrix $\Pi$ allows one to compare with the each spinor
$\xi^\alpha$ its conjugated spinor $\overline{\xi}^\alpha$ by the following
rule:
\begin{equation}\label{6.33'}
\overline{\xi}^\alpha=\Pi^\alpha_{\dot{\alpha}}\xi^{\dot{\alpha}},
\end{equation}
here $\xi^{\dot{\alpha}}=(\xi^\alpha)^\cdot$. In accordance with Theorem
\ref{tpseudo} for the matrix $\Pi^\alpha_{\dot{\beta}}$
we have $\dot{\Pi}=\Pi^{-1}$ or
$\dot{\Pi}=-\Pi^{-1}$, where $\Pi^{-1}=\Pi^{\dot{\alpha}}_\beta$.
Then a twice conjugated spinor looks like
\[
\overline{\xi}^\alpha=\overline{\Pi^\alpha_{\dot{\beta}}\xi^{\dot{\beta}}}=
\Pi^\alpha_{\dot{\alpha}}(\Pi^\alpha_{\dot{\beta}}\xi^{\dot{\beta}})^\cdot=
\Pi^\alpha_{\dot{\alpha}}(\pm\Pi^{\dot{\alpha}}_\beta)\xi^\beta=
\pm\xi^\alpha.
\]
Therefore, the twice conjugated spinor coincides with the initial spinor
in the case of the real subalgebra of $\C_2$ with the ring
$\K\simeq\R$ (the algebras $\cl_{1,1}$ and $\cl_{2,0}$), and also in the case
of $\K\simeq\BH$ (the algebra $\cl_{0,2}\simeq\BH$) at
$a-b\equiv 0,1\pmod{4}$. Since for the algebra $\cl_{0,2}\simeq\BH$ we have
always $a-b\equiv 0\pmod{4}$, then a property of the reciprocal conjugacy
of the spinors
$\xi^\alpha$ ($\alpha=1,2$) is an invariant fact for the fundamental
representation of the group $\spin_+(1,3)$ (this property is very important in
physics, because it is an algebraic expression of the requirement
$C^2=1$). Further, since the `vector' (spintensor) of the
finite-dimensional representation of the group
$\spin_+(1,3)$ is defined by the tensor product
$\xi^{\alpha_1\alpha_2\cdots\alpha_k}=\sum\xi^{\alpha_1}\otimes
\xi^{\alpha_2}\otimes\cdots\otimes\xi^{\alpha_k}$, then its conjugated
spintensor takes the form
\begin{equation}\label{6.33''}
\overline{\xi}^{\alpha_1\alpha_2\cdots\alpha_k}=
\sum\Pi^{\alpha_1}_{\dot{\alpha}_1}\Pi^{\alpha_2}_{\dot{\alpha}_2}\cdots
\Pi^{\alpha_k}_{\dot{\alpha}_k}\xi^{\dot{\alpha}_1\dot{\alpha}_2\cdots
\dot{\alpha}_k}.
\end{equation}\begin{sloppypar}\noindent
It is obvious that a condition of reciprocal conjugacy
$\overline{\overline{\xi}}\!{}^{\alpha_1\alpha_2\cdots\alpha_k}=
\xi^{\alpha_1\alpha_2
\cdots\alpha_k}$ is also fulfilled for (\ref{6.33''}), since for the each matrix
$\Pi^{\alpha_i}_{\dot{\alpha}_i}$ in (\ref{6.33''}) we have
$\dot{\Pi}=\Pi^{-1}$ (all the matrices $\Pi^{\alpha_i}_{\dot{\alpha}_i}$
are defined for the algebra $\C_2$). \end{sloppypar}

Further, in accordance with Theorem \ref{tpseudo} Clifford
algebras over the field $\F=\C$ correspond to \textbf{\emph{charged
particles}} such as electron, proton and so on. In general case all
the elements of $C^{a,b,c,d,e,f,g}$ (resp. $\sExt(\cl_{p,q})$)
depend on the phase factors. Let us suppose
\begin{equation}\label{GenCPT}
P=\eta_p\sW,\quad T=\eta_t\sE,\quad C=\eta_c\Pi,
\end{equation}
where $\eta_p,\,\eta_t,\,\eta_c\in\C^\ast=\C-\{0\}$ are phase factors. Taking into
account (\ref{GenCPT}), we obtain
\begin{multline}
\sExt(\cl_{p,q})\simeq\{1,\,P,\,T,\,PT,\,C,\,CP,\,CT,\,CPT\}\simeq\\
\simeq\{\boldsymbol{1}_{(p+q)/2},\,\eta_p\sW,\,\eta_t\sE,\,\eta_p\eta_t\sE\sW,\,\eta_c\Pi,\,\eta_c\eta_p\Pi\sW,\,
\eta_c\eta_t\Pi\sE,\,\eta_c\eta_p\eta_t\Pi\sE\sW\}\simeq\\
\simeq\{\boldsymbol{1}_{(p+q)/2},\,\eta_p\sW,\,\eta_t\sE,\,\eta_p\eta_t\sC,\,\eta_c\Pi,\,\eta_c\eta_p\sK,\,
\eta_c\eta_t\sS,\,\eta_c\eta_p\eta_t\sF\}.\nonumber
\end{multline}
The multiplication table of this general group is given in Tab.\,1.
\begin{figure}[ht]
{\scriptsize
\begin{center}{\renewcommand{\arraystretch}{1.6}
\begin{tabular}{|c||c|c|c|c|c|c|c|c|}\hline
  & $\boldsymbol{1}_{(p+q)/2}$ & $\eta_p\sW$ & $\eta_t\sE$ & $\eta_{pt}\sC$ & $\eta_c\Pi$ &
$\eta_{cp}\sK$ & $\eta_{ct}\sS$ & $\eta_{cpt}\sF$\\
\hline\hline $\boldsymbol{1}_{(p+q)/2}$ & $\boldsymbol{1}_{(p+q)/2}$ & $\eta_p\sW$ &
$\eta_t\sE$ &
$\eta_{pt}\sC$ & $\eta_c\Pi$ & $\eta_{cp}\sK$ & $\eta_{ct}\sS$ & $\eta_{cpt}\sF$\\
\hline $\eta_p\sW$ & $\eta_p\sW$ & $\eta^2_p\sW^2$ &
$\eta_{pt}\sW\sE$ &
$\eta^2_p\eta_t\sW\sC$ & $\eta_{pc}\sW\Pi$ & $\eta_c\eta^2_p\sW\sK$ & $\eta_{cpt}\sW\sS$ & $\eta_{ct}\eta^2_p\sW\sF$\\
\hline $\eta_t\sE$ & $\eta_t\sE$ & $\eta_{pt}\sE\sW$ &
$\eta^2_t\sE^2$ & $\eta_p\eta^2_t\sE\sC$
& $\eta_{ct}\sE\Pi$ & $\eta_{cpt}\sE\sK$ & $\eta_c\eta^2_t\sE\sS$ & $\eta_{cp}\eta^2_t\sE\sF$\\
\hline $\eta_{pt}\sC$ & $\eta_{pt}\sC$ & $\eta^2_p\eta_t\sC\sW$ &
$\eta_p\eta^2_t\sC\sE$ &
 $\eta^2_{pt}\sC^2$ & $\eta_{cpt}\sC\Pi$ & $\eta_{ct}\eta^2_p\sC\sK$ & $\eta_{cp}\eta^2_t\sC\sS$ &
$\eta_c\eta^2_{pt}\sC\sF$\\ \hline $\eta_c\Pi$ & $\eta_c\Pi$ &
$\eta_{cp}\Pi\sW$ & $\eta_{ct}\Pi\sE$ & $\eta_{cpt}\Pi\sC$ &
$\eta^2_c\Pi^2$ & $\eta^2_c\eta_p\Pi\sK$ & $\eta^2_c\eta_t\Pi\sS$ & $\eta^2_c\eta_{pt}\Pi\sF$\\
\hline $\eta_{cp}\sK$ & $\eta_{cp}\sK$ & $\eta_c\eta^2_p\sK\sW$ &
$\eta_{cpt}\sK\sE$ & $\eta_{ct}\eta^2_p\sK\sC$
& $\eta^2_c\eta_p\sK\Pi$ & $\eta^2_{cp}\sK^2$ & $\eta^2_c\eta_{pt}\sK\sS$ & $\eta^2_{cp}\eta_t\sK\sF$\\
\hline $\eta_{ct}\sS$ & $\eta_{ct}\sS$ & $\eta_{cpt}\sS\sW$ &
$\eta_c\eta^2_t\sS\sE$ & $\eta_{cp}\eta^2_t\sS\sC$ &
$\eta^2_c\eta_t\sS\Pi$ & $\eta^2_c\eta_{pt}\sS\sK$ &
$\eta^2_{ct}\sS^2$ & $\eta^2_{ct}\eta_p\sS\sF$\\
\hline $\eta_{cpt}\sF$ & $\eta_{cpt}\sF$ & $\eta_{ct}\eta^2_p\sF\sW$
& $\eta_{cp}\eta^2_t\sF\sE$ & $\eta_c\eta^2_{pt}\sF\sC$ &
$\eta^2_c\eta_{pt}\sF\Pi$ & $\eta^2_{cp}\eta_t\sF\sK$ &
$\eta^2_{ct}\eta_p\sF\sS$ & $\eta^2_{cpt}\sF^2$\\
\hline
\end{tabular}
}
\end{center}
} \hspace{0.3cm}
\begin{center}\begin{minipage}{32pc}
{\small \textbf{Tab.\,1:} The multiplication table of general
generating group $\sExt(\cl_{p,q})$.}
\end{minipage}
\end{center}
\end{figure}
The Tab.\,1 presents \emph{a general generating matrix} for any
possible $CPT$ groups of the fields of any spin.
\subsubsection{Particle-antiparticle interchange and truly neutral particles}
First of all, the transformation $C$ (the pseudoautomorphism
$\cA\rightarrow\overline{\cA}$) for the algebras $\cl_{p,q}$ over
the field $\F=\R$ and the ring $\K\simeq\BH$ (the types $p-q\equiv
4,6\pmod{8}$) corresponds to \textbf{\emph{particle-antiparticle
interchange}} $C^\prime$ (see \cite{Var03,Var04}). As is known,
neutral particles are described within real representations of the
Lorentz group. There are two classes of neutral particles: 1)
particles which have antiparticles such as neutrons, neutrinos and
so on; 2) particles which coincide with their antiparticles (for
example, photons). The first class is described by the the algebras
$\cl_{p,q}$ over the field $\F=\R$ with the rings $\K\simeq\BH$ and
$\K\simeq\BH\oplus\BH$ (the types $p-q\equiv 4,6\pmod{8}$ and
$p-q\equiv 5\pmod{8}$), and the second class (\textbf{\emph{truly
neutral particles}}) is described by the algebras $\cl_{p,q}$ over
the field $\F=\R$ with the rings $\K\simeq\R$ and
$\K\simeq\R\oplus\R$ (the types $p-q\equiv 0,2\pmod{8}$ and
$p-q\equiv 1\pmod{8}$) (for more details see
\cite{Var03,Var04,Var14,Var12}).

In accordance with Theorem \ref{tpseudo} for the algebras $\cl_{p,q}$ over the real field
$\F=\R$ with real division rings $\K\simeq\R$ and
$\K\simeq\R\oplus\R$ (types $p-q\equiv 0,1,2\pmod{8}$) the
pseudoautomorphism $\cA\rightarrow\overline{\cA}$ (charge
conjugation $C$) is reduced to the identical transformation $\Id$,
therefore, such algebras correspond to \textbf{\emph{truly neutral
particles}} (for example, photons, $K^0$-mesons and so on).

\section{Abstract Hilbert space}
In 1932, von Neumann \cite{Neu32} introduced an abstract Hilbert space with the purpose of understanding the basic mathematical principles of quantum mechanics. Let us consider in brief the main notions of this construction.

Let $\left|\sA\right\rangle$, $\left|\sB\right\rangle$, $\left|\sC\right\rangle$, $\ldots$ are the vectors of a complex Euclidean space\footnote{The space $\bsH$ can be understood as finite-dimensional space $\bsH_n$ and also as infinite-dimensional space $\bsH_\infty$ \cite{Neu32}.} $\bsH$ satisfying the axioms of Hilbert space (see, for example, \cite{RF70}).

As is known, wave functions can be treated as vectors of $\bsH_\infty$. Following to Dirac's designation \cite{Dir}, we denote the wave functions as $\psi\rangle=\left|\sA\right\rangle$, where $\left|\sA\right\rangle\in\bsH_\infty$. The `nature' of the vectors $\left|\sA\right\rangle\in\bsH_\infty$ is not essential, $\left|\sA\right\rangle$ can represent tensors, functions on the group, representations and so on. In accordance with Wigner interpretation (see section 2.3) we choose $\left|\sA\right\rangle$ as the functions on the Lorentz group, where the each vector $\left|\sA\right\rangle\in\bsH_\infty$ corresponds to a some representation of $\SL(2,\C)$. In this context von Neumann condition looks like
\begin{equation}\label{NC}
\int\limits_{\SL(2,\C)}\left\langle\psi\right|\!U(g)\left|\psi\right\rangle dg\;<\;\infty,
\end{equation}
where $dg$ is a Haar measure on the group $\SL(2,\C)$, $U(g)$ is a representation of $\SL(2,\C)$\footnote{In the framework of refined algebraic
quantization \cite{ALMMT95,Mar95}, the inner product of
states is defined using the technique of group averaging. Group
averaging uses the integral
\[
\int_G\langle\phi_1|U(g)|\phi_2\rangle dg
\]
over the gauge group $G$, where $dg$ is a so-called symmetric Haar
measure on $G$, $U(g)$ is a representation of $G$, and $\phi_1$ and
$\phi_2$ are state vectors from an auxiliary Hilbert space
$\cH_{aux}$. Convergent group averaging gives an algorithm for
construction of a complete set of observables of a quantum system
\cite{Hig91}--\cite{Var10}. This topics is related closely with a quantum field theory on the Poincar\'{e} group \cite{GS01,GS09} and also with a wavelet transform for resolution dependent fields \cite{Alt10,Alt13}.
}.

Further, there exists an infinite sequence of the wave functions $\left.\psi_1\right\rangle$, $\left.\psi_2\right\rangle$, $\ldots$, $\left.\psi_n\right\rangle$, $\ldots$ such that any wave function $\left.\psi\right\rangle$ is represented in the form
\begin{equation}\label{NC2}
\left.\psi\right\rangle=c_1\left.\psi_1\right\rangle+c_2\left.\psi_2\right\rangle+\ldots+
c_n\left.\psi_n\right\rangle+\ldots
\end{equation}
The series (\ref{NC2}) converges in average. Namely, if $\left. s_n\right\rangle=\sum^n_{j=1}c_j\left.\psi_j\right\rangle$, then
\begin{equation}\label{NC3}
\left\|\left.\psi\right\rangle-\left. s_n\right\rangle\right\|=\int\limits_{\SL(2,\C)}
\left|\left.\psi\right\rangle-\left. s_n\right\rangle\right|^2dg\;\longrightarrow\;0\;\;\text{at}\;n\rightarrow\infty.
\end{equation}
Such convergence is called \emph{convergence in average}. Using Schmidt's orthogonalization process, we can always to orthogonalize the wave functions $\left.\psi_1\right\rangle$, $\left.\psi_2\right\rangle$, $\ldots$, $\left.\psi_n\right\rangle$, $\ldots$ such that $\left\langle\psi_i\right.\!\left|\psi_j\right\rangle=\delta_{ij}$.

Finally, the space $\bsH_\infty$, satisfying the von Neumann conditions (\ref{NC}) and (\ref{NC3}), is called \emph{an abstract Hilbert space}.

\subsection{Spin multiplets}
Let us consider further generalizations of the abstract Hilbert space $\bsH_\infty$. In  1927, Pauli \cite{Pau27} introduced the first theory of electron spin. The main idea of this theory lies in a \emph{doubling} of the space of wave functions. Let $\left|\psi_1\right\rangle$ and $\left|\psi_2\right\rangle$ be the vectors of $\bsH_\infty$. Then the doubling space should be defined by the formal linear combinations
\begin{equation}\label{CatState}
c_1\left|\psi_1\right\rangle+c_2\left|\psi_2\right\rangle,
\end{equation}
where $c_1$ and $c_2$ are complex coefficients. Hence it follows that $\bsH_\infty$ should be replaced by the tensor product
\begin{equation}\label{SpinHil}
\bsH^S_2\otimes\bsH_\infty.
\end{equation}
Let $\left|e_1\right\rangle$ and $\left|e_2\right\rangle$ be the basis vectors in $\bsH^S_2$, then any vector
\[
\left|\psi^S\right\rangle=\sum_j\left| x_j\right\rangle\otimes\left|\psi_j\right\rangle
\]
from $\bsH^S_2\otimes\bsH_\infty$ can be represented in the form
\begin{equation}\label{SpinVect}
\left|\psi^S\right\rangle=c_1\left|e_1\right\rangle\otimes\left|\psi_1\right\rangle+
c_2\left|e_2\right\rangle\otimes\left|\psi_2\right\rangle.
\end{equation}
A comparison of (\ref{SpinVect}) with the formal sum (\ref{CatState}) shows that the space (\ref{SpinHil}) presents an adequate mathematical description for the space of wave functions with electron spin.
%Representations of $\spin_+(1,3)\simeq\SU(2)\otimes\SU(2)$ in $\bsH^S_2\otimes\bsH_\infty$ are constructed via the %tensor product
%\[
%P^S(g)=P_2(g)\otimes P_\infty(g),
%\]
%where $g\in\spin_+(1,3)$. In turn, representing operators of the algebra $\mathfrak{su}(2)\otimes\mathfrak{su}(2)$ %are defined as
%\begin{equation}\label{AlgP}
%\widetilde{P}^S(A^c_k)=\widetilde{P}_2(A^c_k)\otimes\boldsymbol{1}_\infty+
%\boldsymbol{1}_2\otimes\widetilde{P}_\infty(A^c_k).
%\end{equation}
%The first summand in (\ref{AlgP}) is called an operator of \emph{orbital momentum} of the particle and denoted as %$\sM_k$; the second summand in (\ref{AlgP}) is called a \emph{spin operator} of the particle and denoted via $\sJ_k$. %Therefore, from (\ref{AlgP}) we have a \emph{full momentum} of the particle,
%\[
%\sS_k=\sM_k+\sJ_k,\quad k=1,2,3.
%\]
\subsubsection{Spin doublets}
The first simplest spin multiplet is a \emph{spin doublet} constructed within $\bsH^S_2\otimes\bsH_\infty$. We have here two spin states: one state belongs to the spin-1/2 line, and other state belongs to the dual spin-1/2 line. The first spin doublet (see Fig.\,2) is
\begin{equation}\label{FDoublet}
\begin{diagram}
\node{\overset{(\frac{1}{2},0)}{\bullet}}\arrow{e,..,-}\node{\overset{(0,\frac{1}{2})}{\bullet}}\\
\node{\tfrac{1}{2}}\arrow{e,-}\node{-\tfrac{1}{2}}
\end{diagram}
\end{equation}
Here the second row means that the representation $(1/2,0)$ describes a particle for example, electron) with the spin value 1/2, and the representation $(0,1/2)$ describes a particle with the spin value -1/2. The doublet (\ref{FDoublet}) corresponds to the Dirac equation and for that reason should be called as a \emph{fundamental doublet}. On the other hand, we can construct the spin doublet using $(1,1/2)$- and $(1/2,1)$-representations:
\[
\begin{diagram}
\node{\overset{(1,\frac{1}{2})}{\bullet}}\arrow{e,..,-}\node{\overset{(\frac{1}{2},1)}{\bullet}}\\
\node{\tfrac{1}{2}}\arrow{e,-}\node{-\tfrac{1}{2}}
\end{diagram}
\]
It is easy to see that we have infinitely many spin doublets, where two different spin states 1/2 and -1/2 belong to spin-1/2 and dual spin-1/2 lines, respectively:
\[
\begin{diagram}
\node{\overset{(\frac{3}{2},1)}{\bullet}}\arrow{e,..,-}\node{\overset{(1,\frac{3}{2})}{\bullet}}\\
\node{\tfrac{1}{2}}\arrow{e,-}\node{-\tfrac{1}{2}}
\end{diagram},
\]
\[
\begin{diagram}
\node{\overset{(2,\frac{3}{2})}{\bullet}}\arrow{e,..,-}\node{\overset{(\frac{3}{2},2)}{\bullet}}\\
\node{\tfrac{1}{2}}\arrow{e,-}\node{-\tfrac{1}{2}}
\end{diagram},
\]
\[
\cdot\;\cdot\;\cdot\;\cdot\;\cdot\;\cdot\;\cdot\;\cdot\;\cdot\;
\]
\[
\begin{diagram}
\node{\overset{(\frac{59}{2},29)}{\bullet}}\arrow{e,..,-}\node{\overset{(29,\frac{59}{2})}{\bullet}}\\
\node{\tfrac{1}{2}}\arrow{e,-}\node{-\tfrac{1}{2}}
\end{diagram},
\]
\[
\cdot\;\cdot\;\cdot\;\cdot\;\cdot\;\cdot\;\cdot\;\cdot\;\cdot\;
\]
\subsubsection{Spin triplets}
The next spin multiplet is a \emph{spin triplet} constructed within the space $\bsH^S_3\otimes\bsH_\infty$. We have here three spin states: two states 1 and -1 belong to the spin-1 and dual spin-1 lines, and third spin state 0 belongs to the spin-0 line. As it follows from Fig.\,1, the first spin triplet (\emph{fundamental triplet}) is
\begin{equation}\label{FTriplet}
\begin{diagram}
\node{\overset{(1,0)}{\bullet}}\arrow{e,..,-}\node{\overset{(\frac{1}{2},\frac{1}{2})}{\bullet}}\arrow{e,..,-}
\node{\overset{(0,1)}{\bullet}}\\
\node{1}\arrow{e,-}\node{0}\arrow{e,-}\node{-1}
\end{diagram}
\end{equation}
It is obvious that there are infinitely many spin triplets. The next spin triplet, which follows after (\ref{FTriplet}), is
\[
\begin{diagram}
\node{\overset{(\frac{3}{2},\frac{1}{2})}{\bullet}}\arrow{e,..,-}\node{\overset{(1,1)}{\bullet}}\arrow{e,..,-}
\node{\overset{(\frac{1}{2},\frac{3}{2})}{\bullet}}\\
\node{1}\arrow{e,-}\node{0}\arrow{e,-}\node{-1}
\end{diagram}
\]
and so on.
\subsubsection{The space $\bsH^S_{2s+1}\otimes\bsH_\infty$}
As in the case of spin doublets and triplets we can construct spin quadruplets and other spin multiplets in a similar manner. For example, the first 6-dimensional multiplet (6-plet), defined in $\bsH^S_6\otimes\bsH_\infty$, is
\[
\begin{diagram}
\node{\overset{(\frac{5}{2},0)}{\bullet}}\arrow{e,..,-}\node{\overset{(2,\frac{1}{2})}{\bullet}}\arrow{e,..,-}
\node{\overset{(\frac{3}{2},1)}{\bullet}}\arrow{e,..,-}\node{\overset{(1,\frac{3}{2})}{\bullet}}\arrow{e,..,-}
\node{\overset{(\frac{1}{2},2)}{\bullet}}\arrow{e,..,-}\node{\overset{(0,\frac{5}{2})}{\bullet}}\\
\node{\tfrac{5}{2}}\arrow{e,-}\node{\tfrac{3}{2}}\arrow{e,-}\node{\tfrac{1}{2}}\arrow{e,-}\node{-\tfrac{1}{2}}\arrow{e,-}
\node{-\tfrac{3}{2}}\arrow{e,-}\node{-\tfrac{5}{2}}
\end{diagram}
\]
Generalizing this construction, we come to the following abstract Hilbert space:
\begin{equation}\label{SHil}
\bsH^S_{2s+1}\otimes\bsH_\infty,
\end{equation}
where $s=0,\frac{1}{2},1,\frac{3}{2},2,\ldots$, and $s=|l-\dot{l}|$. Of course, we have infinitely many \emph{spin singlets} in $\bsH^S_{2s+1}\otimes\bsH_\infty$. All these singlets belong to the spin-0 line and defined by representations $(0,0)$, $(1/2,1/2)$, $(1,1)$, $\ldots$, $(s,s)$, $\ldots$. The singlet, defined by the representation $(0,0)$, is called a \emph{fundamental singlet}.

Further, when $s$ is odd we have \emph{fermionic multiplets} in $\bsH^S_{2s+1}\otimes\bsH_\infty$ and, correspondingly, \emph{bosonic multiplets} when $s$ is even. All the spaces $\bsH^S_{2s+1}\otimes\bsH_\infty$ are nonseparable Hilbert spaces. All fermionic and bosonic multiplets in $\bsH^S_{2s+1}\otimes\bsH_\infty$ have their antiparticle counterparts which compound \textbf{\emph{antimatter}} (see Fig.\,7).
\begin{figure}[ht]
\[
\dgARROWPARTS=28
\dgARROWLENGTH=0.5em
\dgHORIZPAD=1.2em %1.5em
\dgVERTPAD=6.2ex %2ex
\begin{diagram}
%\node{\overset{(0,4)}{\bullet}}\arrow{n,..,-}\arrow[2]{e,..,-}\arrow{w,..,-}
%\node[2]{\overset{(\frac{1}{2},\frac{7}{2})}{\bullet}}\arrow{n,..,-}\arrow[2]{e,..,-}\arrow[2]{s,..,-}
%\node[2]{\overset{(1,3)}{\bullet}}
%\arrow{n,..,-}\arrow[2]{e,..,-}\arrow[2]{s,..,-}
%\node[2]{\overset{(\frac{3}{2},\frac{5}{2})}{\bullet}}\arrow{n,..,-}\arrow[2]{e,..,-}\arrow[2]{s,..,-}
%\node[2]{\overset{(2,2)}{\bullet}}\arrow[2]{s,-}\arrow{n,..,-}\arrow[2]{e,..,-}
%\node[2]{\overset{(\frac{5}{2},\frac{3}{2})}{\bullet}}\arrow{n,..,-}\arrow[2]{e,..,-}\arrow[2]{s,..,-}
%\node[2]{\overset{(3,1)}{\bullet}}\arrow{n,..,-}\arrow[2]{e,..,-}\arrow[2]{s,..,-}
%\node[2]{\overset{(\frac{7}{2},\frac{1}{2})}{\bullet}}\arrow{n,..,-}\arrow[2]{e,..,-}\arrow[2]{s,..,-}
%\node[2]{\overset{(4,0)}{\bullet}}\arrow{n,..,-}\arrow{e,..,-}\\
\node{\overset{(0,\frac{7}{2})}{\bullet}}\arrow[2]{e,..,-}\arrow[14]{s,l,20,..,-}{-\tfrac{7}{2}}
\node[2]{\overset{(\frac{1}{2},3)}{\bullet}}\arrow[2]{e,..,-}\arrow[2]{s,..,-}
\node[2]{\overset{(1,\frac{5}{2})}{\bullet}}\arrow[2]{e,..,-}\arrow[2]{s,..,-}
\node[2]{\overset{(\frac{3}{2},2)}{\bullet}}\arrow[2]{e,..,-}\arrow[2]{s,..,-}
\node[2]{\overset{(2,\frac{3}{2})}{\bullet}}\arrow[2]{e,..,-}\arrow[2]{s,..,-}
\node[2]{\overset{(\frac{5}{2},1)}{\bullet}}\arrow[2]{e,..,-}\arrow[2]{s,..,-}
\node[2]{\overset{(3,\frac{1}{2})}{\bullet}}\arrow[2]{e,..,-}\arrow[2]{s,..,-}
\node[2]{\overset{(\frac{7}{2},0)}{\bullet}}\arrow[14]{s,r,..,-}{\tfrac{7}{2}}\\
\node[2]{\overset{(0,3)}{\bullet}}\arrow[2]{e,..,-}\arrow[12]{s,l,..,-}{-3}\arrow[2]{n,..,-}
\node[2]{\overset{(\frac{1}{2},\frac{5}{2})}{\bullet}}\arrow[2]{e,..,-}\arrow[2]{s,..,-}\arrow[2]{n,..,-}
\node[2]{\overset{(1,2)}{\bullet}}\arrow[2]{e,..,-}\arrow[2]{s,..,-}\arrow[2]{n,..,-}
\node[2]{\overset{(\frac{3}{2},\frac{3}{2})}{\bullet}}\arrow[2]{s,-}\arrow[2]{e,..,-}\arrow[2]{n,-}
\node[2]{\overset{(2,1)}{\bullet}}\arrow[2]{e,..,-}\arrow[2]{s,..,-}\arrow[2]{n,..,-}
\node[2]{\overset{(\frac{5}{2},\frac{1}{2})}{\bullet}}\arrow[2]{e,..,-}\arrow[2]{s,..,-}\arrow[2]{n,..,-}
\node[2]{\overset{(3,0)}{\bullet}}\arrow[2]{n,..,-}\arrow[12]{s,r,..,-}{3}\\
\node[3]{\overset{(0,\frac{5}{2})}{\bullet}}\arrow[2]{e,..,-}\arrow[10]{s,l,..,-}{-\tfrac{5}{2}}
\node[2]{\overset{(\frac{1}{2},2)}{\bullet}}\arrow[2]{e,..,-}\arrow[2]{s,..,-}
\node[2]{\overset{(1,\frac{3}{2})}{\bullet}}\arrow[2]{e,..,-}\arrow[2]{s,..,-}
\node[2]{\overset{(\frac{3}{2},1)}{\bullet}}\arrow[2]{e,..,-}\arrow[2]{s,..,-}
\node[2]{\overset{(2,\frac{1}{2})}{\bullet}}\arrow[2]{e,..,-}\arrow[2]{s,..,-}
\node[2]{\overset{(\frac{5}{2},0)}{\bullet}}\arrow[10]{s,r,..,-}{\tfrac{5}{2}}\\
\node[4]{\overset{(0,2)}{\bullet}}\arrow[2]{e,..,-}\arrow[8]{s,l,..,-}{-2}
\node[2]{\overset{(\frac{1}{2},\frac{3}{2})}{\bullet}}\arrow[2]{e,..,-}\arrow[2]{s,..,-}
\node[2]{\overset{(1,1)}{\bullet}}\arrow[2]{s,-}\arrow[2]{e,..,-}
\node[2]{\overset{(\frac{3}{2},\frac{1}{2})}{\bullet}}\arrow[2]{e,..,-}\arrow[2]{s,..,-}
\node[2]{\overset{(2,0)}{\bullet}}\arrow[8]{s,r,..,-}{2}\\
\node[5]{\overset{(0,\frac{3}{2})}{\bullet}}\arrow[2]{e,..,-}\arrow[6]{s,l,..,-}{-\tfrac{3}{2}}
\node[2]{\overset{(\frac{1}{2},1)}{\bullet}}\arrow[2]{e,..,-}\arrow[2]{s,..,-}
\node[2]{\overset{(1,\frac{1}{2})}{\bullet}}\arrow[2]{e,..,-}\arrow[2]{s,..,-}
\node[2]{\overset{(\frac{3}{2},0)}{\bullet}}\arrow[6]{s,r,..,-}{\tfrac{3}{2}}\\
\node[6]{\overset{(0,1)}{\bullet}}\arrow[2]{e,..,-}\arrow[4]{s,l,..,-}{-1}
\node[2]{\overset{(\frac{1}{2},\frac{1}{2})}{\bullet}}\arrow[2]{s,-}\arrow[2]{e,..,-}
\node[2]{\overset{(1,0)}{\bullet}}\arrow[4]{s,r,..,-}{1}\\
\node[7]{\overset{(0,\frac{1}{2})}{\bullet}}\arrow[2]{e,..,-}\arrow[2]{s,l,..,-}{-\tfrac{1}{2}}
\node[2]{\overset{(\frac{1}{2},0)}{\bullet}}\arrow[2]{s,r,..,-}{\tfrac{1}{2}}\\
\node[8]{\overset{(0,0)}{\bullet}}\arrow[2]{s,-}\arrow[9]{w,-}\arrow[9]{e}\\
\node[7]{\overset{(0,\frac{1}{2})}{\bullet}}\arrow[2]{e,..,-}\arrow[2]{s,..,-}
\node[2]{\overset{(\frac{1}{2},0)}{\bullet}}\arrow[2]{s,..,-}\\
\node[6]{\overset{(0,1)}{\bullet}}\arrow[2]{e,..,-}\arrow[2]{s,..,-}
\node[2]{\overset{(\frac{1}{2},\frac{1}{2})}{\bullet}}\arrow[2]{s,-}\arrow[2]{e,..,-}
\node[2]{\overset{(1,0)}{\bullet}}\arrow[2]{s,..,-}\\
\node[5]{\overset{(0,\frac{3}{2})}{\bullet}}\arrow[2]{e,..,-}\arrow[2]{s,..,-}
\node[2]{\overset{(\frac{1}{2},1)}{\bullet}}\arrow[2]{e,..,-}\arrow[2]{s,..,-}
\node[2]{\overset{(1,\frac{1}{2})}{\bullet}}\arrow[2]{e,..,-}\arrow[2]{s,..,-}
\node[2]{\overset{(\frac{3}{2},0)}{\bullet}}\arrow[2]{s,..,-}\\
\node[4]{\overset{(0,2)}{\bullet}}\arrow[2]{e,..,-}\arrow[2]{s,..,-}
\node[2]{\overset{(\frac{1}{2},\frac{3}{2})}{\bullet}}\arrow[2]{e,..,-}\arrow[2]{s,..,-}
\node[2]{\overset{(1,1)}{\bullet}}\arrow[2]{s,-}\arrow[2]{e,..,-}
\node[2]{\overset{(\frac{3}{2},\frac{1}{2})}{\bullet}}\arrow[2]{e,..,-}\arrow[2]{s,..,-}
\node[2]{\overset{(2,0)}{\bullet}}\arrow[2]{s,..,-}\\
\node[3]{\overset{(0,\frac{5}{2})}{\bullet}}\arrow[2]{e,..,-}\arrow[2]{s,..,-}
\node[2]{\overset{(\frac{1}{2},2)}{\bullet}}\arrow[2]{e,..,-}\arrow[2]{s,..,-}
\node[2]{\overset{(1,\frac{3}{2})}{\bullet}}\arrow[2]{e,..,-}\arrow[2]{s,..,-}
\node[2]{\overset{(\frac{3}{2},1)}{\bullet}}\arrow[2]{e,..,-}\arrow[2]{s,..,-}
\node[2]{\overset{(2,\frac{1}{2})}{\bullet}}\arrow[2]{e,..,-}\arrow[2]{s,..,-}
\node[2]{\overset{(\frac{5}{2},0)}{\bullet}}\arrow[2]{s,..,-}\\
\node[2]{\overset{(0,3)}{\bullet}}\arrow[2]{e,..,-}\arrow[2]{s,..,-}
\node[2]{\overset{(\frac{1}{2},\frac{5}{2})}{\bullet}}\arrow[2]{e,..,-}\arrow[2]{s,..,-}
\node[2]{\overset{(1,2)}{\bullet}}\arrow[2]{e,..,-}\arrow[2]{s,..,-}
\node[2]{\overset{(\frac{3}{2},\frac{3}{2})}{\bullet}}\arrow[2]{s,-}\arrow[2]{e,..,-}
\node[2]{\overset{(2,1)}{\bullet}}\arrow[2]{e,..,-}\arrow[2]{s,..,-}
\node[2]{\overset{(\frac{5}{2},\frac{1}{2})}{\bullet}}\arrow[2]{e,..,-}\arrow[2]{s,..,-}
\node[2]{\overset{(3,0)}{\bullet}}\arrow[2]{s,..,-}\\
\node{\overset{(0,\frac{7}{2})}{\bullet}}\arrow[2]{e,..,-}
\node[2]{\overset{(\frac{1}{2},3)}{\bullet}}\arrow[2]{e,..,-}
\node[2]{\overset{(1,\frac{5}{2})}{\bullet}}\arrow[2]{e,..,-}
\node[2]{\overset{(\frac{3}{2},2)}{\bullet}}\arrow[2]{e,..,-}
\node[2]{\overset{(2,\frac{3}{2})}{\bullet}}\arrow[2]{e,..,-}
\node[2]{\overset{(\frac{5}{2},1)}{\bullet}}\arrow[2]{e,..,-}
\node[2]{\overset{(3,\frac{1}{2})}{\bullet}}\arrow[2]{e,..,-}
\node[2]{\overset{(\frac{7}{2},0)}{\bullet}}
%\node{\overset{(0,4)}{\bullet}}\arrow[2]{s,..,-}
%\node[2]{\overset{(\frac{1}{2},\frac{7}{2})}{\bullet}}\arrow[2]{e,..,-}\arrow[2]{s,..,-}
%\node[2]{\overset{(1,3)}{\bullet}}\arrow[2]{e,..,-}\arrow[2]{s,..,-}
%\node[2]{\overset{(\frac{3}{2},\frac{5}{2})}{\bullet}}\arrow[2]{e,..,-}\arrow[2]{s,..,-}
%\node[2]{\overset{(2,2)}{\bullet}}\arrow[2]{e,..,-}\arrow[2]{s,..,-}
%\node[2]{\overset{(\frac{5}{2},\frac{3}{2})}{\bullet}}\arrow[2]{e,..,-}\arrow[2]{s,..,-}
%\node[2]{\overset{(3,1)}{\bullet}}\arrow[2]{e,..,-}\arrow[2]{s,..,-}
%\node[2]{\overset{(\frac{7}{2},\frac{1}{2})}{\bullet}}\arrow[2]{e,..,-}\arrow[2]{s,..,-}
%\node[2]{\overset{(4,0)}{\bullet}}\arrow[2]{s,..,-}
\end{diagram}
\]
%\vspace{0.5cm}
\begin{center}{\small {\bf Fig.\,7:} Matter and antimatter spin multiplets in $\bsH^S_{2s+1}\otimes\bsH_\infty$.}\end{center}
\end{figure}
\subsubsection{Many states or many particles?}
We can imagine that \emph{one and the same particle} (for example, electron) has two spin states with the spin +1/2 or -1/2. However, the electron without definite value of the spin is never observed in nature and presents itself an abstract notion. For that reason from an alternative viewpoint it follows that there exist \emph{two} elementary particles: the electron with the spin +1/2 and the electron with the spin -1/2, whereas a `simple electron' does not exist in nature. It is obvious that the same proposition holds for other spin multiplets.

\subsubsection{Charge multiplets}
%\subsubsection{Charge doublet}
In 1932, Heisenberg \cite{Hei32} proposed to consider proton and
neutron as two different states of the one and the same particle
(nucleon). The Heisenberg theory of proton-neutron states (\emph{a
charge doublet}) formally coincides with the theory of electron spin
states proposed by Pauli. The main object of the Heisenberg theory
is an abstract Hilbert space of the type
\[
\bsH^Q_2\otimes\bsH_\infty,
\]
where $\bsH^Q_2$ is \emph{a charge space} associated with the
fundamental representation of the group $\SU(2)$.
%Operators of the
%momentum type, related with $\bsH^Q_2$, are defined by the matrices
%\[
%I_1=\begin{bmatrix} 0 & \frac{1}{2}\\
%\frac{1}{2} & 0
%\end{bmatrix},\quad
%I_2=\begin{bmatrix} 0 & -\frac{i}{2}\\
%\frac{i}{2} & 0
%\end{bmatrix},\quad
%I_3=\begin{bmatrix} \frac{1}{2} & 0\\
%0 & -\frac{1}{2}
%\end{bmatrix}.
%\]
%$I_k$ are called \emph{operators of isotopic spin}. Further, \emph{a
%charge operator} $Q$ is defined as
%\[
%Q=I_3+\frac{1}{2}\boldsymbol{1}_2=\begin{bmatrix} 1 & 0\\
%0 & 0
%\end{bmatrix}.
%\]
%Eigenvectors of $Q$ have coordinates $(1,0)$ and $(0,1)$, that is,
%these eigenvectors coincide with the basis vectors $\left| e_1\right\rangle$
%and $\left| e_2\right\rangle$:
%\[
%Q\left| e_1\right\rangle=1\cdot\left| e_1\right\rangle,\quad Q\left| e_2\right\rangle=0\cdot\left| e_2\right\rangle,
%\]
%where 1 and 0 are eigenvalues of $Q$. Operators
%\[
%I_+=I_1+iI_2=\begin{bmatrix} 0 & 1\\
%0 & 0
%\end{bmatrix},\quad
%I_-=I_1-iI_2=\begin{bmatrix} 0 & 0\\
%1 & 0
%\end{bmatrix}
%\]
%convert vectors of proton and neutron states into each other:
%\[
%I_+\left| e_2\right\rangle=\left| e_1\right\rangle,\quad I_-\left| e_1\right\rangle=\left| e_2\right\rangle.
%\]

%\subsubsection{Charge triplet}
In 1938, Kemmer \cite{Kem38} generalized the Heisenberg theory to
the case of a particle with three different charge states $1$, $0$,
$-1$. \emph{A charge triplet} is constructed within an abstract
Hilbert space of the type
\[
\bsH^Q_3\otimes\bsH_\infty,
\]
where $\bsH^Q_3$ is \emph{a charge space} associated with the
representation $\boldsymbol{\tau}_{1,0}$ of the group $\SU(2)$.

%Operators of the momentum type, related with $\bsH^Q_3$, are defined
%by the matrices
%\[
%I_1=\begin{bmatrix} 0 & \frac{\sqrt{2}}{2} & 0\\
%\frac{\sqrt{2}}{2} & 0 & \frac{\sqrt{2}}{2}\\
%0 & \frac{\sqrt{2}}{2} & 0
%\end{bmatrix},\quad
%I_2=\begin{bmatrix} 0 & -i\frac{\sqrt{2}}{2} & 0\\
%i\frac{\sqrt{2}}{2} & 0 & -i\frac{\sqrt{2}}{2}\\
%0 & i\frac{\sqrt{2}}{2} & 0
%\end{bmatrix},\quad
%I_3=\begin{bmatrix} 1 & 0 & 0\\
%0 & 0 & 0\\
%0 & 0 & -1
%\end{bmatrix}.
%\]
%In turn, \emph{a charge operator} $Q$ is defined as
%\[
%Q=I_3=\begin{bmatrix} 1 & 0 & 0\\
%0 & 0 & 0\\
%0 & 0 & -1
%\end{bmatrix}.
%\]
%Eigenvectors of $Q$ have coordinates $(1,0,0)$, $(0,0,0)$ and
%$(0,0,-1)$, that is, these eigenvectors coincide with the basis
%vectors $\left| e_1\right\rangle$, $\left| e_2\right\rangle$ and
%$\left| e_3\right\rangle$:
%\[
%Q\left| e_1\right\rangle=1\cdot\left| e_1\right\rangle,\quad Q\left|
%e_2\right\rangle=0\cdot\left| e_2\right\rangle,\quad Q\left|
%e_3\right\rangle=-1\cdot\left| e_3\right\rangle,
%\]
%where $1$, $0$, $-1$ are eigenvalues of $Q$. Further, operators
%\[
%I_+=I_1+iI_2=\begin{bmatrix} 0 & \sqrt{2} & 0\\
%0 & 0 & \sqrt{2}\\
%0 & 0 & 0
%\end{bmatrix},\quad
%I_-=I_1-iI_2=\begin{bmatrix} 0 & 0 & 0\\
%\sqrt{2} & 0 & 0\\
%0 & \sqrt{2} & 0
%\end{bmatrix}
%\]
%satisfy the relations
%\[
%\left[Q,I_+\right]=I_+,\quad\left[Q,I_-\right]=-I_-,\quad\left[I_+,I_-\right]=2Q.
%\]

\subsection{The space $\bsH^S\otimes\bsH^Q\otimes\bsH_\infty$}
The spaces $\bsH^S_{2s+1}\otimes\bsH_\infty$ (spin multiplets) and $\bsH^Q\otimes\bsH_\infty$ (charge multiplets), considered in the previous sections, lead naturally to the following generalization of the abstract Hilbert space. Let
\begin{equation}\label{SQHil}
\bsH^S\otimes\bsH^Q\otimes\bsH_\infty
\end{equation}
be a tensor product of $\bsH_\infty$ and a \emph{spin-charge space} $\bsH^S\otimes\bsH^Q$. State vectors of (\ref{SQHil}) describe particles of the spin $s=|l-\dot{l}|$ and charge $Q$ with the mass $m$ defined by the formula (\ref{MGY}). All the totality of state vectors of $\bsH^S\otimes\bsH^Q\otimes\bsH_\infty$ is divided into six classes according to the orbits $\boldsymbol{O}$ in the Wigner interpretation. Moreover, state vectors are grouped into spin lines: spin-0 line, spin-1/2 line, spin-1 line and so on (bosonic and fermionic lines). The each state vector presents itself an irreducible representation $\boldsymbol{\tau}_{l\dot{l}}$ of $\spin_+(1,3)$ which acts in the space $\Sym_{(k,r)}$ (Hilbert space of elementary particle\footnote{Recall that a \emph{superposition} of the state vectors forms an irreducible unitary representation $U$ (quantum elementary particle system) of the group $\spin_+(1,3)\simeq\SL(2,\C)$ which acts in the Hilbert space $\sH_\infty$. At the reduction of the superposition we have $U\rightarrow\boldsymbol{\tau}_{l\dot{l}}$ and $\sH_\infty\rightarrow\Sym_{(k,r)}$. For example, in the case of electron we have two spin states: the state 1/2 described by the representation $\boldsymbol{\tau}_{1/2,0}$ on the spin-1/2 line and the state -1/2 described by $\boldsymbol{\tau}_{0,1/2}$ on the dual spin-1/2 line. The representations $\boldsymbol{\tau}_{1/2,0}$ and $\boldsymbol{\tau}_{0,1/2}$ act in the spaces $\Sym_{(1,0)}$ and $\Sym_{(0,1)}$, respectively. The superposition of these two spin states leads to a unitary representation $U^{m,+,1/2}$ of the orbit $\boldsymbol{O}^+_m$ which acts in the Hilbert space $\sH^{m,+,1/2}_\infty\simeq\bsH^S_2\otimes\bsH_\infty$. At the reduction we have $U^{m,+,1/2}\rightarrow\boldsymbol{\tau}_{1/2,0}$ or $U^{m,+,1/2}\rightarrow\boldsymbol{\tau}_{0,1/2}$ and $\sH^{m,+,1/2}_\infty\rightarrow\Sym_{(1,0)}$ or $\sH^{m,+,1/2}_\infty\rightarrow\Sym_{(0,1)}$.}). The charge $Q$ takes three values $-1$, $0$, $+1$\footnote{Of course, in the case of charge quadruplet (for example, $\boldsymbol{\Delta}$-quadruplet of the spin 3/2) we have four values $-1$, $0$, $1$, $2$.}, where the values $-1$, $+1$ correspond to charged particles, and the value $0$ corresponds to neutral (or truly neutral) particles. In the underlying spinor structure charged particles are described by the complex representations of $\spin_+(1,3)$, for which the pseudoatomorphism $\cA\rightarrow\overline{\cA}$ is not trivial ($\F=\C$) and an action of $\cA\rightarrow\overline{\cA}$ replaces complex representations (charge state $-1$) by complex conjugate representations (charge state $+1$). The neutral particles (charge state $0$) are described by the real representations of $\spin_+(1,3)$, for which the transformation $\cA\rightarrow\overline{\cA}$ is also not trivial ($\F=\R$, $\K\simeq\BH$), that is, we have here particle-antiparticle interchange. In turn, truly neutral particles are described by the real representations of $\spin_+(1,3)$ for which the action of the pseudoautomorphism $\cA\rightarrow\overline{\cA}$ is trivial ($\F=\R$, $\K\simeq\R$). With the aim to distinguish this case from the neutral particles (state 0) we denote this charge state as $\overline{0}$. Therefore, the spinor structure with the help of $\cA\rightarrow\overline{\cA}$ allows us to separate real representations for neutral (charge state 0) and truly neutral (charge state $\overline{0}$) particles.

Vectors of $\bsH^S\otimes\bsH^Q\otimes\bsH_\infty$ have the form
\begin{equation}\label{VectH}
\left|\sA\right\rangle=\left.\left|\boldsymbol{\tau}_{l\dot{l}},\,\Sym_{(k,r)},\,\cl_{p,q},\,\dS_{(p+q)/2},\,C^{a,b,c,d,e,f,g},\,
\right.\ldots\right\rangle,
\end{equation}
where $\boldsymbol{\tau}_{l\dot{l}}$ is a representation of the
proper orthocronous Lorentz group, $\Sym_{(k,r)}$ is a
representation space of $\boldsymbol{\tau}_{l\dot{l}}$ with the degree (\ref{Degree}), $\cl_{p,q}$
is a Clifford algebra associated with
$\boldsymbol{\tau}_{l\dot{l}}$, $\dS_{(p+q)/2}$ is a spinspace
associated with $\cl_{p,q}$, $C^{a,b,c,d,e,f,g}$ is a $CPT$ group
defined within $\cl_{p,q}$. It is obvious that the main object,
defining the vectors of the type (\ref{VectH}), is the
representation $\boldsymbol{\tau}_{l\dot{l}}$. Objects
$\cl_{p,q}$, $\dS_{(p+q)/2}$ and $C^{a,b,c,d,e,f,g}$ belong to spinor structure. In
other words, the representation $\boldsymbol{\tau}_{l\dot{l}}$
corresponds to the each vector of $\bsH^S\otimes\bsH^Q\otimes\bsH_\infty$, and a set of all possible
representations $\boldsymbol{\tau}_{l\dot{l}}$ generates the all
abstract Hilbert space $\bsH^S\otimes\bsH^Q\otimes\bsH_\infty$ with the vectors (\ref{VectH}).
\subsubsection{Nucleon doublet}
Let us suppose that vectors $\left| e_1\right\rangle$ and $\left| e_2\right\rangle$
of the charge doublet have the form
\begin{eqnarray}
\left| e_1\right\rangle&=&\left.\left|\boldsymbol{\tau}^c_{l\dot{l}},\,\Sym_{(k,r)},\,
\C\otimes\cl_{p,q},\,\dS_{2^{n/2}},\,
C^{a,b,c,d,e,f,g}\,\right.\right\rangle,\nonumber\\
\left| e_2\right\rangle&=&\left.\left|\boldsymbol{\tau}^r_{l\dot{l}},\,\Sym_{(k,r)},\,
\cl_{p,q},\,\dS_{2^{n/2}},\,
C^{a,b,c,d,e,f,g}\,\right.\right\rangle,\nonumber
\end{eqnarray}
where in the case of $\left| e_1\right\rangle$ the complex representation
$\boldsymbol{\tau}^c_{l\dot{l}}$
belongs to spin-1/2 line (see Fig.\,2),
$\Sym_{(k,r)}$ is a representation space of
$\boldsymbol{\tau}^c_{l\dot{l}}$ with the degree (\ref{Degree}),
$\C\otimes\cl_{p,q}$ is a
Clifford algebra associated with
$\boldsymbol{\tau}^c_{l\dot{l}}$,
$\dS_{2^{n/2}}$ is a spinspace over the
field $\F=\C$, $n=p+q$. The vector $\left| e_1\right\rangle$ describes
\emph{a charged fermion of the spin-1/2}. In the case of $\left| e_2\right\rangle$ we have a real representation
$\boldsymbol{\tau}^r_{l\dot{l}}$
belonging to the spin-1/2 line. In contrast to $\left| e_1\right\rangle$,
$\cl_{p,q}$ is a Clifford algebra over the
field $\F=\R$, where $\cl_{p,q}$ is a Clifford algebra over $\F=\R$
with the quaternionic division ring $\K\simeq\BH$, the types
$p-q\equiv 4,6\pmod{8}$. For that reason the vector $\left| e_2\right\rangle$ describes \emph{a neutral fermion of the spin-1/2 which
admits particle-antiparticle conjugation}.

Returning to 3540-dimensional proton representation space
$\boldsymbol{\tau}^c_{\frac{59}{2},29}$,
considered in the section 2.2.2, we can define nucleon doublet, that is,
proton $\left| e_1\right\rangle=\bP$ and neutron $\left|
e_2\right\rangle=\bN$ states. The representation
$\boldsymbol{\tau}^c_{\frac{59}{2},29}$
acts in the space $\Sym_{(59,58)}$ of the degree 3540. Let
$\C\otimes\cl_{119,115}\simeq\C_{234}$
be a Clifford algebra associated with the proton state $\left|
e_1\right\rangle=\bP$. The real subalgebra $\cl_{119,115}$ has the
quaternionic division ring $\K\simeq\BH$, type $p-q\equiv
4\pmod{8}$, and the parity with $P^2=1$. Further, let
$\dS_{2^{117}}$ be a complex spinspace
associated with $\left| e_1\right\rangle=\bP$ ($CPT$ group of $\bP$
acts in this spinspace). In turn, the neutron state $\left|
e_2\right\rangle=\bN$ is described by a real representation
$\boldsymbol{\tau}^r_{\frac{59}{2},29}$
belonging also to spin-1/2 line with the Clifford algebra
$\cl_{119,115}$ and a quaternionic
spinspace $\dS_{2^{117}}(\BH)$. Thus, for the
vectors of the nucleon doublet we have
\begin{eqnarray}
\bP&=&\left.\left|\boldsymbol{\tau}^c_{\frac{59}{2},29},\,\Sym_{(59,58)},\,
\C_{234},\,\dS_{2^{117}},\,
P^2=1\,\right.\right\rangle,\nonumber\\
\bN&=&\left.\left|\boldsymbol{\tau}^r_{\frac{59}{2},29},\,\Sym_{(59,58)},\,
\cl_{119,115},\,\dS_{2^{117}},\,
P^2=1\,\right.\right\rangle.\nonumber
\end{eqnarray}
\subsubsection{$\boldsymbol{\Sigma}$-triplet}
In this case we take the vectors of the charge triplet in the form
\begin{eqnarray}
\left|
e_1\right\rangle&=&\left.\left|\boldsymbol{\tau}^c_{l\dot{l}},\,\Sym_{(k,r)},\,
\C\otimes\cl_{p,q},\,\dS_{2^{n/2}},\,
C^{a,b,c,d,e,f,g}\,\right.\right\rangle,\nonumber\\
\left|
e_2\right\rangle&=&\left.\left|\boldsymbol{\tau}^r_{l\dot{l}},\,\Sym_{(k,r)},\,
\cl_{p,q},\,\dS_{2^{n/2}},\,
C^{a,b,c,d,e,f,g}\,\right.\right\rangle,\nonumber\\
\left|
e_3\right\rangle&=&\left.\left|\boldsymbol{\tau}^c_{\dot{l}l},\,\Sym_{(r,k)},\,
\overset{\ast}{\C}\otimes\cl_{p,q},\,\hat{\dS}_{2^{n/2}},\,
C^{a,b,c,d,e,f,g}\,\right.\right\rangle.\nonumber
\end{eqnarray}

According to the mass formula
(\ref{MGY}) and the interlocking scheme (Fig.\,2) the next representation on the spin-1/2 line after
$\boldsymbol{\tau}^c_{\frac{59}{2},29}$
(nucleon doublet) is a 4556-dimensional complex representation
$\boldsymbol{\tau}^c_{\frac{67}{2},33}$, since $m_\Sigma/m_e\approx 2280$.
This representation can be identified with the
$\boldsymbol{\Sigma}$-triplet. We have here three charge states:
$\left| e_1\right\rangle=\boldsymbol{\Sigma}^+$, $\left|
e_2\right\rangle=\boldsymbol{\Sigma}^0$ and $\left|
e_3\right\rangle=\boldsymbol{\Sigma}^-$. The representation
$\boldsymbol{\tau}^c_{\frac{67}{2},33}$
acts in the space $\Sym_{(67,66)}$ of the degree 4556. Let
$\C\otimes\cl_{135,131}\simeq\C_{266}$
be a Clifford algebra associated with the state $\left|
e_1\right\rangle=\boldsymbol{\Sigma}^+$. The real subalgebra
$\cl_{135,131}$ has the quaternionic division ring $\K\simeq\BH$
(the type $p-q\equiv 4\pmod{8}$) and, therefore, the parity with
$P^2=1$. Further, let $\dS_{2^{133}}$ be a
complex spinspace associated with $\left|
e_1\right\rangle=\boldsymbol{\Sigma}^+$ and also with $\left|
e_3\right\rangle=\boldsymbol{\Sigma}^-$ ($CPT$ groups of
$\boldsymbol{\Sigma}^+$ and $\boldsymbol{\Sigma}^-$ act in this
spinspace). In turn, the state $\left|
e_2\right\rangle=\boldsymbol{\Sigma}^0$ is described by a real
representation
$\boldsymbol{\tau}^r_{\frac{67}{2},33}$
belonging also to spin-1/2 line with the Clifford algebra
$\cl_{135,131}$ and a quaternionic
spinspace $\dS_{2^{133}}(\BH)$. Thus, for the
vectors of the $\boldsymbol{\Sigma}$-triplet we have
\begin{eqnarray}
\boldsymbol{\Sigma}^+&=&\left.\left|\boldsymbol{\tau}^c_{\frac{67}{2},33},\,\Sym_{(67,66)},\,
\C_{266},\,\dS_{2^{133}},\,
P^2=1\,\right.\right\rangle,\nonumber\\
\boldsymbol{\Sigma}^0&=&\left.\left|\boldsymbol{\tau}^r_{\frac{67}{2},33},\,\Sym_{(67,66)},\,
\cl_{135,131},\,\dS_{2^{133}},\,
P^2=1\,\right.\right\rangle,\nonumber\\
\boldsymbol{\Sigma}^-&=&\left.\left|\boldsymbol{\tau}^c_{33,\frac{67}{2}},\,\Sym_{(66,67)},\,
\overset{\ast}{\C}_{266},\,\hat{\dS}_{2^{133}},\,
P^2=1\,\right.\right\rangle.\nonumber
\end{eqnarray}
\subsubsection{$\boldsymbol{\pi}$-triplet}
On the other hand, we have $\boldsymbol{\pi}$-triplet on the spin-0
line. As is known, $m_\pi/m_e\approx 270$, therefore, we assume that
$\boldsymbol{\pi}$-triplet can be described within 529-dimensional
complex representation
$\boldsymbol{\tau}^c_{11,11}$
belonging to spin-0 line. In this case we have three charge states:
$\left| e_1\right\rangle=\boldsymbol{\pi}^+$, $\left|
e_2\right\rangle=\boldsymbol{\pi}^0$, $\left|
e_3\right\rangle=\boldsymbol{\pi}^-$. The representation
$\boldsymbol{\tau}^c_{11,11}$ acts
in the space $\Sym_{(22,22)}$ of the degree 529. Let
$\C\otimes\cl_{45,43}\simeq\C_{88}$
be a Clifford algebra associated with the state $\left|
e_1\right\rangle=\boldsymbol{\pi}^+$. The real subalgebra
$\cl_{45,43}$ has the real division ring $\K\simeq\R$, the type
$p-q\equiv 2\pmod{8}$, and, therefore, we have here the parity with
$P^2=-1$. Further, let $\dS_{2^{44}}$ be a
complex spinspace associated with $\left|
e_1\right\rangle=\boldsymbol{\pi}^+$ and also with $\left|
e_3\right\rangle=\boldsymbol{\pi}^-$ ($CPT$ groups of
$\boldsymbol{\pi}^+$ and $\boldsymbol{\pi}^-$ act in this
spinspace). In turn, the state $\left|
e_2\right\rangle=\boldsymbol{\pi}^0$ is described by a real
representation
$\boldsymbol{\tau}^r_{11,11}$
belonging also to spin-0 line with the Clifford algebra
$\cl_{45,43}$ and a real spinspace
$\dS_{2^{44}}(\R)$. In contrast to
$\boldsymbol{\Sigma}$-triplet (state $\left|
e_2\right\rangle=\boldsymbol{\Sigma}^0$), the state $\left|
e_2\right\rangle=\boldsymbol{\pi}^0$ in $\boldsymbol{\pi}$-triplet
is described within the algebra $\cl_{23,21}$ over the field $\F=\R$
with the real ring $\K\simeq\R$ that corresponds to \emph{truly
neutral particles} (see section 3.1.2). Thus, for the vectors of the
$\boldsymbol{\pi}$-triplet we have
\begin{eqnarray}
\boldsymbol{\pi}^+&=&\left.\left|\boldsymbol{\tau}^c_{11,11},\,\Sym_{(22,22)},\,
\C_{88},\,\dS_{2^{44}},\,
P^2=-1\,\right.\right\rangle,\nonumber\\
\boldsymbol{\pi}^0&=&\left.\left|\boldsymbol{\tau}^r_{11,11},\,\Sym_{(22,22)},\,
\cl_{45,43},\,\dS_{2^{44}},\,
P^2=-1\,\right.\right\rangle,\nonumber\\
\boldsymbol{\pi}^-&=&\left.\left|\boldsymbol{\tau}^c_{11,11},\,\Sym_{(22,22)},\,
\overset{\ast}{\C}_{88},\,\hat{\dS}_{2^{44}},\,
P^2=-1\,\right.\right\rangle.\nonumber
\end{eqnarray}

\subsubsection{Superselection rules}
We consider further a general structure of the abstract Hilbert space $\bsH^S\otimes\bsH^Q\otimes\bsH_\infty$. Let $\left|\boldsymbol{\Psi}\right\rangle$ be the vector of $\bsH^S\otimes\bsH^Q\otimes\bsH_\infty$, then $e^{i\alpha}\left|\boldsymbol{\Psi}\right\rangle$, where $\alpha$ runs all real numbers and $\sqrt{\left\langle\boldsymbol{\Psi}\right.\!\left|\boldsymbol{\Psi}\right\rangle}=1$, is called a \emph{unit ray}. All the states of physical (quantum) system are described by unit rays. We assume that a basic correspondence between physical states and elements of the space $\bsH^S\otimes\bsH^Q\otimes\bsH_\infty$ includes a \emph{superposition principle} of quantum theory, that is, there exists such a collection of basic states that arbitrary states can be constructed from them with the help of linear superpositions.

However, as is known \cite{WWW52} not all unit rays are physically realizable. There exist physical restrictions (superselection rules) on execution of superposition principle. In 1952, Wigner, Wightman and Wick \cite{WWW52} showed that existence of superselection rules is related with the measurability of relative phase of the superposition. It means that a pure state cannot be realized in the form of superposition of some states, for example, there is no a pure state consisting of fermion and boson (superselection rule on the spin). In the space $\bsH^S\otimes\bsH^Q\otimes\bsH_\infty$ there are superselection rules on the spin, parity, baryon number, lepton number\footnote{At this moment it is not possible to enumerate all the superselection rules for $\bsH^S\otimes\bsH^Q\otimes\bsH_\infty$.}. We divide the space $\bsH^S\otimes\bsH^Q\otimes\bsH_\infty$ on the subsets (\emph{coherent subspaces}) according to superselection rules. The superposition principle is executed in the each coherent subspace. For example, spin lines in $\bsH^S\otimes\bsH^Q\otimes\bsH_\infty$ form coherent subspaces corresponding to superselection rule on the spin.
\subsubsection{Group action on $\bsH^S\otimes\bsH^Q\otimes\bsH_\infty$}
We assume that one and the same quantum system can be described by the two different ways in one and the same coherent subspace of $\bsH^S\otimes\bsH^Q\otimes\bsH_\infty$ one time by the rays $\boldsymbol{\Psi}_1$, $\boldsymbol{\Psi}_2$, $\ldots$ and other time by the rays $\boldsymbol{\Psi}^\prime_1$, $\boldsymbol{\Psi}^\prime_2$, $\ldots$. One can say that we have here a symmetry of the quantum system when one and the same physical state is described with the help of $\boldsymbol{\Psi}_1$ in the first case and with the help of $\boldsymbol{\Psi}^\prime_1$ in the second case such that probabilities of transitions are the same. Therefore, we have a mapping $\hat{T}$ between the rays $\boldsymbol{\Psi}_1$ and $\boldsymbol{\Psi}^\prime_1$. Since only the absolute values are invariant, then the transformation $\hat{T}$ in $\boldsymbol{\Psi}_1$, $\boldsymbol{\Psi}_2$, $\ldots$ should be unitary or antiunitary. These two possibilities are realized in the case when a coherent subspace (or all the space $\bsH^S\otimes\bsH^Q\otimes\bsH_\infty$) is defined over the complex field $\F=\C$, since the complex field has two (and only two) automorphisms preserving absolute values: an identical automorphism and complex conjugation. When the coherent subspace (or all $\bsH^S\otimes\bsH^Q\otimes\bsH_\infty$) is defined over the real field $\F=\R$ we have only unitary transformations $\hat{T}$, since the real field has only one identical automorphism.

Let $\left|\boldsymbol{\psi}_1\right\rangle$, $\left|\boldsymbol{\psi}_2\right\rangle$, $\ldots$ be the unit vectors from the rays $\boldsymbol{\Psi}_1$, $\boldsymbol{\Psi}_2$, $\ldots$ and let $\left|\boldsymbol{\psi}^\prime_1\right\rangle$, $\left|\boldsymbol{\psi}^\prime_2\right\rangle$, $\ldots$ be the unit vectors from the rays $\boldsymbol{\Psi}^\prime_1$, $\boldsymbol{\Psi}^\prime_2$, $\ldots$ such that a correspondence $\left|\boldsymbol{\psi}_1\right\rangle\leftrightarrow\left|\boldsymbol{\psi}^\prime_1\right\rangle$, $\left|\boldsymbol{\psi}_1\right\rangle\leftrightarrow\left|\boldsymbol{\psi}^\prime_1\right\rangle$, $\ldots$ is unitary or antiunitary. The first collection corresponds to the states $\{s\}$ and the second collection corresponds to transformed states $\{gs\}$. We choose the vectors $\left|\boldsymbol{\psi}_1\right\rangle\in\boldsymbol{\Psi}_1$, $\left|\boldsymbol{\psi}_2\right\rangle\in\boldsymbol{\Psi}_2$, $\ldots$ and $\left|\boldsymbol{\psi}^\prime_1\right\rangle\in\boldsymbol{\Psi}^\prime_1$, $\left|\boldsymbol{\psi}^\prime_2\right\rangle\in\boldsymbol{\Psi}^\prime_2$, $\ldots$ such that
\begin{equation}\label{Action}
\left|\boldsymbol{\psi}^\prime_1\right\rangle=T_g\left|\boldsymbol{\psi}_1\right\rangle,\quad
\left|\boldsymbol{\psi}^\prime_2\right\rangle=T_g\left|\boldsymbol{\psi}_2\right\rangle,\quad\ldots
\end{equation}
It means that if $\left|\boldsymbol{\psi}_1\right\rangle$ is the vector associated with the ray $\boldsymbol{\Psi}_1$, then $T_g\left|\boldsymbol{\psi}_1\right\rangle$ is the vector associated with the ray $\boldsymbol{\Psi}^\prime_1$. If there exist two operators $T_g$ and $T_{g^\prime}$ with the property (\ref{Action}), then they can be distinguished by only a constant factor. Therefore,
\begin{equation}\label{Product}
T_{gg^\prime}=\omega(g,g^\prime)T_gT_{g^\prime},
\end{equation}
where $\omega(g,g^\prime)$ is a phase factor. Representations of the type (\ref{Product}) are called \emph{ray (projective) representations}. It means also that we have here a correspondence between physical states and rays in the abstract Hilbert space $\bsH^S\otimes\bsH^Q\otimes\bsH_\infty$. Hence it follows that the ray representation $T$ of a \emph{topological group} $G$ is a continuous homomorphism $T:\;G\rightarrow L(\hat{H})$, where $L(\hat{H})$ is a set of linear operators in the projective space $\hat{H}$ endowed with a factor-topology according to the mapping $\hat{H}\rightarrow\bsH^S\otimes\bsH^Q\otimes\bsH_\infty$, that is, $\left|\boldsymbol{\psi}\right\rangle\rightarrow\boldsymbol{\Psi}$. However, when $\omega(g,g^\prime)\neq 1$ we cannot to apply the mathematical theory of usual group representations. With the aim to avoid this obstacle we construct a more large group $\bcE$ in such manner that usual representations of $\bcE$ give all nonequivalent ray representations (\ref{Product}) of the group $G$. This problem can be solved by the \emph{lifting} of projective representations of $G$ to usual representations of the group $\bcE$. Let $\bcK$ be an Abelian group generated by the multiplication of nonequivalent phases $\omega(g,g^\prime)$ satisfying the condition
\[
\omega(g,g^\prime)\omega(gg^\prime,g^{\prime\prime})=\omega(g^\prime,g^{\prime\prime})\omega(g,g^\prime,g^{\prime\prime}).
\]
Let us consider the pairs $(\omega,x)$, $\omega\in\bcK$, $x\in G$, in particular, $\bcK=\{(\omega,e)\}$, $G=\{(e,x)\}$. The pairs $(\omega,x)$ form a group with the following multiplication law: $(\omega_1,x_1)(\omega_2,x_2)=(\omega_1\omega(x_1,x_2)\omega_2,x_1x_2)$. The group $\bcE=\{(\omega,x)\}$ is called a \emph{central extension} of the group $G$ via the group $\bcK$. Vector representations of the group $\bcE$ contain all the ray representations of the group $G$. Hence it follows that a symmetry group $G$ of physical system induces a unitary or antiunitary representation $T$ of invertible mappings of the space $\bsH^S\otimes\bsH^Q\otimes\bsH_\infty$ into itself, which is a representation of the central extension $\bcE$ of $G$.

Below we consider a symmetry group $G$ as one from the sequence of unitary unimodular groups $\SU(2)$, $\SU(3)$, $\ldots$, $\SU(N)$, $\ldots$ (groups of internal symmetries) which act in the space $\bsH^S\otimes\bsH^Q\otimes\bsH_\infty$.

\section{$\SU(3)$ symmetry}
In 1961, Gell-Mann \cite{Gel61} and Ne'eman \cite{Nee61} proposed a
wide generalization of charge multiplets. The main idea of this
generalization lies in the assumption that the charge multiplets of
the group $\SU(2)$ can be unified within a more large group, for
example, the group $\SU(3)$. In this context the isospin group
$\SU(2)$ is understood as a subgroup of $\SU(3)$,
$\SU(2)\subset\SU(3)$. In accordance with $\SU(3)$-theory, baryons
and mesons are described within irreducible representations
(supermultiplets) of the group $\SU(3)$.

As is known, hadrons are divided into charge multiplets, and the
each hadron is described by a following number collection:
$(B,\,s,\,P,\,Q,\,Y,\,I)$, where $B$ is a baryon number, $s$ is a
spin, $P$ is a parity, $Q$ is a charge, $Y$ is a hypercharge
(doubled mean value of the of the all particles in the multiplet),
$I$ is an isospin. The number of particles in the charge multiplet is $M=2I+1$. The spin $s$ and parity $P$ are external parameters with respect to $\SU(3)$-theory.

\subsection{Representations of $\SU(3)$}
Let $G=\SU(3)$ be the group of internal symmetries acting in the Hilbert space $\bsH^S\otimes\bsH^Q\otimes\bsH_\infty$ by means of a central extension $\bcE$ (see section 4.3.5). A parameter number
of $\SU(3)$ is equal to $3^2-1=8$. Operators from $\SU(3)$ act on
the vectors (\ref{VectH}) of $\bsH^S\otimes\bsH^Q\otimes\bsH_\infty$.

As is known, Young schemes in the case of the group $\SU(3)$ have
the form
\begin{equation}\label{Young}
\unitlength=0.35mm
%\linethickness{0.4pt}
\begin{picture}(150.00,30.00)
\put(0,0){\line(1,0){10}} \put(0,10){\line(1,0){10}}
\put(0,0){\line(0,1){10}} \put(10,0){\line(0,1){10}}
\put(10,0){\line(1,0){10}}\put(10,10){\line(1,0){10}}\put(20,0){\line(0,1){10}}
\put(20,0){\line(1,0){10}}\put(20,10){\line(1,0){10}}\put(30,0){\line(0,1){10}}
\put(30,0){\line(1,0){10}}\put(30,10){\line(1,0){10}}\put(40,0){\line(0,1){10}}
\put(0,10){\line(0,1){10}}\put(10,10){\line(0,1){10}}\put(0,20){\line(1,0){10}}
\put(10,20){\line(1,0){10}}\put(20,10){\line(0,1){10}}
\put(20,20){\line(1,0){10}}\put(30,10){\line(0,1){10}}
\put(30,20){\line(1,0){10}}\put(40,10){\line(0,1){10}}
\put(40,10){\line(1,0){10}}\put(40,20){\line(1,0){10}}\put(50,10){\line(0,1){10}}
\put(50,10){\line(1,0){10}}\put(50,20){\line(1,0){10}}\put(60,10){\line(0,1){10}}
\put(60,10){\line(1,0){10}}\put(60,20){\line(1,0){10}}\put(70,10){\line(0,1){10}}
\put(70,10){\line(1,0){10}}\put(70,20){\line(1,0){10}}\put(80,10){\line(0,1){10}}
\end{picture}
\end{equation}
Here we have $p+q$ squares in the first row and $q$ squares in the
second row. Let $C(p+2q,0)$ be a space of tensors of
the rank $p+2q$. The each Young scheme of the type (\ref{Young})
corresponds to subspace $C_{p,q}$ of $C(p+2q,0)$ consisting of the
tensors
\begin{equation}\label{Tensor}
T^{\{\alpha_1\ldots\alpha_p\}[\gamma_1\delta_1][\gamma_2\delta_2]\ldots[\gamma_q\delta_q]}
\end{equation}
with the following properties: 1) $T$ is symmetric with respect to
the indices $\alpha_1,\ldots,\alpha_p$; 2) $T$ is antisymmetric with
respect to the each pair of the indices from $[\gamma_i,\delta_i]$;
3) $T$ is symmetric with respect to the pairs $[\gamma_i,\delta_i]$.

Further, there is an isomorphic mapping \cite{RF70}
\[
\varphi:\; C_{p,q}\longrightarrow\Sym_{(p,q)},
\]
where $\Sym_{(p,q)}$ is a space of bisymmetric tensors of the type
\begin{equation}\label{Bitensor}
T^{\{\alpha_1\ldots\alpha_p\}}_{\{\beta_1\ldots\beta_p\}}.
\end{equation}
Coordinates of (\ref{Bitensor}) are constructed from the tensors
(\ref{Tensor}) of $C_{p,q}$ via the formula
\[
T^{\alpha_1\ldots\alpha_p}_{\beta_1\ldots\beta_p}=2^{-\frac{q}{2}}\phi_{\beta_1\gamma_1\delta_1}
\phi_{\beta_2\gamma_2\delta_2}\cdots\phi_{\beta_q\gamma_q\delta_q}
T^{\{\alpha_1\ldots\alpha_p\}[\gamma_1\delta_1][\gamma_2\delta_2]\ldots[\gamma_q\delta_q]},
\]
where $\phi_{\rho\sigma\tau}$ is a pseudotensor with the following
properties:
\[
\phi_{\rho\sigma\tau}=\phi^{\rho\sigma\tau}=\left\{\begin{array}{rcl}
0,&& \text{if the indices}\; \rho,\,\sigma,\,\tau\;\text{are not different};\nonumber\\
1,&& \text{if the substitution} \begin{pmatrix} 1 & 2 & 3\\
\rho & \sigma & \tau\end{pmatrix} \text{is even};\nonumber\\
-1,&& \text{if the substitution} \begin{pmatrix} 1 & 2 & 3\\
\rho & \sigma & \tau\end{pmatrix} \text{is odd}.\nonumber
\end{array}\right.
\]
Tensors (\ref{Bitensor}) with additional condition
\[
T^{\alpha\alpha_1\ldots\alpha_p}_{\alpha\beta_1\ldots\beta_p}=0
\]
form a space $\Sym^0_{(p,q)}$ of traceless bisymmetric tensors. All
irreducible representations of the group $\SU(3)$ are defined by the
traceless bisymmetric tensors in the spaces $\Sym^0_{(p,q)}$, where
a degree of the irreducible representation is given by the formula
\begin{equation}\label{Degree2}
N(p,q)=\frac{1}{2}(p+1)(q+1)(p+q+2).
\end{equation}
Degrees $N(p,q)$ ($p,q=0,1,\ldots,6$) are given in the Tab.\,2.
\begin{figure}[ht]
\begin{center}
{\renewcommand{\arraystretch}{1.2}
\begin{tabular}{c|cccccccl}
q  &  0 & 1 & 2 & 3 & 4 & 5 & 6 & \ldots\\ \hline
p &      &   &   &   &   &   &   &   \\
0 & 1 & 3 & 6 & 10 & 15 & 21 & 28 &
$\ldots$\\
1& 3 & 8 & 15 & 24 & 35 & 48 & 63 &
$\ldots$\\
2& 6 & 15 & 27 & 42 & 60 & 81 & 105 &
$\ldots$\\
3& 10 & 24 & 42 & 64 & 90 & 120 & 154 &
$\ldots$\\
4& 15 & 35 & 60 & 90 & 125 & 165 &
 210 &$\ldots$\\
5& 21 & 48 & 81 & 120 & 165 & 216 &
273 &$\ldots$\\
6& 28 & 63 & 105 & 154 & 210 & 273 &
343 &$\ldots$\\
$\vdots$&$\vdots$&$\vdots$&$\vdots$&$\vdots$&$\vdots$&$\vdots$&
$\vdots$
\end{tabular}}
\end{center}
\hspace{0.3cm}
\begin{center}\begin{minipage}{32pc}
{\small \textbf{Tab.\,2:} Degrees of irreducible representations of the group $\SU(3)$.}
\end{minipage}
\end{center}
\end{figure}

As is known, an algebra $\mathfrak{su}(3)$ of the group $\SU(3)$
consists of traceless hermitean operators acting in the space
$\C^3$. With the aim to fix the subalgebra $\mathfrak{su}(2)$ in
$\mathfrak{su}(3)$ we express the units of $\mathfrak{su}(2)$ via
the units of $\mathfrak{su}(3)$. It is more convenient to choose the
units of the algebra $\mathfrak{su}(3)$ in an `external' Okubo basis
\cite{Ok62}:
\[
A^1_1=\begin{bmatrix} \frac{2}{3} & 0 & 0\\
0 & -\frac{1}{3} & 0\\
0 & 0 & -\frac{1}{3}
\end{bmatrix},\quad A^2_1=\begin{bmatrix} 0 & 1 & 0\\
0 & 0 & 0\\
0 & 0 & 0
\end{bmatrix},\quad A^3_1=\begin{bmatrix} 0 & 0 & 1\\
0 & 0 & 0\\
0 & 0 & 0
\end{bmatrix},
\]
\[
A^1_2=\begin{bmatrix} 0 & 0 & 0\\
1 & 0 & 0\\
0 & 0 & 0
\end{bmatrix},\quad A^2_2=\begin{bmatrix} -\frac{1}{3} & 0 & 0\\
0 & \frac{2}{3} & 0\\
0 & 0 & -\frac{1}{3}
\end{bmatrix},\quad A^3_2=\begin{bmatrix} 0 & 0 & 0\\
0 & 0 & 1\\
0 & 0 & 0
\end{bmatrix},
\]
\begin{equation}\label{Okubo3}
A^1_3=\begin{bmatrix} 0 & 0 & 0\\
0 & 0 & 0\\
1 & 0 & 0
\end{bmatrix},\quad A^2_3=\begin{bmatrix} 0 & 0 & 0\\
0 & 0 & 0\\
0 & 1 & 0
\end{bmatrix},\quad A^3_3=\begin{bmatrix} -\frac{1}{3} & 0 & 0\\
0 & -\frac{1}{3} & 0\\
0 & 0 & \frac{2}{3}
\end{bmatrix}.
\end{equation}
Diagonal matrices $A^i_i$ from (\ref{Okubo3}) are hermitean and
satisfy the relation
\[
A^1_1+A^2_2+A^3_3=0.
\]
Commutation relations for the Okubo operators $A^i_k$ are
\begin{equation}\label{Commut}
\left[A^i_k,\,A^l_m\right]=\delta^i_mA^l_k-\delta^l_kA^i_m=
\left(\delta^i_m\delta^l_r\delta^s_k-\delta^l_k\delta^i_l\delta^s_m\right)A^i_k.
\end{equation}

%A relation between Okubo matrices and well-known Gell-Mann matrices
%$\lambda_i$ \cite{Gel61} is given by the formulas
%\[
%\lambda_1=A^2_1+A^1_2,\quad\lambda_4=A^3_1+A^1_3,\quad\lambda_6=A^3_2+A^2_3,
%\]
%\[
%\lambda_2=\frac{1}{i}\left(A^2_1-A^1_2\right),\quad\lambda_5=\frac{1}{i}\left(A^3_1-A^1_3\right),\quad
%\lambda_7=\frac{1}{i}\left(A^3_2-A^2_3\right),
%\]
%\[
%\lambda_3=A^1_1-A^2_2,\quad\lambda_8=-\sqrt{3}A^3_3,
%\]
%where
%\[
%\lambda_1=\begin{bmatrix} 0 & 1 & 0\\
%1 & 0 & 0\\
%0 & 0 & 0
%\end{bmatrix},\quad\lambda_2=\begin{bmatrix} 0 & -i & 0\\
%i & 0 & 0\\
%0 & 0 & 0
%\end{bmatrix},\quad\lambda_3=\begin{bmatrix} 1 & 0 & 0\\
%0 & -1 & 0\\
%0 & 0 & 0
%\end{bmatrix},
%\]
%\[
%\lambda_4=\begin{bmatrix} 0 & 0 & 1\\
%0 & 0 & 0\\
%1 & 0 & 0
%\end{bmatrix},\quad\lambda_5=\begin{bmatrix} 0 & 0 & -i\\
%0 & 0 & 0\\
%i & 0 & 0
%\end{bmatrix},\quad\lambda_6=\begin{bmatrix} 0 & 0 & 0\\
%0 & 0 & 1\\
%0 & 1 & 0
%\end{bmatrix},
%\]
%\[
%\lambda_7=\begin{bmatrix} 0 & 0 & 0\\
%0 & 0 & -i\\
%0 & i & 0
%\end{bmatrix},\quad\lambda_8=\begin{bmatrix} \frac{1}{\sqrt{3}} & 0 & 0\\
%0 & \frac{1}{\sqrt{3}} & 0\\
%0 & 0 & -\frac{2}{\sqrt{3}}
%\end{bmatrix}.
%\]
Let $a^i_k$ be Okubo operators of the subalgebra $\mathfrak{su}(2)$
and let $A^i_k$ be Okubo operators of the algebra
$\mathfrak{su}(3)$. The operators $a^i_k$ are
\begin{equation}\label{Okubo2}
a^1_1=\begin{bmatrix} \frac{1}{2} & 0\\
0 & -\frac{1}{2}\end{bmatrix},\quad a^2_1=\begin{bmatrix} 0 & 0\\
0 & 0\end{bmatrix},\quad a^1_2=\begin{bmatrix} 0 & 0\\
1 & 0\end{bmatrix},\quad a^2_2=\begin{bmatrix} -\frac{1}{2} & 0\\
0 & \frac{1}{2}\end{bmatrix},
\end{equation}
and their relations with the Pauli matrices are defined as $a^2_1=\sigma_1+i\sigma_2$, $a^1_2=\sigma_1-i\sigma_2$, $a^1_1=-a^2_2=\sigma_3$.
Further, let $\widetilde{P^0}$ be an irreducible
representation of the algebra $\mathfrak{su}(3)$ of the degree $N$
and let
\[
\widetilde{P^0}(a^i_k)=a^i_k(N),\quad\widetilde{P^0}(A^i_k)=A^i_k(N).
\]
The operators $a^i_k$ act in the space $\C^2=\Sym_{(1,0)}$, and the operators
$A^i_k$ act in $\C^3=\Sym^0_{(1,0)}$. In turn, the operators $a^i_k(N)$ and
$A^i_k(N)$ act in the representation space $\C^N$.

Returning to the algebra $\mathfrak{su}(3)$, we take
\[
a^1_1=A^1_1+\frac{1}{2}A^3_3,\quad
a^2_2=A^2_2+\frac{1}{2}A^3_3,\quad a^1_2=A^1_2,\quad a^2_1=A^2_1,
\]
or
\[
a^i_j=A^i_j-\frac{1}{2}\delta^i_jA^k_k,
\]
where the indices $i$, $j$, $k$ take the values $1,2$. At this
point,
\[
a^1_1+a^2_2=A^1_1+A^2_2+A^3_3=0.
\]
Further, using the relations (\ref{Commut}), we find
\[
\left[a^i_j,\,a^k_l\right]=\left[A^i_j-\frac{1}{2}\delta^i_jA^r_r,\,A^k_l-\frac{1}{2}\delta^k_lA^s_s\right]=
\left[A^i_j,\,A^k_l\right]=\delta^i_lA^k_j-\delta^k_jA^i_l=\delta^i_la^k_j-\delta^k_ja^i_l,
\]
where $i,\,j,\,k,\,l,\,r,\,s=1,\,2$. It is easy to see that $a^i_j$
satisfy the commutation relations for $2\times 2$ Okubo matrices
(\ref{Okubo2}). Therefore, the operators $a^i_j$ generate the
subalgebra $\mathfrak{su}(2)\subset\mathfrak{su}(3)$.

Since the rank of $\mathfrak{su}(3)$ is equal to 2, then the algebra
$\mathfrak{su}(3)$ contains \emph{two linearly independent
operators}, for example, $A^1_1$ and $A^3_3$. Therefore, any
operator from $\mathfrak{su}(3)$ can be represented as a linear
combination of $A^1_1$ and $A^3_3$. Hence it follows that in the
case of $\mathfrak{su}(3)$ an analogue of the operator $I_3$ (the
isospin operator of $\mathfrak{su}(2)$) has the form $A=\alpha
A^1_1+\beta A^3_3$, where $\alpha$ and $\beta$ are constant
coefficients. Further, for the operator $\widetilde{P^0}(A)$, which
acts in the space $\C^8=\Sym^0_{(1,1)}$, we have
$\widetilde{P^0}(A)=\alpha\widetilde{P^0}(A^1_1)+\beta\widetilde{P^0}(A^3_3)$
and, therefore, \emph{a charge operator of the octet} $F_{1/2}$ is
defined as
\[
Q(8)=\alpha A^1_1(8)+\beta A^3_3(8)+\gamma\boldsymbol{1}_8,
\]
where the constant $\gamma$ defines a shift of eigenvalues of $Q$. This fixation of the subalgebra $\mathfrak{su}(2)$ in $\mathfrak{su}(3)$ leads to $I_3=A^1_1+\frac{1}{2}A^3_3$ and called $I$-\emph{spin}. However, in common with $I$-spin there are two different fixations of $\mathfrak{su}(2)$ in $\mathfrak{su}(3)$ which lead to $U_3=A^3_3+\frac{1}{2}A^1_1$ ($U$-\emph{spin}) and $V_3=A^1_1+\frac{1}{2}A^2_2$ ($V$-\emph{spin}). The choice of $\mathfrak{su}(2)$ with respect to $U$-spin is used in the Gell-Mann--Okubo mass formula (see section 7).

Further, hadrons are classified in $\SU(3)$-theory into supermultiplets consisting of the particles of one and the same baryon number, spin and parity. The each supermultiplet corresponds to some irreducible representation of the group $\SU(3)$. At this point, the number of particles, belonging to supermultiplet, is equal to a degree of the representation (see Tab.\,2). \emph{The each vector of the space $\Sym^0_{(p,q)}$ of the irreducible representation corresponds to a state (particle) of the supermultiplet}. The operators of charge $Q(N)$ and hypercharge $Y(N)$ are defined on the space $\Sym^0_{(p,q)}$, where $N=N(p,q)$ is the degree of representation defined by the formula (\ref{Degree2}). Supermultiplets correspond to such irreducible representations of $\SU(3)$, for which all the eigenvalues of the operators $Q(N)$ and $Y(N)$ are integer. Hence it follows that hadron supermultiplets correspond to such representations $\Sym^0_{(p,q)}$ of $\SU(3)$ for which $p-q\equiv 0\pmod{3}$. From the Tab.\,2 we see that `admissible' hadron supermultiplets have degrees 1, 8, 10, 27, 28, 35, 55, 64, 80, 81, 91, 125, 136, 143, 154, $\ldots$. There is a $\SU(3)/\SU(2)$-reduction of the given supermultiplet into charge multiplets of the group $\SU(2)$. Namely, an irreducible representation $\Sym^0_{(p,q)}$, defining the supermultiplet, induces a reducible representation on the subgroup $\SU(2)\subset\SU(3)$.

\section{Supermultiplets of $\SU(3)$ and $\SU(3)/\SU(2)$-reduction}
In this section we will consider in details supermultiplets of the
group $\SU(3)$ (octets $F_{1/2}$, $B_0$, $B_1$) based on the
eight-dimensional regular representation $\Sym^0_{(1,1)}$ and their
reductions into isotopic multiplets of the subgroup $\SU(2)$. As is
known \cite{RF70}, $\SU(3)/\SU(2)$-reduction of $\Sym^0_{(1,1)}$ is given by the
following expression:
\begin{equation}\label{Reduction}
\Sym^0_{(1,1)}=\Phi_3\oplus\Phi_2\oplus\overset{\ast}{\Phi}_2\oplus\Phi_0,
\end{equation}
where $\Phi_3$, $\Phi_2$, $\overset{\ast}{\Phi}_2$, $\Phi_0$ are
charge multiplets of $\SU(2)$, $\Phi_3$ is a triplet, $\Phi_2$ and
$\overset{\ast}{\Phi}_2$ are doublets, $\Phi_0$ is a singlet.

Below we consider $\SU(3)/\SU(2)$-reductions and mass spectrum of the octets $F_{1/2}$, $B_0$, $B_1$ (eightfold way) with respect to charge multiplets.
\subsection{Octet $F_{1/2}$} $F_{1/2}$ is a fermionic supermultiplet
of $\SU(3)$ containing baryons of the spin 1/2. Therefore, all the
particles of $F_{1/2}$ are described by the vectors of the abstract
Hilbert space belonging to spin-1/2 line with positive parity $P^2=1$. In accordance with
(\ref{Reduction}), $\SU(3)/\SU(2)$-reduction of the octet $F_{1/2}$
leads to the following charge multiplets:
\[
{\renewcommand{\arraystretch}{1.5} \Phi_3:\;\left\{
\begin{array}{cll}
\boldsymbol{\Sigma}^+&=&\left.\left|\boldsymbol{\tau}^c_{\frac{67}{2},33},\,\Sym_{(67,66)},\,
\C_{266},\,\dS_{2^{133}},\,
P^2=1\,\right.\right\rangle,\\
\boldsymbol{\Sigma}^0&=&\left.\left|\boldsymbol{\tau}^r_{\frac{67}{2},33},\,\Sym_{(67,66)},\,
\cl_{135,131},\,\dS_{2^{133}},\,
P^2=1\,\right.\right\rangle,\\
\boldsymbol{\Sigma}^-&=&\left.\left|\boldsymbol{\tau}^c_{\frac{67}{2},33},\,\Sym_{(67,66)},\,
\overset{\ast}{\C}_{266},\,\hat{\dS}_{2^{133}},\,
P^2=1\,\right.\right\rangle.
\end{array}\right.
}
\]
\[
{\renewcommand{\arraystretch}{1.5} \Phi_2:\;\left\{
\begin{array}{cll}
\bP&=&\left.\left|\boldsymbol{\tau}^c_{\frac{59}{2},29},\,\Sym_{(59,58)},\,
\C_{234},\,\dS_{2^{117}},\,
P^2=1\,\right.\right\rangle,\\
\bN&=&\left.\left|\boldsymbol{\tau}^r_{\frac{59}{2},29},\,\Sym_{(59,58)},\,
\cl_{119,115},\,\dS_{2^{117}},\,
P^2=1\,\right.\right\rangle.
\end{array}\right.
}
\]
\[
{\renewcommand{\arraystretch}{1.5} \overset{\ast}{\Phi}_2:\;\left\{
\begin{array}{cll}
\boldsymbol{\Xi}^-&=&\left.\left|\boldsymbol{\tau}^c_{\frac{71}{2},35},\,\Sym_{(71,70)},\,
\C_{282},\,\dS_{2^{141}},\,
P^2=1\,\right.\right\rangle,\\
\boldsymbol{\Xi}^0&=&\left.\left|\boldsymbol{\tau}^r_{\frac{71}{2},35},\,\Sym_{(71,70)},\,
\cl_{143,139},\,\dS_{2^{141}},\,
P^2=1\,\right.\right\rangle.
\end{array}\right.
}
\]
\[
\Phi_0:\;\boldsymbol{\Lambda}=\left.\left|\boldsymbol{\tau}^r_{\frac{65}{2},32},\,\Sym_{(65,64)},\,
\cl_{131,127},\,\dS_{2^{129}},\,
P^2=1\,\right.\right\rangle.
\]
Here $\Phi_3$ is the $\boldsymbol{\Sigma}$-triplet considered in the section 4.3.2, $\Phi_2$ is the nucleon doublet defined in the section 4.3.1. $\overset{\ast}{\Phi}_2$ is a $\boldsymbol{\Xi}$-doublet, $\Phi_0$ is a $\boldsymbol{\Lambda}$-singlet.

$\boldsymbol{\Xi}$-doublet is constructed within the complex representation $\boldsymbol{\tau}^c_{\frac{75}{2},35}$ of the orbit $\boldsymbol{O}^+_{m_\Xi}$ with the degree 5112, since $m_\Xi/m_e\approx  2520$. This representation belongs to spin-1/2 line with positive parity $P^2=1$ and acts in the space $\Sym_{(71,70)}$. The algebra $\C_{282}\simeq\C\otimes\cl_{143,139}$ and complex spinspace $\dS_{2^{141}}$ are associated with the state $\left| e_1\right\rangle=\boldsymbol{\Xi}^-$ in the spinor structure. The real subalgebra $\cl_{143,139}$ has the quaternionic division ring $\K\simeq\BH$, type $p-q\equiv 4\pmod{8}$, and, therefore, $P^2=1$. The state $\left| e_2\right\rangle=\boldsymbol{\Xi}^0$ is described by a real representation $\boldsymbol{\tau}^r_{\frac{75}{2},35}$ belonging also to spin-1/2 line with the Clifford algebra $\cl_{133,139}$ and a quaternionic spinspace $\dS_{2^{141}}(\BH)$.

$\boldsymbol{\Lambda}$-singlet is defined within the real representation $\boldsymbol{\tau}^r_{\frac{65}{2},32}$ of the orbit $\boldsymbol{O}^+_{m_\Lambda}$ with the degree 4290, since $m_\Lambda/m_e\approx 2140$. This representation belongs to spin-1/2 line and acts in the space $\Sym_{(65,64)}$. The real algebra $\cl_{131,127}$ (type $p-q\equiv 4\pmod{8}$, $\K\simeq\BH$, $P^2=1$) and quaternionic spinspace $\dS_{2^{129}}(\BH)$ are associated with the $\boldsymbol{\Lambda}$-singlet in the underlying spinor structure.

Charge multiplets, considered above, compound eight-dimensional regular representation of $\SU(3)$\footnote{At this point we do not use the quark structure of $F_{1/2}$, since this structure is a derivative construction of $\SU(3)$-symmetry. The quark scheme in itself is a reformulation of $\SU(3)$ group representations in terms of tensor products of the vectors of fundamental representations $\Sym^0_{(1,0)}$ and $\Sym^0_{(0,1)}$. So, quarks $u$, $d$, $s$ are described within $\Sym^0_{(1,0)}$, and antiquarks $\overline{u}$, $\overline{d}$, $\overline{s}$ within $\Sym^0_{(0,1)}$. Quarks and antiquarks have fractional charges $Q$ and hypercharges $Y$. The each hadron supermultiplet can be constructed from the quarks and antiquarks in the tensor space $\C^{k,r}$ which corresponds to a standard representation of $\SU(3)$. The space $\C^{k,r}$ is a tensor product of $k$ spaces $\C^3$ and $r$ spaces $\overset{\ast}{\C}{}^3$. The quark composition of a separate particle, belonging to a given supermultiplet of $\SU(3)$, is constructed as follows. $I$-basis is constructed from the eigenvectors of $Q$ and $Y$ in the space of irreducible representation of the given supermultiplet. These basis vectors present particles of the supermultiplet, the each of them belongs to $\C^{k,r}$ and, therefore, is expressed via the polynomial on basis vectors $e_1$, $e_2$, $e_3$ of $\C^3$ and basis vectors $\tilde{e}_1$, $\tilde{e}_2$, $\tilde{e}_3$ of $\overset{\ast}{\C}{}^3$ with the degree $k+r$. The substitution of $e_1$, $e_2$, $e_3$, $\tilde{e}_1$, $\tilde{e}_2$, $\tilde{e}_3$ by $u$, $d$, $s$, $\overline{u}$, $\overline{d}$, $\overline{s}$ leads to a quark composition of the particle. It is assumed that quarks and antiquarks have the spin $1/2$ (however, spin is an external parameter with respect to $\SU(3)$-theory). Hence it follows that a maximal spin of the particle, consisting of $k$ quarks and $r$ antiquarks, is equal to $(k+r)/2$. When $k+r$ is odd we have fermions and bosons when $k+r$ is even.

In this connection it is interesting to note that $k+r$ tensor products of $\C_2$ and $\overset{\ast}{\C}_2$ biquaternion algebras in (\ref{TenAlg}), which generate the underlying spinor structure, lead to a fermionic representation of $\spin_+(1,3)$ when $k+r$ is odd and to a bosonic representation when $k+r$ is even (see spin-lines considered in the section 2.1.1). Due to the difference between dimensions of basic constituents in tensor products ($n=2$ for spinors and $n=3$ for quarks) which define spinor and quark structures, we can assume that \textbf{\emph{spinors are more fundamental than quarks}}.}.
%\begin{figure}[h]
%\unitlength=0.70mm
%\begin{center}
%\begin{picture}(100,90)
%\put(35,76){$\bullet$} \put(35,80){\bN}%\put(36,70){$ddu$}
%\put(40,77){\line(1,0){22}}\put(66,76){$\bullet$}
%\put(59,70){$uud$}
%\put(65,80){\bP}
%\put(69.5,73){\line(1,-2){9.5}}\put(80,50){$\bullet$}
%\put(81,54){$\boldsymbol{\Sigma}^+$}%\put(71,50){$uus$}
%\put(68.5,27.5){\line(1,2){10}}\put(65,25){$\bullet$}%\put(59,30){$ssu$}
%\put(65,20){$\boldsymbol{\Xi}^0$}\put(40,26){\line(1,0){22}}\put(35,25){$\bullet$}
%\put(36,30){$ssd$}
%\put(35,20){$\boldsymbol{\Xi}^-$}
%\put(23.5,48){\line(1,-2){10}}\put(20,50){$\bullet$}
%\put(25,50){$dds$}
%\put(16,53){$\boldsymbol{\Sigma}^-$}
%\put(23,54){\line(1,2){10}}\put(50,55){$\bullet$}
%\put(50,58){$\boldsymbol{\Sigma}^0$}%\put(53,55){$uds$}
%\put(50,45){$\bullet$}\put(50,40){$\boldsymbol{\Lambda}$}%\put(53,45){$uds$}
%\put(0,20){\vector(0,1){70}}\put(4,90){\textbf{Strangeness}}
%\put(100,90){\textbf{Mass} (MeV)}\put(105,76){939}
%\put(107,70){\vector(0,-1){10}}\put(105,54){1192}
%\put(118,54){$\Sigma$}\put(105,46){1115} \put(118,46){$\Lambda$}
%\put(107,42){\vector(0,-1){10}}\put(105,25){1318}
%\put(-1.5,76){-}\put(-4,76){0}
%\put(-1.5,50){-}\put(-6,50){-1}
%\put(-1.5,25){-}\put(-6,25){-2}
%\end{picture}
%\end{center}
%\end{figure}
\subsection{Octet $B_0$}
$B_0$ is a bosonic supermultiplet of $\SU(3)$ containing mesons of the spin 0 with the negative parity $P^2=-1$. Hence it follows that all the particles of $B_0$ are described by the vectors of the abstract Hilbert space belonging to spin-0 line. In accordance with basic mass levels defined by the mass formula (\ref{MGY}), $\SU(3)/\SU(2)$-reduction of the octet $B_0$ leads to the following charge multiplets:
\[
{\renewcommand{\arraystretch}{1.5} \Phi_3:\;\left\{
\begin{array}{cll}
\boldsymbol{\pi}^+&=&\left.\left|\boldsymbol{\tau}^c_{11,11},\,\Sym_{(22,22)},\,
\C_{88},\,\dS_{2^{44}},\,P^2=-1\,\right.\right\rangle,\\
\boldsymbol{\pi}^0&=&\left.\left|\boldsymbol{\tau}^r_{11,11},\,\Sym_{(22,22)},\,
\cl_{45,43},\,\dS_{2^{44}},\,P^2=-1\,\right.\right\rangle,\\
\boldsymbol{\pi}^-&=&\left.\left|\boldsymbol{\tau}^c_{11,11},\,\Sym_{(22,22)},\,
\overset{\ast}{\C}_{88},\,\hat{\dS}_{2^{44}},\,P^2=-1\,\right.\right\rangle.
\end{array}\right.
}
\]
\[
{\renewcommand{\arraystretch}{1.5} \Phi_2:\;\left\{
\begin{array}{cll}
\boldsymbol{K}^-&=&\left.\left|\boldsymbol{\tau}^c_{\frac{43}{2},\frac{43}{2}},\,\Sym_{(43,43)},\,
\C_{172},\,\dS_{2^{86}},\,P^2=-1\,\right.\right\rangle,\\
\overline{\boldsymbol{K}}^0&=&\left.\left|\boldsymbol{\tau}^r_{\frac{43}{2},\frac{43}{2}},\,\Sym_{(43,43)},\,
\hat{\cl}_{89,83},\,\hat{\dS}_{2^{86}},\,P^2=-1\,\right.\right\rangle.
\end{array}\right.
}
\]
\[
{\renewcommand{\arraystretch}{1.5} \overset{\ast}{\Phi}_2:\;\left\{
\begin{array}{cll}
\boldsymbol{K}^0&=&\left.\left|\boldsymbol{\tau}^r_{\frac{43}{2},\frac{43}{2}},\,\Sym_{(43,43)},\,
\cl_{89,83},\,\dS_{2^{86}},\,P^2=-1\,\right.\right\rangle,\\
\boldsymbol{K}^+&=&\left.\left|\boldsymbol{\tau}^c_{\frac{43}{2},\frac{43}{2}},\,\Sym_{(43,43)},\,
\overset{\ast}{\C}_{172},\,\hat{\dS}_{2^{86}},\,P^2=-1\,\right.\right\rangle.
\end{array}\right.
}
\]
\[
\Phi_0:\;\boldsymbol{\eta}=\left.\left|\boldsymbol{\tau}^r_{\frac{45}{2},\frac{45}{2}},\,\Sym_{(45,45)},\,
\cl_{46,44},\,\dS_{2^{90}},\,P^2=-1\,\right.\right\rangle.
\]
Here $\Phi_3$ is the $\boldsymbol{\pi}$-triplet considered in the section 4.3.3. $\Phi_2$ and $\overset{\ast}{\Phi}_2$ are $\boldsymbol{K}_1$- and $\boldsymbol{K}_2$-doublets, $\Phi_0$ is a $\boldsymbol{\eta}$-singlet. $\Phi_2$ and $\overset{\ast}{\Phi}_2$ are particle-antiparticle counterparts with respect to each other.

The $\boldsymbol{K}_1$-doublet is constructed within the representation $\boldsymbol{\tau}_{\frac{43}{2},\frac{43}{2}}$ of the orbit $\boldsymbol{O}^+_{m_K}$ with the degree 1936, since $m_K/m_e\approx 972$. This representation belongs to spin-0 line and acts in the space $\Sym_{(43,43)}$. The state $\left| e_1\right\rangle=\boldsymbol{K}^-$ is described by the complex representation $\boldsymbol{\tau}^c_{\frac{43}{2},\frac{43}{2}}$ with the algebra $\C_{172}\simeq\C\otimes\cl_{89,83}$ and complex spinspace $\dS_{2^{86}}$ in the spinor structure. The real subalgebra $\cl_{89,83}$ has the quaternionic division ring $\K\simeq\BH$, type $p-q\equiv 6\pmod{8}$, and, therefore, $P^2=-1$. In turn, the state $\left| e_2\right\rangle=\overline{\boldsymbol{K}}^0$ is described by the real representation $\boldsymbol{\tau}^r_{\frac{43}{2},\frac{43}{2}}$ with the algebra $\cl_{89,83}$ and the quaternionic spinspace $\dS_{2^{86}}(\BH)$. The $\boldsymbol{K}_2$-doublet has the same construction within the representation $\boldsymbol{\tau}_{\frac{43}{2},\frac{43}{2}}$ of the orbit $\boldsymbol{O}^-_{m_K}$.

The $\boldsymbol{\eta}$-singlet is defined within the real representation $\boldsymbol{\tau}^r_{\frac{45}{2},\frac{45}{2}}$ of the orbit $\boldsymbol{O}^0_{m_\eta}$ with the degree 2116, since $m_\eta/m_e\approx 1076$. This representation belongs to spin-0 line and acts in $\Sym_{(45,45)}$. Since $\boldsymbol{\eta}$-state presents a truly neutral particle (the orbit $\boldsymbol{O}^0_{m_\eta}\sim\boldsymbol{O}^+_{m_\eta}\simeq\boldsymbol{O}^-_{m_\eta}$), then the real algebra $\cl_{46,44}$ with the real division ring $\K\simeq\R$ (type $p-q\equiv 2\pmod{8}$) and real spinspace $\dS_{2^{90}}(\R)$ are associated with the $\boldsymbol{\eta}$-singlet in the spinor structure.
\subsection{Octet $B_1$}
The next supermultiplet of the group $\SU(3)$ in eightfold way is the octet $B_1$. $B_1$ describes mesons of the spin 1 (vector bosons) with negative parity ($P^2=-1$). In this case we see that all the particles of $B_1$ are defined by the vectors of $\bsH^S\otimes\bsH^Q\otimes\bsH_\infty$ belonging to spin-1 line. In accordance with the mass formula (\ref{MGY}), $\SU(3)/\SU(2)$-reduction of the octet $B_1$ leads to the following charge multiplets:
\[
{\renewcommand{\arraystretch}{1.5} \Phi_3:\;\left\{
\begin{array}{cll}
\boldsymbol{\rho}^-&=&\left.\left|\boldsymbol{\tau}^c_{\frac{55}{2},\frac{53}{2}},\,\Sym_{(55,53)},\,
\C_{216},\,\dS_{2^{108}},\,P^2=-1\,\right.\right\rangle,\\
\boldsymbol{\rho}^0&=&\left.\left|\boldsymbol{\tau}^r_{\frac{55}{2},\frac{53}{2}},\,\Sym_{(55,53)},\,
\cl_{109,107},\,\dS_{2^{108}},\,P^2=-1\,\right.\right\rangle,\\
\boldsymbol{\rho}^+&=&\left.\left|\boldsymbol{\tau}^c_{\frac{53}{2},\frac{55}{2}},\,\Sym_{(55,53)},\,
\overset{\ast}{\C}_{216},\,\hat{\dS}_{2^{108}},\,P^2=-1\,\right.\right\rangle.
\end{array}\right.
}
\]
\[
{\renewcommand{\arraystretch}{1.5} \Phi_2:\;\left\{
\begin{array}{cll}
{}^\ast\boldsymbol{K}^-&=&\left.\left|\boldsymbol{\tau}^c_{\frac{59}{2},\frac{57}{2}},\,\Sym_{(59,57)},\,
\C_{232},\,\dS_{2^{116}},\,P^2=-1\,\right.\right\rangle,\\
{}^\ast\overline{\boldsymbol{K}}^0&=&\left.\left|\boldsymbol{\tau}^r_{\frac{57}{2},\frac{59}{2}},\,\Sym_{(57,59)},\,
\hat{\cl}_{119,113},\,\hat{\dS}_{2^{116}},\,P^2=-1\,\right.\right\rangle.
\end{array}\right.
}
\]
\[
{\renewcommand{\arraystretch}{1.5} \overset{\ast}{\Phi}_2:\;\left\{
\begin{array}{cll}
{}^\ast\boldsymbol{K}^0&=&\left.\left|\boldsymbol{\tau}^r_{\frac{59}{2},\frac{57}{2}},\,\Sym_{(59,57)},\,
\cl_{119,113},\,\dS_{2^{116}},\,P^2=-1\,\right.\right\rangle,\\
{}^\ast\boldsymbol{K}^+&=&\left.\left|\boldsymbol{\tau}^c_{\frac{57}{2},\frac{59}{2}},\,\Sym_{(57,59)},\,
\overset{\ast}{\C}_{232},\,\hat{\dS}_{2^{116}},\,P^2=-1\,\right.\right\rangle.
\end{array}\right.
}
\]
\[
\Phi_0:\;\boldsymbol{\varphi}=\left.\left|\boldsymbol{\tau}^r_{28,27},\,\Sym_{(56,54)},\,
\cl_{110,108},\,\dS_{2^{109}},\,P^2=-1\,\right.\right\rangle.
\]
Here $\boldsymbol{\rho}$-triplet is constructed within a representation $\boldsymbol{\tau}_{\frac{55}{2},\frac{53}{2}}$ of the degree 3024, since $m_\rho/m_e\approx 1496$. This representation belongs to spin-1 line and acts in the space $\Sym_{(55,53)}$. The state $\left| e_1\right\rangle=\boldsymbol{\rho}^-$ is defined within the complex representation $\boldsymbol{\tau}^c_{\frac{55}{2},\frac{53}{2}}$ of the orbit $\boldsymbol{O}^+_{m_\rho}$ with the associated algebra $\C_{216}$ and complex spinspace $\dS_{2^{108}}$. Analogously, the state $\left| e_3\right\rangle=\boldsymbol{\rho}^+$ is defined within $\boldsymbol{\tau}^c_{\frac{53}{2},\frac{55}{2}}$ of the orbit $\boldsymbol{O}^-_{m_\rho}$ with
$\overset{\ast}{\C}_{216}$ and $\hat{\dS}_{2^{108}}$ (the states $\left| e_1\right\rangle=\boldsymbol{\rho}^-$ and $\left| e_3\right\rangle=\boldsymbol{\rho}^+$ are particle-antiparticle counterparts with respect to each other). In its turn, the state $\left| e_2\right\rangle=\boldsymbol{\rho}^0$ is defined within the real representation $\boldsymbol{\tau}^r_{\frac{55}{2},\frac{53}{2}}$ of the orbit $\boldsymbol{O}^0_{m_\rho}$ (truly neutral particle). In this case we have the real Clifford algebra $\cl_{109,107}$ with the real division ring $\K\simeq\R$, the type $p-q\equiv 2\pmod{8}$, and, therefore, $P^2=-1$.

Further, ${}^\ast\boldsymbol{K}_1$- and ${}^\ast\boldsymbol{K}_2$-doublets are particle-antiparticle counterparts with respect to each other. The ${}^\ast\boldsymbol{K}_1$-doublet is constructed within $\boldsymbol{\tau}_{\frac{59}{2},\frac{57}{2}}$ of the orbit $\boldsymbol{O}^+_{m_{{}^\ast K}}$ with the degree 3480, since $m_{{}^\ast K}/m_e\approx 1747$. $\boldsymbol{\tau}_{\frac{59}{2},\frac{57}{2}}$ belongs to spin-1 line and acts in $\Sym_{(59,57)}$. The state $\left| e_1\right\rangle={}^\ast\boldsymbol{K}^-$ is defined within the complex representation $\boldsymbol{\tau}^c_{\frac{59}{2},\frac{57}{2}}$ with the associated algebra $\C_{232}\simeq\C\otimes\cl_{119,113}$ and complex spinspace $\dS_{2^{116}}$ in the spinor structure. The real subalgebra $\cl_{119,113}$ has the quaternionic division ring $\K\simeq\BH$, type $p-q\equiv 6\pmod{8}$ and, therefore, $P^2=-1$. The state $\left| e_2\right\rangle={}^\ast\overline{\boldsymbol{K}}^0$ of the ${}^\ast\boldsymbol{K}_1$-doublet is described by the real representation $\boldsymbol{\tau}^r_{\frac{57}{2},\frac{59}{2}}$ with the algebra $\hat{\cl}_{119,113}$ and quaternionic spinspace $\hat{\dS}_{2^{116}}(\BH)$. The ${}^\ast\boldsymbol{K}_2$-doublet has the same construction within the representation $\boldsymbol{\tau}_{\frac{59}{2},\frac{57}{2}}$ of the orbit $\boldsymbol{O}^-_{m_{{}^\ast K}}$.

The $\boldsymbol{\varphi}$-singlet is defined within the real representation $\boldsymbol{\tau}^r_{28,27}$ of the orbit $\boldsymbol{O}^0_{m_\varphi}$ (truly neutral particle) with the degree 3135, since $m_\varphi/m_e\approx 1533$. This representation belongs to spin-1 line and acts in $\Sym_{(56,54)}$. Since $\boldsymbol{\varphi}$-state presents a truly neutral particle, then we have the real spinspace $\dS_{2^{109}}(\R)$, and the associated algebra $\cl_{110,108}$ has the real division ring $\K\simeq\R$, $p-q\equiv 2\pmod{8}$.

In contrast with bosonic supermultiplets $B_0$ and $B_1$, the fermionic supermultiplet $F_{1/2}$ has an antiparticle counterpart $\overline{F}_{1/2}$. Moreover, $F_{1/2}$ and $\overline{F}_{1/2}$ are coherent subspaces of $\bsH^S\otimes\bsH^Q\otimes\bsH_\infty$ on the spin ($s=1/2$) and parity ($P^2=1$). On the other hand, $F_{1/2}$ and $\overline{F}_{1/2}$ form different coherent subspaces of $\bsH^S\otimes\bsH^Q\otimes\bsH_\infty$ with respect to baryon number. In turn, bosonic supermultiplets $B_0$ and $B_1$ are the same coherent subspaces of $\bsH^S\otimes\bsH^Q\otimes\bsH_\infty$ on the baryon number ($B=0$) and parity ($P^2=-1$), whereas $B_0$ and $B_1$ are different coherent subspaces on the spin.

\section{Mass spectrum}
As is known, in $\SU(3)$-theory a mass distribution of the particles
in supermultiplets is described by Gell-Mann--Okubo formula
\cite{Gel61,Ok62}. According to the fundamental viewpoint,
Gell-Mann--Okubo mass formula is analogous to a Zeeman-effect
description in atomic spectra \cite{RF70}.

As follows from a group theoretical description of Zeeman effect,
\emph{an energy operator} has the form
\begin{equation}\label{Energy}
H=H_0+H_1,
\end{equation}
where
\begin{equation}\label{Zeeman}
H_1=d^\alpha_\beta\cH^\beta_\alpha,
\end{equation}
and
\[
d^\alpha_\beta=-\frac{e\hbar}{2m}a^\alpha_\beta
\]
is \emph{a magnetic moment} of the particle, $a^\alpha_\beta$ are
Okubo operators of the representation of the group $\SU(2)$
corresponding to an eigenvalue of $H_0$. The field $\cH^\beta_\alpha$
is related with a homogeneous magnetic field
$\boldsymbol{\cH}=\rot\bA$ by the following formulas:
\[
\cH^1_1=\frac{1}{2}\cH^3,\quad\cH^2_1=\frac{1}{2}\left(\cH^1+i\cH^2\right),\quad
\cH^1_2=\frac{1}{2}\left(\cH^1-i\cH^2\right),\quad\cH^2_2=-\frac{1}{2}\cH^3.
\]

In 1964, Okubo and Ryan \cite{OR64} proposed to describe mass
spectrum of the particles in any supermultiplet of $\SU(3)$-theory
by the formula of type
\begin{equation}\label{MassQ}
m^2=m^2_0+\delta m^2.
\end{equation}
At this point, terms in (\ref{MassQ}) have the same properties as in
(\ref{Energy}). Namely, the operator $m^2_0$ is symmetric with
respect to the group $\SU(3)$, and the operator $\delta m^2$ has an
expression of the type (\ref{Zeeman}). When
\begin{equation}\label{MassRel}
\frac{\delta m^2}{m^2_0}\ll 1
\end{equation}
is fulfilled, then in the decomposition
\[
m=m_0\left(1+\frac{\delta
m^2}{m^2_0}\right)^{1/2}=m_0+\frac{1}{2m_0}\delta m^2+\ldots
\]
we can remain only the first two terms:
\begin{equation}\label{MassL}
m=m_0+\frac{1}{2m_0}\delta m^2.
\end{equation}

Further, by analogy with Zeeman effect the formula (\ref{MassQ}) can
be written as
\[
m^2=m^2_0+D^a_bZ^b_a,
\]
where $D^a_b$ ($a,b=1,2,3$) is \emph{a tensor-operator of the
unitary moment} belonging to a regular representation of the group
$\SU(3)$, $Z^b_a$ is a tensor with the scalar components (so called
`unitary field' which is analogous to external magnetic field in
Zeeman effect) belonging also to a regular representation of
$\SU(3)$. $m^2_0$ is proportional to the unit operator. In general,
$m_0$ is not described by $\SU(3)$-theory. It is an external
parameter with respect to $\SU(3)$-theory, and concrete value of
$m_0$ depends on the selected supermultiplet of $\SU(3)$ (below we will show that $m_0$ is defined by the mass formula (\ref{MGY})). The
unitary moment $D^a_b$ is expressed via the Okubo operators $A^a_b$
of the same irreducible representation of $\SU(3)$ by the formula
\[
D^a_b=\lambda\delta^a_b\boldsymbol{1}+\mu A^a_b+\nu A^a_cA^c_b+\rho
A^a_cA^c_dA^d_b+\ldots,
\]
where $\lambda$, $\mu$, $\nu$, $\rho$ are constants.

The \emph{unitary field}\footnote{The physical sense of the unitary field is unknown (see \cite{RF70}). The field $Z^b_a$ is not one and the same for the all supermultiplets of $\SU(3)$-theory. However, $Z$-fields of different supermultiplets are distinguished by only two real parameters. $Z$-field of type (\ref{UField}) takes place also at the $\SU(6)/\SU(3)$-reduction in the flavor-spin $\SU(6)$-theory. In some sense, $Z$-field can be identified with a \emph{nonlocal quantum substrate} in the decoherence theory \cite{Zur98}. In this context $Z$-field can be understood as a mathematical description of the decoherence process (localization) of the particles, that is, it is a reduction of the initial quantum substrate into localized particles at the given energy level.} has the form
\begin{equation}\label{UField}
Z=C\begin{bmatrix} 1/3 & 0 & 0\\
0 & 1/3 & 0\\
0 & 0 & -2/3
\end{bmatrix}+C^\prime\begin{bmatrix} 2/3 & 0 & 0\\
0 & -1/3 & 0\\
0 & 0 & -1/3
\end{bmatrix},
\end{equation}
where $C^\prime\ll C$. The first term  in (\ref{UField}) splits the
supermultiplet of $\SU(3)$ into $I$-multiplets of the subgroup
$\SU(2)$ with respect to different values of the hypercharge $Y$.
The second term in (\ref{UField}) generates a charge splitting of
the $I$-multiplets. Hence it follows that
\begin{equation}\label{MassQ2}
m^2=m^2_0+\delta m^2+\delta{m^2}^\prime,
\end{equation}
\begin{equation}\label{MassQ3}
\delta m^2=\xi A^3_3+\eta A^3_cA^c_3,
\end{equation}
\begin{equation}\label{MassQ4}
\delta{m^2}^\prime=\xi^\prime A^1_1+\eta^\prime A^1_cA^c_1,
\end{equation}
where
\[
\xi^\prime=\theta\xi,\quad\eta^\prime=\theta\eta,\quad |\theta|\ll
1.
\]
Expressing the operator of hypercharge mass splitting $\delta m^2$
(\ref{MassQ3}) and the operator of charge splitting
$\delta{m^2}^\prime$ via Casimir operators of $\SU(3)$ and
substituting the results to (\ref{MassQ2}), we come to a well-known
Gell-Mann--Okubo mass formula
\begin{equation}\label{GMOQ}
m^2=m^2_0+\alpha+\beta Y+\gamma\left[I(I+1)-\frac{1}{4}Y^2\right]
+\alpha^\prime-\beta^\prime
Q+\gamma^\prime\left[U(U+1)-\frac{1}{4}Q^2\right],
\end{equation}
where
\[
\frac{\alpha^\prime}{\alpha}=\frac{\beta^\prime}{\beta}=\frac{\gamma^\prime}{\gamma}=\theta,\quad
|\theta|\ll 1.
\]
In the case when the condition (\ref{MassRel}) is fulfilled, the
quadratic mass formula (\ref{MassQ}) can be replaced by the linear
mass formula (\ref{MassL}) and from (\ref{GMOQ}) we have
\begin{equation}\label{GMOL}
m=m_0+\alpha+\beta Y+\gamma\left[I(I+1)-\frac{1}{4}Y^2\right]
+\alpha^\prime-\beta^\prime
Q+\gamma^\prime\left[U(U+1)-\frac{1}{4}Q^2\right],
\end{equation}

Let us consider in details mass splitting of the supermultiplets
$F_{1/2}$, $B_0$ and $B_1$.
\subsection{Octet $F_{1/2}$} First of all, we consider the mass
splitting of $F_{1/2}$ into multiplets of $\SU(2)$ defined by the
first term (\ref{MassQ3}) in (\ref{MassQ2}). In this case the
unitary field has the form
\[
Z=C\begin{bmatrix} 1/3 & 0 & 0\\
0 & 1/3 & 0\\
0 & 0 & -2/3
\end{bmatrix}
\]
and we can use the linear formula (\ref{GMOL}) at
$\alpha^\prime=\beta^\prime=\gamma^\prime=0$:
\begin{equation}\label{GMOL_1}
m=m_0+\alpha+\beta Y+\gamma\left[I(I+1)-\frac{1}{4}Y^2\right].
\end{equation}
Since at this step we neglect the mass splitting within multiplets,
therefore, from (\ref{GMOL_1}) we obtain particle masses containing
in the Tab.\,3.
%\begin{figure}{ht}
\begin{center}{\renewcommand{\arraystretch}{1.4}
\begin{tabular}{|c||c|c|c|l|}\hline
 & $I$ & $Y$ & $m_{exp}$ & $m_{th}$\\ \hline\hline
$\boldsymbol{\Xi}$ & $\frac{1}{2}$ & $-1$ & $1318$ &
$m_0+\alpha-\beta+\frac{1}{2}\gamma$\\
$\boldsymbol{\Sigma}$ & $1$ & $0$ & $1192$ & $m_0+\alpha+2\gamma$\\
$\boldsymbol{\Lambda}$ & $0$ & $0$ & $1115$ & $m_0+\alpha$\\
$\boldsymbol{N}$ & $\frac{1}{2}$ & $1$ & $939$ &
$m_0+\alpha+\beta+\frac{1}{2}\gamma$\\
\hline
\end{tabular}
}
\end{center}
\hspace{0.3cm}
\begin{center}\begin{minipage}{22pc}
{\small \textbf{Tab.\,3:} The hypercharge mass splitting of the octet $F_{1/2}$.}
\end{minipage}
\end{center}
%\end{figure}
In the Tab.\,3 $\boldsymbol{N}$ is the nucleon doublet
($\boldsymbol{N}^+=\bP$, $\boldsymbol{N}^0=\bN$), and $\boldsymbol{\Sigma}$-triplet, $\boldsymbol{\Xi}$- and $\boldsymbol{N}$-doublets are defined as in the section 6.1. Excluding unknown
parameters, we come to the following relations between masses:
\[
m_\Xi+m_N=2m_0+2\alpha+\gamma=\frac{3}{2}m_\Lambda+\frac{1}{2}m_\sigma,\quad
m_\Xi+m_N=\frac{1}{2}(3m_\Lambda+m_\Sigma).
\]

On the other hand, since $m_0$ is the external parameter with
respect to $\SU(3)$-theory we assume that $m_0$ is described by the mass formula (\ref{MGY}) which defines a
relation between the mass and spin. Within the supermultiplet of
$\SU(3)$ \emph{the parameter $m_0$ is an average value of the all masses
corresponding to charge multiplets}. In case of the baryon octet
$F_{1/2}$ we have
\[
m_0=\frac{1}{4}(m_N+m_\Lambda+m_\Sigma+m_\Xi),
\]
where \emph{\textbf{basic mass terms}} $m_N$, $m_\Lambda$,
$m_\Sigma$, $m_\Xi$ are defined by the mass
formula (\ref{MGY}).

Coming to charge splitting of $F_{1/2}$, defined by the second term
(\ref{MassQ4}) in (\ref{MassQ2}), we use the full linear formula
(\ref{GMOL}). In this case the unitary field is described by
(\ref{UField}). Taking into account values of the $U$-spin, we
calculate theoretical masses of the all particles belonging to the
baryon octet $F_{1/2}$. The results are given in the Tab.\,4.
%\begin{figure}{ht}
\begin{center}{\renewcommand{\arraystretch}{1.4}
\begin{tabular}{|c||c|c|l|}\hline
 & $Q$ & $m_{exp}$ & $m_{th}$\\ \hline\hline
$\boldsymbol{\Xi}^-$ & $-1$ & $1320,8$ & $m_0+\alpha-\beta+\frac{1}{2}\gamma+\alpha^\prime+\beta^\prime+\frac{1}{2}\gamma^\prime$\\
$\boldsymbol{\Xi}^0$ & $0$ & $1314,3$ & $m_0+\alpha-\beta+\frac{1}{2}\gamma+\alpha^\prime+2\gamma^\prime$\\
\hline
$\boldsymbol{\Sigma}^-$ & $-1$ & $1197,1$ & $m_0+\alpha+2\gamma+\alpha^\prime+\beta^\prime+\frac{1}{2}\gamma^\prime$\\
$\boldsymbol{\Sigma}^0$ & $0$ & $1192,4$ & $m_0+\alpha+2\gamma+\alpha^\prime+2\gamma^\prime$\\
$\boldsymbol{\Sigma}^+$ & $1$ & $1189,4$ & $m_0+\alpha+2\gamma+\alpha^\prime-\beta^\prime+\frac{1}{2}\gamma^\prime$\\
\hline
$\boldsymbol{\Lambda}$ & $0$ & $1115,4$ & $m_0+\alpha+\alpha^\prime$\\
\hline
$\bN$ & $0$ & $939,5$ & $m_0+\alpha+\beta+\frac{1}{2}\gamma+\alpha^\prime+2\gamma^\prime$\\
$\bP$ & $1$ & $938,3$ & $m_0+\alpha+\beta+\frac{1}{2}\gamma+\alpha^\prime-\beta^\prime+2\gamma^\prime$\\
\hline
\end{tabular}
}
\hspace{0.3cm}
\begin{center}\begin{minipage}{22pc}
{\small \textbf{Tab.\,4:} The charge splitting of the octet $F_{1/2}$.}
\end{minipage}
\end{center}
\end{center}
%\end{figure}
\subsection{Octet $B_0$}
Let us consider now the first bosonic octet $B_0$. $B_0$ describes mesons of the spin 0. As in the case of $F_{1/2}$, the octet $B_0$ is defined within the regular representation $\Sym^0_{(1,1)}$, but in contrast to $F_{1/2}$ the condition (\ref{MassRel}) is not fulfilled in the case of $B_0$. Therefore, we must use here the quadratic Gell-Mann--Okubo mass formula (\ref{GMOQ}). At the first step we have the hypercharge mass splitting of $B_0$ into multiplets of $\SU(2)$ defined by the quadratic formula (\ref{GMOQ}) at $\alpha^\prime=\beta^\prime=\gamma^\prime=0$:
\begin{equation}\label{GMOQ_1}
m^2=m^2_0+\alpha+\beta Y+\gamma\left[I(I+1)-\frac{1}{4}Y\right].
\end{equation}
For the octet $B_0$ at this step we have the Tab.\,5.
%\begin{figure}{ht}
\begin{center}{\renewcommand{\arraystretch}{1.4}
\begin{tabular}{|c||c|c|c|l|}\hline
 & $I$ & $Y$ & $m_{exp}$ & $m_{th}$\\ \hline\hline
$\boldsymbol{\eta}$ & $0$ & $0$ & $549$ &
$m^2_0+\alpha$\\
$\boldsymbol{K}_1=\left(\boldsymbol{K}^-,\overline{\boldsymbol{K}}^0\right)$ & $\frac{1}{2}$ & $-1$ & $496$ & $m^2_0+\alpha-\beta+\frac{1}{2}\gamma$\\
$\boldsymbol{K}_2=\left(\boldsymbol{K}^0,\boldsymbol{K}^+\right)$ & $\frac{1}{2}$ & $1$ & $496$ & $m^2_0+\alpha+\beta+\frac{1}{2}\gamma$\\
$\boldsymbol{\pi}$ & $1$ & $0$ & $138$ &
$m^2_0+\alpha+2\gamma$\\
\hline
\end{tabular}
}
\hspace{0.3cm}
\begin{center}\begin{minipage}{22pc}
{\small \textbf{Tab.\,5:} The hypercharge mass splitting of the octet $B_{0}$.}
\end{minipage}
\end{center}
\end{center}
%\end{figure}
The charge multiplets $\boldsymbol{K}_1$ and $\boldsymbol{K}_2$ contain particles and, correspondingly, their antiparticles. Therefore, masses of these multiplets should be equal to each other. Hence it follows that $\beta=0$. From the Tab.\,5 we have
\[
m^2_\pi-m^2_\eta=2\gamma,\quad m^2_K-m^2_\eta=\frac{1}{2}\gamma.
\]
Whence
\[
3m^2_\eta+m^2_\pi=4m^2_K.
\]
As in the case of $F_{1/2}$, the external parameter $m_0$ is described by the mass formula (\ref{MGY}). In the case of $B_0$ we have
\[
m^2_0=\frac{1}{4}\left(2m^2_K+m^2_\eta+m^2\pi\right).
\]

Further, coming to charge splitting of $B_0$, defined by the second term (\ref{MassQ4}) in (\ref{MassQ2}), we use the full quadratic formula (\ref{GMOQ}). Taking into account values of the $U$-spin, we calculate theoretical masses of the all particles belonging to the meson octet $B_0$ (see Tab.\,6).
%\begin{figure}{ht}
\begin{center}{\renewcommand{\arraystretch}{1.4}
\begin{tabular}{|c||c|c|l|}\hline
 & $Q$ & $m_{exp}$ & $m_{th}$\\ \hline\hline
$\boldsymbol{\eta}$ & $0$ & $548,7$ & $m^2_0+\alpha+\alpha^\prime$\\
\hline
$\boldsymbol{K}^-$ & $-1$ & $493,8$ & $m^2_0+\alpha+\frac{1}{2}\gamma+\alpha^\prime+\beta^\prime+\frac{1}{2}\gamma^\prime$\\
$\overline{\boldsymbol{K}}^0$ & $0$ & $498,0$ & $m^2_0+\alpha+\frac{1}{2}\gamma+\alpha^\prime+\frac{3}{4}\gamma^\prime$\\
\hline
$\boldsymbol{K}^0$ & $0$ & $498,0$ & $m^2_0+\alpha+\frac{1}{2}\gamma+\alpha^\prime+\frac{3}{4}\gamma^\prime$\\
$\boldsymbol{K}^+$ & $1$ & $493,8$ & $m^2_0+\alpha+\frac{1}{2}\gamma+\alpha^\prime-\beta^\prime+\frac{1}{2}\gamma^\prime$\\
\hline
$\boldsymbol{\pi}^-$ & $-1$ & $139,6$ & $m^2_0+\alpha+2\gamma+\alpha^\prime+\beta^\prime+\frac{7}{4}\gamma^\prime$\\
$\boldsymbol{\pi}^0$ & $0$ & $135,0$ & $m^2_0+\alpha+2\gamma+\alpha^\prime+2\gamma^\prime$\\
$\boldsymbol{\pi}^+$ & $1$ & $139,6$ & $m^2_0+\alpha+2\gamma+\alpha^\prime-\beta^\prime+\frac{7}{4}\gamma^\prime$\\
\hline
\end{tabular}
}
\hspace{0.3cm}
\begin{center}\begin{minipage}{22pc}
{\small \textbf{Tab.\,6:} The charge splitting of the octet $B_{0}$.}
\end{minipage}
\end{center}
\end{center}
%\end{figure}
\subsection{Octet $B_1$}
The next bosonic octet $B_1$ describes mesons of the spin 1 (vector bosons). As in the case of $B_0$, the condition (\ref{MassRel}) is not fulfilled for the octet $B_1$. Therefore, in this case we must use the quadratic formula (\ref{GMOQ}). The hypercharge splitting of $B_1$ into multiplets of $\SU(2)$ is defined by the formula (\ref{GMOQ_1}). For the octet $B_1$ at this step we have the Tab.\,7.
%\begin{figure}{ht}
\begin{center}{\renewcommand{\arraystretch}{1.4}
\begin{tabular}{|c||c|c|c|l|}\hline
 & $I$ & $Y$ & $m_{exp}$ & $m_{th}$\\ \hline\hline
${}^\ast\boldsymbol{K}_1=\left({}^\ast\boldsymbol{K}^-,{}^\ast\overline{\boldsymbol{K}}^0\right)$ & $\frac{1}{2}$ & $-1$ & $892$ &
$m^2_0+\alpha-\beta+\frac{1}{2}\gamma$\\
${}^\ast\boldsymbol{K}_2=\left({}^\ast\boldsymbol{K}^0,{}^\ast\overline{\boldsymbol{K}}^+\right)$ & $\frac{1}{2}$ & $1$ & $892$ & $m^2_0+\alpha+\beta+\frac{1}{2}\gamma$\\
$\boldsymbol{\varphi}$ & $0$ & $0$ & $782$ & $m^2_0+\alpha$\\
$\boldsymbol{\rho}$ & $1$ & $0$ & $770$ &
$m^2_0+\alpha+2\gamma$\\
\hline
\end{tabular}
}
\hspace{0.3cm}
\begin{center}\begin{minipage}{22pc}
{\small \textbf{Tab.\,7:} The hypercharge mass splitting of the octet $B_{1}$.}
\end{minipage}
\end{center}
\end{center}
%\end{figure}
As in the case of the octet $B_0$, the charge doublets ${}^\ast\boldsymbol{K}_1$ and ${}^\ast\boldsymbol{K}_2$ contain particles and, correspondingly, their antiparticles. Therefore, $\beta=0$. From the Tab.\,7 we have
\[
m^2_\rho-m^2_\varphi=2\gamma,\quad m^2_{{}^\ast K}-m^2_\varphi=\frac{1}{2}\gamma
\]
and
\[
3m^2_\varphi+m^2_\rho=4m^2_{{}^\ast K}.
\]
For the external parameter $m_0$ we have
\[
m^2_0=\frac{1}{4}\left(2m^2_{{}^\ast K}+m^2_\varphi+m^2_\rho\right).
\]
The charge splitting of $B_1$ leads to the Tab.\,8.
%\begin{figure}{ht}
\begin{center}{\renewcommand{\arraystretch}{1.4}
\begin{tabular}{|c||c|c|l|}\hline
 & $Q$ & $m_{exp}$ & $m_{th}$\\ \hline\hline
$\boldsymbol{\varphi}$ & $0$ & $782$ & $m^2_0+\alpha+\alpha^\prime$\\
\hline
${}^\ast\boldsymbol{K}^-$ & $-1$ & $891,66$ & $m^2_0+\alpha+\frac{1}{2}\gamma+\alpha^\prime+\beta^\prime+\frac{1}{2}\gamma^\prime$\\
${}^\ast\overline{\boldsymbol{K}}^0$ & $0$ & $895,81$ & $m^2_0+\alpha+\frac{1}{2}\gamma+\alpha^\prime+\frac{3}{4}\gamma^\prime$\\
\hline
${}^\ast\boldsymbol{K}^0$ & $0$ & $895,81$ & $m^2_0+\alpha+\frac{1}{2}\gamma+\alpha^\prime+\frac{3}{4}\gamma^\prime$\\
${}^\ast\boldsymbol{K}^+$ & $1$ & $891,66$ & $m^2_0+\alpha+\frac{1}{2}\gamma+\alpha^\prime-\beta^\prime+\frac{1}{2}\gamma^\prime$\\
\hline
$\boldsymbol{\rho}^-$ & $-1$ & $766,5$ & $m^2_0+\alpha+2\gamma+\alpha^\prime+\beta^\prime+\frac{7}{4}\gamma^\prime$\\
$\boldsymbol{\rho}^0$ & $0$ & $769$ & $m^2_0+\alpha+2\gamma+\alpha^\prime+2\gamma^\prime$\\
$\boldsymbol{\rho}^+$ & $1$ & $766,5$ & $m^2_0+\alpha+2\gamma+\alpha^\prime-\beta^\prime+\frac{7}{4}\gamma^\prime$\\
\hline
\end{tabular}
}
\hspace{0.3cm}
\begin{center}\begin{minipage}{22pc}
{\small \textbf{Tab.\,8:} The charge splitting of the octet $B_{1}$.}
\end{minipage}
\end{center}
\end{center}
%\end{figure}

\section{Summary}
We have presented a group theoretical approach for unification of spinor structure and internal symmetries based on the generalized definition of the spin and abstract Hilbert space. The main idea of this description is a correspondence between Wigner interpretation of elementary particles and quark phenomenologies of $\SU(N)$-models. This correspondence is realized on the ground of the abstract Hilbert space $\bsH^S\otimes\bsH^Q\otimes\bsH_\infty$. This description allows one to take a new look at the problem of mass spectrum of elementary particles. Complex momentum and underlying spinor structure play an essential role in this description. Complex momentum presents itself a quantum mechanical energy operator which generates basic energy levels described by the irreducible representations of the group $\SL(2,\C)$ (the group $\spin_+(1,3)$ in the spinor structure). Basic energy (mass) levels correspond to elementary particles which grouped into spin multiplets according to interlocking schemes and defined as vectors in the space $\bsH^S\otimes\bsH^Q\otimes\bsH_\infty$. The following mass (hypercharge and charge) splitting of the basic mass levels is generated by the action of $\SU(3)$ in $\bsH^S\otimes\bsH^Q\otimes\bsH_\infty$. The action of $\SU(3)$ is analogous to Zeeman effect in atomic spectra and by means of $\SU(3)/\SU(2)$ supermultiplet reductions it leads to different mass levels within charge multiplets. Schematically, mass spectrum of a given supermultiplet of $\SU(3)$ in the $\SU(3)/\SU(2)$-reduction can be defined as follows. Basic mass levels are described by the formula
\[
m^{(s)}_i=\mu^0\left(l+\frac{1}{2}\right)\left(\dot{l}+\frac{1}{2}\right),\quad i=1,\,\ldots,\,N,
\]
where $s=|l-\dot{l}|$, $\mu^0$ is a minimal rest mass, $N$ is a number of charge multiplets of $\SU(2)$ in the given supermultiplet of the group $\SU(3)$. Masses of particles belonging to the supermultiplet are
\[
m_j=m_0+\alpha+\beta Y_j+\gamma\left[I_j(I_j+1)-\frac{1}{4}Y^2_j\right]
+\alpha^\prime-\beta^\prime
Q_j+\gamma^\prime\left[U_j(U_j+1)-\frac{1}{4}Q^2_j\right],\quad j=1,\,\ldots,\,M,
\]
where
\[
m_0\equiv\frac{m^{(s)}_1+m^{(s)}_2+\,\ldots\,+m^{(s)}_N}{N},
\]
$M$ is a number of particles incoming to the given supermultiplet, $Q_j$ and $Y_j$ are charges and hypercharges of the particles, $I_j$ and $U_j$ are isotopic spins.

At this point, all the quark phenomenology of $\SU(3)$-model is included naturally into this more general framework. It is of interest to consider $\SU(4)$ quark model within this scheme (mainly with respect to charmed baryons). However, as it mentioned in Introduction, $\SU(5)$ and $\SU(6)$ flavor symmetries are strongly broken due to large masses of $b$ and $t$ quarks. For that reason multiplets of flavor $\SU(5)$- and $\SU(6)$-models are not observed in nature. On the other hand, we have a wide variety of hypermultiplets in the flavor-spin $\SU(6)$-theory. It is of great interest to consider $\SU(6)/\SU(3)$ and $\SU(6)/\SU(4)$ hypermultiplet reductions within presented scheme, where $\SU(4)$ is a Wigner subgroup. It is of interest also to consider $\SU(6)\otimes\GO(3)$ model.

\end{document}